\documentclass[sigconf]{acmart}
\AtBeginDocument{%
  \providecommand\BibTeX{{%
    Bib\TeX}}}

 \copyrightyear{2023} 
\acmYear{2023} 
\setcopyright{acmlicensed}\acmConference[CCS '23]{Proceedings of the 2023 ACM SIGSAC Conference on Computer and Communications Security}{November 26--30, 2023}{Copenhagen, Denmark}
\acmBooktitle{Proceedings of the 2023 ACM SIGSAC Conference on Computer and Communications Security (CCS '23), November 26--30, 2023, Copenhagen, Denmark}
\acmPrice{15.00}
\acmDOI{10.1145/3576915.3623170}
\acmISBN{979-8-4007-0050-7/23/11}

\usepackage[T1]{fontenc}
\usepackage{amsmath,amsfonts}
\usepackage[inline]{enumitem}
\usepackage{algorithmic}
\usepackage{graphicx}
\usepackage{textcomp}
\usepackage{xcolor}
\usepackage{caption}

\def\BibTeX{{\rm B\kern-.05em{\sc i\kern-.025em b}\kern-.08em
    T\kern-.1667em\lower.7ex\hbox{E}\kern-.125emX}}

\usepackage{booktabs}   
\usepackage{subcaption} 

\usepackage{mathtools}
\usepackage{amsthm}
\usepackage{enumitem}
\usepackage{algorithm2e}

\usepackage{eufrak}

\theoremstyle{plain}
\newtheorem{theorem}{Theorem}
\newtheorem{lemma}[theorem]{Lemma}
\newtheorem{corollary}[theorem]{Corollary}
\newtheorem{proposition}[theorem]{Proposition}
\newtheorem{claim}{Claim}
\theoremstyle{definition}
\newtheorem{definition}{Definition}
\newtheorem{example}{Example}

\theoremstyle{remark}
\newtheorem*{remark}{Remark}

\theoremstyle{remark}
\newtheorem*{notation}{Notation}

\newcommand{\exec}{\rho}

\DeclarePairedDelimiter{\abs}{\lvert}{\rvert}
\newcommand{\integers}{\mathbb{Z}}
\newcommand{\Rats}{\mathbb{Q}}
\newcommand{\Reals}{\mathbb{R}}
\newcommand{\RealsInf}{\ensuremath{\Reals_\infty}}
\newcommand{\Nats}{\mathbb{N}}

\newcommand{\Lap}[1]{\mathsf{Lap}{(#1)}}

\newcommand{\euler}{e}
\newcommand{\eulerv}[1]{\ensuremath{\euler^{#1}}}
\newcommand{\cC}{\mathcal{C}}

\newcommand{\cX}{\mathcal{X}}

\newcommand{\cM}{\mathcal{M}}

\newcommand{\Prob}{\mathsf{Prob}}

\newcommand{\st}  {\mathbin{|}}
\newcommand{\rcomment}[1]{#1}
\newcommand{\set}[1]{\{#1\}}
\newcommand{\setpred}[2] {\set{#1\: |\: #2}}

\newcommand{\card}[1]{\mathbin{|}#1 \mathbin{|}}
\newcommand{\true}{\mathsf{true}}
\newcommand{\false}{\mathsf{false}}
\newcommand{\cA}{\mathcal{A}}

\newcommand{\rv}{\mathsf{r}}

\newcommand{\prbfn}[1] {\Prob[#1]}

\newcommand{\tuple}[1] {\langle #1 \rangle}
\newcommand{\s}[1] {\mathsf{#1}}

\newcommand{\pto} {\hookrightarrow}
\newcommand{\rmv}[1] {}
\newcommand{\dom} {\mathsf{dom}}

\newcommand{\diptext} {DiP}

\newcommand{\dipautos} {{\diptext} automaton}

\newcommand{\dipautop}{{\diptext}A}
\newcommand{\newdipautop} {{\diptext} automata}

\newcommand{\dipa} {{\diptext}A}
\newcommand{\qinit} {q_{\mathsf{init}}}
\newcommand{\inalph} {\Sigma}
\newcommand{\outalph} {\Gamma}
\newcommand{\parf} {P}
\newcommand{\vars} {X}
\newcommand{\svar} {\mathsf{insample}}
\newcommand{\rvar} {\mathsf{x}}
\newcommand{\rvarx} {\mathsf{x}}
\newcommand{\rvary} {\mathsf{y}}
\newcommand{\states}{Q}
\newcommand{\instates}{Q_{\mathsf{in}}}
\newcommand{\epsstates}{Q_{\mathsf{non}}}

\newcommand{\cnds} {{\mathcal G}}
\newcommand{\getest} {\svar \geq \rvar}
\newcommand{\lttest} {\svar < \rvar}
\newcommand{\getestx} {\svar \geq \rvarx}
\newcommand{\getestxi} {\svar \geq \rvarx_i}
\newcommand{\lttestx} {\svar < \rvarx}
\newcommand{\lttestxi} {\svar < \rvarx_i}
\newcommand{\getesty} {\svar \geq \rvary}

\newcommand{\defaut} {(\states, \outalph, \qinit, \vars, \parf, \transf)}
\newcommand{\emptystr}{\lambda}
\newcommand{\noinp}{\tau}

\newcommand{\pathprob}[1] {\mathsf{Pr}[#1]}
\newcommand{\ith}[2][i]{#2[#1]}
\newcommand{\subseq}[3]{#1[#2:#3]}
\newcommand{\suffix}[2][j]{#2[#1:]}
\newcommand{\psuffix}[2]{#2[#1:]}
\newcommand{\Detcond} {Determinism}

\newcommand{\Outcond} {Output Distinction}
\newcommand{\outcond} {output distinction}
\newcommand{\Initcond} {Initialization}
\newcommand{\initcond} {initialization}
\newcommand{\Noninpcond} {Non-input transition}

\newcommand{\trns}[1]  {\xrightarrow{#1}}
\newcommand{\defexec} {q_0 \trns{a_0,o_0} q_1 \trns{a_1,o_1} q_2 \cdots q_{n-1} \trns{a_{n-1},o_{n-1}} q_n}
\newcommand{\len}[1] {\card{#1}}
\newcommand{\fstst} {\mathsf{first}}
\newcommand{\lstst} {\mathsf{last}}
\newcommand{\tl} {\mathsf{tail}}
\newcommand{\inseq} {\mathsf{inseq}}
\newcommand{\outseq} {\mathsf{outseq}}
\newcommand{\trname} {\mathsf{trans}}
\newcommand{\stname} {\mathsf{state}}
\newcommand{\grdname}[1] {\mathsf{guard}{(#1)}}
\newcommand{\grd} {\mathsf{guard}}
\newcommand{\src} {\mathsf{src}}
\newcommand{\trg} {\mathsf{trg}}
\newcommand{\otpt} {\mathsf{out}}
\newcommand{\rn} {\rho}
\newcommand{\inpsq} {\sigma}
\newcommand{\outsq} {\gamma}
\newcommand{\pth} {\kappa}
\newcommand{\runinpout} {computation}
\newcommand{\Uruninpout} {Computation}
\newcommand{\runs} {\mathsf{Runs}}

\newcommand{\rangeauto} {\cA_{\s{range}}}

\newcommand{\ba}{{\bf a}}
\newcommand{\bb}{{\bf b}}
\newcommand{\transf}{\delta}

\newcommand{\lcycle}{\ensuremath{\mathsf{L}}-cycle}
\newcommand{\gcycle}{\ensuremath{\mathsf{G}}-cycle}

\newcommand{\agpath}{\ensuremath{\mathsf{AG}}-path}
\newcommand{\critical}{{leaking}}

\newcommand{\criticalpath}{{leaking path}}
\newcommand{\criticalcycle}{{leaking cycle}}
\newcommand{\criticalpair}{{leaking pair}}

\newcommand{\Criticalcycle}{{Leaking cycle}}
\newcommand{\Criticalpair}{{Leaking pair}}

\newcommand{\violatingc}{{disclosing cycle}}
\newcommand{\violatingp}{{privacy violating path}}
\newcommand{\Violatingc}{{Disclosing cycle}}
\newcommand{\Violatingp}{{Privacy violating path}}

\newcommand{\execl}[1] {q_0 \trns{a_0,o_0} q_1 \trns{a_1,o_1}  q_2 \cdots q_{#1-1} \trns{a_{#1-1},o_{#1-1}} q_{#1}}

\newcommand{\execlb}[1] {q_0 \trns{b_0,o_0} q_1 \trns{b_1,o_1}  q_2 \cdots q_{#1-1} \trns{b_{#1-1},o_{#1-1}} q_{#1}}

\newcommand{\abst}{\ensuremath{\mathsf{abstract}}}

\newcommand{\absexec}{\eta}
\newcommand{\inalphaseq}{\alpha}

\newcommand{\inbetaseq}{\beta}
\newcommand{\inalpha}{a}
\newcommand{\inbeta}{b}
\newcommand{\outgammaseq}{\gamma}

\newcommand{\wt}{\mathsf{wt}}
\newcommand{\weight}[1]{\wt(#1)}

\newcommand{\newd}{\mathfrak{D}}

\newcommand{\toolaut}{{\sf DiPAut}}
\newcommand{\CheckDP}{{\sf CheckDP}}
\newcommand{\PSI}{{\sf PSI}}
\makeatletter
\newcommand{\removelatexerror}{\let\@latex@error\@gobble}
\makeatother

\newcommand*{\AppendixTrue}{}

\newcommand{\rc}{\rcomment{This paragraph needs to be rewritten as this is a copy of LICS21:}}

\newcommand{\augs}{\ensuremath{\mathsf{aug}}}
\newcommand{\aug}[1]{\ensuremath{\augs(#1)}}

\newcommand{\restr}[2]{\ensuremath{{#1}|_{#2}}}
\newcommand{\intersect}{\ensuremath{\mathbin{\cap}}}
\newcommand{\union}{\ensuremath{\mathbin{\cup}}}

\newcommand{\Rvars}{\mathsf{stor}}
\newcommand{\defautaug} {(\aug{\states},\outalph, \aug{\qinit}, \vars, \aug{\parf}, \aug{\transf})}

\newcommand{\lt}{{\mathsf{lt}}}

\newcommand{\eqs}{{\mathsf{eq}}}
\newcommand{\avars}{\mathsf{assignv}}
\newcommand{\navars}{\mathsf{nonassignv}}
\newcommand{\svars}{\mathsf{sm\_vars}}
\newcommand{\lvars}{\mathsf{lg\_vars}}
\newcommand{\smvars}{\mathsf{smallv}}
\newcommand{\lgvars}{\mathsf{largev}}
\newcommand{\gcyclevars}{\mathsf{gcycle\_vars}}
\newcommand{\lcyclevars}{\mathsf{lcycle\_vars}}
\newcommand{\id}{\mathsf{id}}

\newcommand{\proj}[1]{\mathsf{proj}(#1)}

\newcommand{\after}{\mathsf{after}}
\newcommand{\before}{\mathsf{before}}
\newcommand{\ltb}{\lt_\before}

\newcommand{\lta}{\lt_\after}

\newcommand{\sM}{\ensuremath{\mathsf{M}}}
\newcommand{\used}{\ensuremath{\mathsf{usedv}}}

\newcommand{\val}{\ensuremath{\eta}}

\newcommand{\fsttrns} {\mathsf{fst\_trans}}
\newcommand{\inp}{\mathsf{inp}}
\newcommand{\low}{\ensuremath{\ell}}
\newcommand{\up}{\ensuremath{u}}
\newcommand{\lavert}{\ensuremath{\mathsf{lastassign}}}

\newcommand{\lcyclevertex}{\mathsf{lcycle\_node}}
\newcommand{\gcyclevertex}{\mathsf{gcycle\_node}}

\newcommand{\pathtwo}[2]{\pth_{#1,#2}}
\newcommand{\ignore}[1]{}
\newcommand{\pspace}{\textsf{PSPACE}}
\newcommand{\Alg}{\mathsf{Alg}}

\newcommand{\PLY}{\textsc{PLY}}
\newcommand{\igraph}{\textsc{igraph}}
\newcommand{\pyperf}{\textsc{pyperf}}
\newcommand{\Range}{\textsc{Range}}
\newcommand{\NumericRangeOne}{\textsc{Num-Range-1}}
\newcommand{\NumericRangeTwo}{\textsc{Num-Range-2}}
\newcommand{\TRangeTwo}{\textsc{Two-Range-2}}
\newcommand{\TRangeOne}{\textsc{Two-Range-1}}
\newcommand{\minmax}{\textsc{Min-Max}}
\newcommand{\SVT}{\textsc{SVT}}
\newcommand{\Spse}{\textsc{Num-Sparse}}
\newcommand{\DC}{\textsc{DC-Example}}
\newcommand{\LC}{\textsc{LC-Example}}
\newcommand{\Noisy}{\textsc{NoisyMax}}

\newcommand{\timeouta}{T.O.}

\newcommand{\resultdc}{DC}
\newcommand{\resultpv}{PV}
\newcommand{\resultlc}{LC}
\newcommand{\resultlp}{LP}

\newcommand{\memoryerror}{M.E.}

\usepackage{tikz}
\usetikzlibrary{shapes}
\usetikzlibrary{shapes.multipart}
\usetikzlibrary{automata}
\usetikzlibrary{positioning}
\usetikzlibrary{arrows}
\def\mystrut{\vrule height .2cm depth .2cm width 0pt}
\tikzset{
         node distance=3.3cm,
         mynonstate/.style={
           rectangle split,
           rounded corners=0.4cm,
           rectangle split parts=2,
           draw,
           semithick,
           minimum size=1.4cm
         },
         myinstate/.style={
           circle split,
           draw,
           semithick,
         },
         myedgelabel/.style={
           rectangle split,
           rectangle split parts=2
         },
         initial text={},
         every edge/.style={
           draw,
           ->,>=stealth',
           auto,
           semithick
         }
}

\begin{document}

\author{Rohit Chadha}
\email{chadhar@missouri.edu}
\affiliation{%
  \institution{University of Missouri}
  \city{Columbia}
  \state{Missouri}
  \country{USA}
  \postcode{65201}
}

\author{A. Prasad Sistla}
\email{sistla@uic.edu}
\affiliation{%
  \institution{University of Illinois at Chicago}
  \city{Chicago}
  \state{Illinois}
  \country{USA}
  \postcode{60607}
}

\author{Mahesh Viswanathan}
\email{vmahesh@illinois.edu}
\affiliation{%
  \institution{University of Illinois at Urbana-Champaign}
  \city{Urbana}
  \state{Illinois}
  \country{USA}
  \postcode{61820}
}

\author{Bishnu Bhusal}
\email{bhusalb@mail.missouri.edu}
\orcid{0000-0001-7522-5878}
\affiliation{%
  \institution{University of Missouri}
  \city{Columbia}
  \state{Missouri}
  \country{USA}
  \postcode{65201}
}

\title{Deciding Differential Privacy of Online Algorithms with Multiple Variables}
\begin{abstract}

We consider the problem of checking the differential privacy of \emph{online} randomized algorithms that process a stream of inputs and produce outputs corresponding to each input. This paper generalizes  
an automaton model called {\newdipautop}~\cite{ChadhaSV21} to describe such algorithms by allowing multiple real-valued storage variables. A {\dipautos} is a parametric automaton whose behavior depends on the privacy budget $\epsilon$. 
An automaton $\cA$ will be said to be differentially private if, for some $\newd$, the automaton is $\newd\epsilon$-differentially private for all values of  $\epsilon>0$. We identify a precise characterization of the class of all differentially private {\newdipautop}. We show that the problem of determining if a given {\dipautos} belongs to this class is {\pspace}-complete. Our {\pspace} algorithm also computes a value for $\newd$ when the given automaton is differentially private. 
The algorithm has been implemented,  and experiments demonstrating its effectiveness are presented.

\end{abstract}



\maketitle

\section{Introduction}

Differential privacy~\cite{dmns06,DR14} is a popular requirement that is demanded of algortihms that analyze data containing sensitive personal information of individuals. A data analysis that meets the high bar of differential privacy guarantees the privacy of individuals. However, ensuring differential privacy is difficult, subtle and error-prone --- relatively minor tweaks to correct algorithms can lead to the loss of privacy as demonstrated by the examples in~\cite{DNRRV09,lyu2016understanding}. Though the problem of checking the differential privacy of a program is in general undecidable~\cite{BartheCJS020}, the importance of the problem has led to extensive investigation in the last 15 years; see Section~\ref{sec:related} for a short overview of work in this space.

In this paper, we look at the problem of verifying the differential privacy of online algorithms. An online algorithm is one that processes an unbounded (but finite) stream of inputs, samples from distributions, and produces outputs in response to the inputs. The stream of inputs is a sequence of real numbers that are answers to queries to a database. 
A novel approach using automata to describe and study such algorithms was proposed in ~\cite{ChadhaSV21}.
It was shown that checking differential privacy of algorithms described by such automaton is in linear time. Remarkably the verification procedure in~\cite{ChadhaSV21} checks some properties of the underlying graph of the automaton and does not explicitly reason about probabilities. However, the automaton model in~\cite{ChadhaSV21} has one serious limitation --- only one storage variable is available, and hence only one previously sampled value can be remembered.



\subsubsection*{Contributions}
We extend the line of research initiated in~\cite{ChadhaSV21} by generalizing the automata model in~\cite{ChadhaSV21} to allow for multiple real-valued storage variables. A {\dipautos} ({\dipa} for short)~\footnote{Even though the automata model in this paper has the same name as the one in~\cite{ChadhaSV21}, the generalization significantly extends the expressive power of the model.} is a parametric automaton (depending on privacy budget $\epsilon$) with finitely many control states that process an unbounded (but finite) stream of real values that represent answers to queries asked of a database. A {\dipa} can sample real values from Laplace distributions whose mean may depend on the value read, and {\dipa} has finitely many real-valued variables in which they can store values they sample in each step (which in turn depend on the input read). Transitions depend on the current control state, the values stored, and the input read, which influences the values sampled. In response to an input, they produce an output that is either a symbol from a finite set or a real number.

We show that, even in the case of automata with multiple storage variables, the problem of determining whether a given {\dipa} $\cA$ is $\newd\epsilon$-differentially private for some constant $\newd>0$ (independent of $\epsilon$) and all $\epsilon>0$, can be reduced to checking graph-theoretic conditions. These conditions demand the absence of certain paths, cycles, and interactions among them. However, unlike the single variable automata case~\cite{ChadhaSV21}, these paths and cycles cannot be captured only by considering the underlying graph of the automata. Instead, we use an auxiliary graph to capture these undesirable paths and cycles precisely. This is a non-trivial extension of~\cite{ChadhaSV21}; for a more detailed comparison with~\cite{ChadhaSV21}, see Section~\ref{sec:related}.
%
An automaton $\cA$ is said to be \emph{well-formed} if it does not have any of these undesirable paths or cycles. We show that a well-formed {\dipa} is differentially private; thus, well-formedness is a sufficient condition to guarantee privacy. 
Conversely, we show that if additionally, for every state of $\cA$, the transitions of $\cA$ from that state have distinct outputs (called \emph{output distinct}), then well-formedness is also necessary to guarantee differential privacy. In other words, a {\dipa} $\cA$, having distinct outputs on transitions from any state, that is differentially private is well-formed. These proofs of necessity and sufficiency require novel ideas that are a significant extension of the techniques presented in~\cite{ChadhaSV21}; once again see Section~\ref{sec:related} for more details.

Next, we show that there is a {\pspace} algorithm
that checks if a {\dipa} $\cA$ is well-formed. This algorithm additionally computes a value for $\newd$ that shows that $\cA$ is $\newd\epsilon$-differentially private for all $\epsilon$. We also show that checking differential privacy of output-distinct {\dipa} is {\pspace}-hard, thus establishing the optimality of our verification algorithm.

We have implemented our algorithm in a tool called {\toolaut}. Our experiments show that the approach scales and that our algorithm produces known estimates for $\newd$.  It successfully proves differential privacy and identifies  violations of privacy in various examples. The tool is evaluated for scalability with respect to both the number of states and variables. Despite the {\pspace}-hardness, the tool is able to perform well in our experiments.  We compare {\toolaut} with {\CheckDP}~\cite{CheckDP}, a state-of-the-art tool to check differential privacy. {\toolaut} significantly outperforms {\CheckDP} in all our experiments.  The tool {\toolaut} is available to download at~\cite{DipAut}.

\subsubsection*{Organization}

The rest of the paper is organized as follows. Section~\ref{sec:prelims} introduces basic notation and definitions used in the paper. Our model of {\dipautos} extended with multiple variables is introduced in Section~\ref{sec:dipauto}. Section~\ref{sec:decidability} defines well-formed {\dipautop}, which is a (almost) precise characterization of differentially private automata. We show that well-formed automata are differentially private in Section~\ref{sec:sufficiency}; and show that checking well-formedness is {\pspace}-complete.
Section~\ref{sec:necessity} shows that differentially private automata that have distinct outputs on transitions are well-formed. {\pspace}-hardness of checking differential privacy is also presented in this section. Experimental results are presented in Section~\ref{sec:experiments}. Closely related work is discussed in Section~\ref{sec:related}.  We discuss on the restrictions placed on the automata and the adjacency relations used in the paper. Finally we present our conclusions (Section~\ref{sec:conclusions}).

\ifdefined\short

{
For lack of space reasons, some proofs are omitted and can be found in ~\cite{ChadhaSVB23}.}
\else
\rcomment{The missing proofs are in the accompanying Appendix. This is the author’s version of the paper. It is posted here for your personal use. Not for redistribution. The definitive version will be published in the Proceedings of the annual ACM Computer and Communications Security Conference (CCS' 2023).}
\fi

\section{Preliminaries}
\label{sec:prelims}

\sloppy

The definitions and notations in this section are borrowed from ~\cite{ChadhaSV21}.  Let $\Nats, \integers,\Rats,\Rats^{\geq 0}, \Reals, \Reals^{>0}$ denote the set of natural numbers, integers, rational numbers, non-negative rationals, real numbers, and positive real numbers, respectively. In addition, $\RealsInf$ will denote the set $\Reals \cup \set{-\infty,\infty}$, where $-\infty$ is the smallest and $\infty$ is the largest element in $\Reals_\infty$. For a real number $x \in \Reals$, $\card{x}$ denotes its absolute value.
\subsubsection*{Sequences}  For a set $\inalph$, $\inalph^*$ denotes the set of all finite sequences/strings over $\inalph$. We use $\emptystr$ to denote the empty sequence/string over $\inalph$. For two sequences/strings $\rho,\sigma \in \inalph^*$, we use their juxtaposition $\rho\sigma$ to indicate the sequence/string obtained by concatenating them in order. Consider $\sigma = a_0a_1\cdots a_{n-1} \in \inalph^*$ (where $a_i \in \inalph$). We use $\len{\sigma}$ to denote its length $n$ and use $\ith{\sigma}$ to denote its $i$th symbol $a_i$. The substring $a_ia_{i+1}\cdots a_{j-1}$ from position $i$ (inclusive) to $j$ (not inclusive) will be denoted as $\subseq{\sigma}{i}{j}$; if $j \leq i$ then $\subseq{\sigma}{i}{j} = \emptystr$. Thus, $\subseq{\sigma}{0}{\len{\sigma}} = \sigma$. The suffix starting at position $j$ will be denoted as $\suffix{\sigma}$, i.e., $\suffix{\sigma} = \subseq{\sigma}{j}{\len{\sigma}}$. For any partial function $f\::A \pto B$, where $A,B$ are some sets, we let $\dom(f)$ be the set of $x\in A$ such that $f(x)$ is defined.



\subsubsection*{Laplace Distribution}
Differential privacy mechanisms often add noise by sampling values from the \emph{Laplace distribution}. The distribution, denoted $\Lap{k, \mu}$, is parameterized by two values: $k \geq 0$ which is called the scaling parameter, and $\mu$ which is the mean. The probability density function of $\Lap{k, \mu}$, denoted $f_{k,\mu}$, is given by
$
f_{k,\mu}(x) = \frac{k}{2}\eulerv{-k\card{x-\mu}},
$ where $\euler$ is the Euler constant.



\subsubsection*{Differential Privacy} 
Differential privacy~\cite{dmns06} is a framework that enables statistical analysis of databases containing sensitive, personal information of individuals while ensuring that the privacy of individuals is not adversely affected by the results of the analysis. In the differential privacy framework, 
a randomized algorithm, $M$, called the \emph{differential privacy mechanism}, 
mediates the interaction between a (possibly dishonest) data analyst asking queries and a database $D$ responding with answers. 
Queries are  deterministic functions and typically include aggregate questions about the data, like the mean etc.
In response to such a sequence of queries, $M$ responds with a series of answers computed using the actual answers from the database and random sampling, resulting in \lq\lq{noisy}\rq\rq\ answers. 
Thus, $M$ provides privacy at the cost of accuracy. Typically, $M$'s  noisy  response depends on a \emph{privacy budget} $\epsilon > 0$. 

Differential privacy captures the privacy guarantees for individuals whose information is in the database $D$. For an individual $i$, let $D \setminus \set{i}$ denote the database where $i$'s information has been removed.
A secure mechanism $M$ ensures that for any individual $i$ in $D$, and any sequence of possible outputs $\overline{o}$, the probability that $M$ outputs $\overline{o}$ on a sequence of queries is approximately the same whether the interaction is with the database $D$ or with $D \setminus \set{i}.$
To capture this definition formally, we need to characterize the inputs on which $M$ is required to behave similarly.
Inputs to a differential privacy mechanism can be seen as answers from the database to a sequence of queries asked by the data analyst. If queries are aggregate queries, then answers to $q$ on $D$ and $D \setminus \set{i}$ (for individual $i$) are likely to be away by at most $1$.~\footnote{The difference in general can be bounded by a constant $\Delta.$} 
This intuition leads to the following 
often-used definition of \emph{adjacency} that characterizes inputs on which the differential privacy mechanism $M$ is expected to behave similarly; for example this definition is used in SVT~\cite{DNRRV09,lyu2016understanding,DR14, AH18,DingWWZK18} and NumericSparse~\cite{DR14}.\footnote{Please see the discussion of SVT on pages 56 and 57 of ~\cite{DR14} and its description on pages 58, 62, and 64. For simplicity, it is assumed that these queries are $1$-sensitive.  So, by considering SVT as an algorithm that works directly on the sequence of the outputs of queries, we get naturally the adjacency relation used here. } We assume that at each step, the differential privacy mechanism either gets a real number as input (answer to an aggregate query) or is asked to respond without an answer from the database which is encoded as $\noinp$. 
\begin{definition}
\label{def:adjacency}
Sequences $\rho,\sigma \in (\Reals\cup \set{\noinp})^*$ are \emph{adjacent} if $\len{\rho} = \len{\sigma}$ and for each $i \leq \len{\rho}$ (a) $\ith{\rho} \in \Reals$ iff $\ith{\sigma} \in \Reals$ and (b) if $\ith{\rho} \in \Reals$ then $\card{\ith{\rho} - \ith{\sigma}} \leq 1$.
\end{definition}

We are now ready to formally define the notion of privacy which uses Definition~\ref{def:adjacency}. In response to a sequence of inputs, a differential privacy mechanism produces a sequence of outputs from the set (say) $\outalph$. Since a differential privacy mechanism $M$ is a randomized algorithm, it will induce a probability distribution on $\outalph^*$. 

\begin{definition}[$\epsilon$-differential privacy]
\label{def:diff-priv}
A randomized algorithm $M$ with input in $(\Reals \cup \set{\noinp})^*$ and output in $\outalph^*$ is said to be \emph{$\epsilon$-differentially private} if for all measurable sets $S \subseteq \outalph^*$ and adjacent $\rho,\sigma \in \Reals^*$ (Definition~\ref{def:adjacency}),
\[
\prbfn{M(\rho) \in S} \leq \eulerv{\epsilon}\, \prbfn{M(\sigma) \in S}.
\]
\end{definition}

\begin{example}
\label{ex:range-query-pgm}

\RestyleAlgo{boxed} 
\begin{algorithm}
\DontPrintSemicolon
\SetAlgoLined

\KwIn{$q[1:N]$}
\KwOut{$out[1:N]$}
\;
$\s{low} \gets \Lap{ \frac{\epsilon}{4} , T_\ell}$\;
$\s{high} \gets \Lap{ \frac{\epsilon}{4}, T_u}$\;
\For{$i\gets 1$ \KwTo $N$}
{
    $\rv\gets \Lap{\frac{\epsilon}{4} , q[i]}$\;
    \uIf{$(\rv \geq \s{low}) \wedge (\rv < \s{high})$}{
      $out[i] \gets \bot$}
    \uElseIf{$(\rv \geq \s{low}) \wedge (\rv \geq \s{high})$}{
      $out[i] \gets \top_1$\;
      exit
    } \ElseIf{$(\rv < \s{low}) \wedge (\rv < \s{high})$}{
      $out[i] \gets \top_2$\;
      exit
    }

}
\caption{Range query algorithm}
\label{fig:range-query}
\end{algorithm}

Consider the following problem. Given a sequence of answers to queries (array $q[1:N]$) and an interval $[T_\ell,T_u)$ given by thresholds $T_\ell$ and $T_u$, determine the first time a query answer lies outside this interval; indicate (through the output) whether the query answer is $\geq T_u$ or $\leq T_\ell$ at this point. A differentially private algorithm to solve this problem is shown as Algorithm~\ref{fig:range-query}. The algorithm starts by adding noise to both $T_\ell$ and $T_u$ to get a perturbed interval defined by numbers $\s{low}$ and $\s{high}$. After that the algorithm perturbs each query answer and stores the result in $\rv$, and then checks if $\rv$ lies between $\s{low}$ and $\s{high}$. If it does, the algorithm outputs $\bot$ and processes the next query answer. Otherwise, if $\rv$ is larger than both $\s{low}$ and $\s{high}$ it outputs $\top_1$ and stops. On the other hand, if $\rv$ is less than both $\s{low}$ and $\s{high}$ then it outputs $\top_2$ and halts. The algorithm's behavior depends on the value of $\epsilon$. It can be shown that for each value of $\epsilon$, the algorithm for that value of $\epsilon$ is $\epsilon$-differentially private.
\end{example}

\section{\dipautop}
\label{sec:dipauto}

{\diptext} ({\bf \textsf{Di}}fferentially {\bf \textsf{P}}rivate) automata ({\dipa}s for short) are an automata-based model introduced in~\cite{ChadhaSV21} to describe some differential privacy mechanisms. They process an input string $\sigma \in (\Reals \cup \set{\noinp})^*$ by sampling values from the Laplace distribution, using real variables to store information during the computation, and producing a sequence of outputs. The model introduced in~\cite{ChadhaSV21} had only \emph{one} storage variable. In this paper, we generalize this model naturally to allow \emph{multiple} real-valued storage variables. However, as discussed in Section~\ref{sec:related}, both the characterization of differentially private algorithms described by them and the proofs of decidability are a non-trivial extension of the single variable model.

\subsection{Syntax}
A {\dipautos} is a \emph{parametric} automaton whose behavior depends on a parameter $\epsilon$ (the privacy budget). It has finitely many control states and finitely many real-valued variables $\rvarx_1, \rvarx_2, \ldots \rvarx_k$ that are used to store information during the computation. At each step, the automaton freshly samples two real values from Laplace distributions whose parameters depend on $\epsilon$, and these sampled values are stored in the (additional) variables $\svar$ and $\svar'$. Given an input $\sigma \in (\Reals \cup \set{\noinp})^*$, a {\dipa} does the following in each step.
\begin{enumerate}
\item Two values are drawn from the distributions $\Lap{d\epsilon, \mu}$ and $\Lap{d'\epsilon,\mu'}$ and stored in the variables $\svar$ and $\svar'$, respectively. The scaling factors $d,d'$ and means $\mu, \mu'$ of these distributions depend on the current state.
\item The states of the automaton are partitioned into \emph{input} states and \emph{non-input} states. At a non-input state, the automaton expects to read $\noinp$ from the input. On the other hand, at an input state, it expects to read a real number, say $a$, and it updates $\svar$ and $\svar'$ by adding $a$ to them. {The properties of the Laplacian distribution imply that the distribution of $\svar+ a$ ($\svar'+a$) is the same as the distribution of  $\Lap{d\epsilon, \mu+a}$ ($\Lap{d\epsilon, \mu'+a}$ respectively).}
\item A transition changes the control state and outputs a value. The value output could either be a symbol from a finite set 
or one of the two real numbers $\svar$ and $\svar'$ that are sampled in this step. 
At an input state, the transition is guarded by a Boolean condition that depends on the result of comparing the sampled value $\svar$ with the stored values $\rvarx_i$ ($1\leq i\leq k$). It is possible that for certain values of $\rvarx_i$ ($1\leq i\leq k$) and $\svar$, no transition is enabled from the current state. In such a case, the computation ends. 
\item Finally, the automaton may choose to store the sampled value $\svar$ in any of the variables $\rvarx_i$ ($1\leq i\leq k$).
\end{enumerate}
We now formally define {\dipautos} capturing the above intuition. First, some necessary notation. Let $\cnds'$ be the set of constraints defined as $\cnds' = \setpred{\getestxi}{1\leq i \leq k}
\union
\setpred{\lttestxi}{1\leq i \leq k}.$
 Let $\cnds''$ be the set of conditions formed by taking conjunctions of two or more constraints in $\cnds'$ such that both $\getestxi$ and $\lttestxi$ don't appear for any $1 \leq i \leq k$. Finally, let $\cnds = \set{\true} \cup \cnds' \cup \cnds''$; these are the constraints that \emph{guard} transitions in a {\dipa}.\footnote{We could also allow guards of the form $\svar>\rvar_i$ and $\svar\leq \rvar_i.$ However, we
chose to keep the presentation simple. As all random variables in a {\dipa} are noisy, the equality happens with probability $0$.
}
\begin{definition}[{\dipa}]
\label{def:dipa}
A \emph{\dipautos} $\cA = \defaut$ where
\begin{itemize}
\item $\states$ is a finite set of states partitioned into two sets: the set of input states $\instates$ and the set of non-input states $\epsstates$,
\item $\outalph$ is a finite output alphabet,
\item $\qinit \in \states$ is the initial state,
\item $\vars = \set{\svar,\svar'}\cup \setpred{\rvarx_i}{1\leq i\leq k}$ is the set of variables; we will use $\Rvars = \setpred{\rvarx_i}{1\leq i\leq k}$ to denote the \emph{storage} variables,
\item $\parf : \states \to \Rats^{\geq 0} \times \Rats \times \Rats^{\geq 0} \times \Rats$ is the parameter function that assigns to each state a 4-tuple $(d,\mu,d',\mu')$, where $\svar$ is sampled from $\Lap{d\epsilon,\mu}$ and $\svar'$ is sampled from $\Lap{d'\epsilon,\mu'}$,
 \item and $\transf: (\states \times \cnds) \pto (\states \times (\outalph \cup \set{\svar,\svar'}) \times \set{\true,\false}^k)$ is the transition (partial) function that given a current state and the result of comparing each $\rvarx_i$ ($1\leq i\leq k$)  with $\svar$, determines the next state, the output, and whether the variables $\rvarx_i$ should be updated to store $\svar$. The output could either be a symbol from $\outalph$ or the values $\svar$ and $\svar'$ that were sampled.
\end{itemize}
In addition, the transition function $\transf$  satisfies the following two conditions.

\vspace*{0.05in}
\noindent
{\bf \Detcond:} For any state $q \in \states$, if $\transf(q,c)$ and $\transf(q,c')$ are defined for $c,c'\in \cnds$ then either $c = c'$ or $c \wedge c'$ is unsatisfiable. That is, from any state, at most one transition is enabled at any time.

\rmv{
\vspace*{0.05in}
\noindent
{\bf \Outcond:} For any state $q \in \states$, for any distinct $c,c'\in\cnds$, if $\transf(q,c)$ is defined to be $(q_1,o_1,b_1)$ and $\transf(q,c')$ is defined to be $(q_2,o_2,b_2)$ then $o_1 \neq o_2$, i.e., distinct transitions from a state have different outputs. Further, output of at most one of the transitions, does not belongs to $\outalph$, i.e., no two transitions from $q$ can both output real values.

\vspace*{0.05in}
\noindent
}

\vspace*{0.05in}
\noindent
{\bf {\Noninpcond}s:} From any $q \in \epsstates$, if $\transf(q,c)$ is defined, then $c = \true$; that is, there is at most one transition from a non-input state which is always enabled.
\end{definition}

\begin{remark}
{Although $\svar'$ is never used in comparisons, it is nevertheless needed to model examples such as $\Spse$ (See~\cite{DR14}). $\svar'$ is often used in algorithms when we want to output the noisy input value in a differentially private fashion. Outputting $\svar$ instead of $\svar'$ can violate differential privacy, as $\svar$ may have been used in other comparisons: See the definition of {\violatingp} (Definition~\ref{def:violating} in Section~\ref{sec:decidability}); also ~\cite{lyu2016understanding}.}
\end{remark}
Before concluding this section, it is useful to introduce some notation and terminology for transitions. A quintuple $t = (q,c,q',o,b)$ denotes a transition of $\cA$ if $\transf(q,c) = (q',o,b)$, where $b = (b_1,b_2,\ldots b_k) \in \set{\true,\false}^k$. For such a transition, $\src(t) = q$ denotes the \emph{source}, $\trg(t) = q'$ the \emph{target}, $\otpt(t) = o \in \Gamma \cup \set{\svar,\svar'}$ the \emph{output}, and $\grd(t) = c$ the \emph{guard}. Based on the guard $c$ and the Booleans $b$, we can associate the following sets of variables with transition $t$.
\begin{align*}
\smvars(t) & = \setpred{\rvar \in \Rvars}{\getestx \mbox{ is a conjunct of }c}\\
\lgvars(t) &= \setpred{\rvar \in \Rvars}{\lttestx \mbox{ is a conjunct of }c}\\  
\used(t) &= \smvars(t) \cup \lgvars(t)\\
\avars(t) &= \setpred{\rvar_i}{b_i=\true}\\
\navars(t) &= \setpred{\rvar_i}{b_i=\false}
\end{align*}
Intuitively, $\smvars(t)$ ($\lgvars(t)$) are the storage variables that lower bound (upper bound) $\svar$ if the guard is satisfied; $\used(t)$ are the storage variables that are referenced in the guard of $t$; $\avars(t)$ are the variables that are set by $t$; and $\navars(t)$ are the variables that are left unchanged by $t$. For any $i$, if $\rvarx_i \in \avars(t)$ then $t$ sets $\rvarx_i = \svar$ during the transition and hence $t$ is an \emph{assignment transition} for variable $\rvarx_i$. Finally, if $\src(t) = q \in \instates$ then $t$ is said to be \emph{input transition} and if $q \in \epsstates$ then $t$ is a \emph{non-input transition}. 

\rmv{
{\rc} It is useful to classify transitions of a {\dipa} into different types. Consider a transition $\transf(q,c) = (q',o,b)$ where $b\:=(b_1,...,b_k)$. If $q \in \instates$ then it is an \emph{input transition} and if $q \in \epsstates$ then it is a \emph{non-input transition}. If $b_i = \true$, for any $i$, $1\leq i\leq k$, then the transition will set $\rvarx_i = \svar$. If $b_i\:=\true$,  then the transition is called an \emph{assignment transition} for the variable $\rvarx_i$; observe that $b_i$ can be $\true$, for multiple values of $i$. We call the transition an {\em assignment transition} if it is an assignment transition for at least one variable. On the other hand, if $b_i = \false$, for all $i$, $1\leq i\leq k$, the transition will be said to be a \emph{non-assignment} transition. A \emph{pure assignment} transition is an assignment transition with $c = \true$. 
}

\begin{example}
\label{ex:range-auto}

The differential privacy mechanism in Example~\ref{ex:range-query-pgm} can be modeled as a {\dipa}. This is shown in Figure~\ref{fig:range-auto}. We will use these conventions when drawing {\dipa}s in this paper. Input states will be represented as circles, while non-input states will be drawn as rectangles. The name of each state is written above the line, while the scaling factor $d$ and mean $\mu$ of the distribution used to sample $\svar$ is written below the line. The parameters $d'$ and $\mu'$ for sampling $\svar'$ are not shown in the figures, but will be mentioned in the caption and text when they are important; they are relevant only when $\svar'$ is output on a transition. Edges will be labeled with the guard of the transition, followed by the output, and a vector of Booleans to indicate which variables $\svar$ is stored in.

\begin{figure}
\begin{center}
\begin{tikzpicture}
\footnotesize
\node[mynonstate, initial] (q0) {\mystrut $q_0$ \nodepart{two} \mystrut $\frac{1}{4},\ 0$};
\node[mynonstate, right of=q0] (q1) {\mystrut $q_1$ \nodepart{two} \mystrut $\frac{1}{4},\ 1$};
\node[myinstate, above of=q1] (q2) {$q_2$ \nodepart{lower} $\frac{1}{4},\ 0$};
\node[myinstate, left of=q2] (q3) {$q_3$ \nodepart{lower} $\frac{1}{4},\ 0$};
\draw (q0) edge node[myedgelabel] {$\true,\: \bot$ \nodepart{two} $(\true,\false)$} (q1);
\draw (q1) edge node[myedgelabel, right] {$\true,\: \bot$ \nodepart{two} $(\false,\true)$} (q2);
\draw (q2) edge[loop right] node[myedgelabel] {$g_1,\: \bot$ \nodepart{two} $(\false,\false)$} (q2)
                 edge[bend right] node[above,myedgelabel] {$g_2,\: \top_1$ \nodepart{two} $(\false,\false)$} (q3)
                 edge[bend left] node[myedgelabel] {$g_3,\: \top_2$ \nodepart{two} $(\false,\false)$} (q3);
\end{tikzpicture}
\end{center}
\captionsetup{font=footnotesize,labelfont=footnotesize}
\caption{{\dipa} $\rangeauto$ modeling Algorithm~\ref{fig:range-query}. Threshold $T_\ell$ is set to $0$ (sampling mean of $\svar$ in $q_0$) and $T_u$ is set to $1$ (sampling mean of $\svar$ in $q_1$). The guards $g_1 = (\svar \geq \rvar_1) \wedge (\svar < \rvar_2)$, $g_2 = (\svar \geq \rvar_1) \wedge (\svar \geq \rvar_2)$, and $g_3 = (\svar < \rvar_1) \wedge (\svar < \rvar_2)$.}
\label{fig:range-auto}
\end{figure}
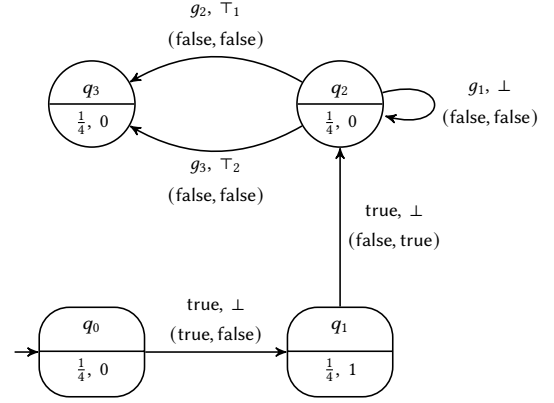

The working of $\rangeauto$ in Fig.~\ref{fig:range-auto} can be explained as follows. Since $\svar'$ is not output in any step, the parameters associated with sampling $\svar'$ are not reported. The thresholds $T_\ell$ and $T_u$ are hard-coded as $0$ and $1$, respectively, as the distribution means for the non-input states $q_0$ and $q_1$. The transition from $q_0$ to $q_1$ perturbs $T_\ell$ ($=0$) and sets this to variable $\rvar_1$; thus, $\rvar_1$ corresponds to the variable $\s{low}$ in Algorithm~\ref{fig:range-query}. The transition from $q_1$ to $q_2$ perturbs $T_u$ ($=1$) and stores it in $\rvar_2$. Thus, $\rvar_2$ corresponds to variable $\s{high}$ in Algorithm~\ref{fig:range-query}. State $q_2$ is an input state. Transitions from $q_2$ perturb the query answer given as input storing it in $\svar$, compare $\svar$ to the values stored in $\rvar_1$ and $\rvar_2$, and output the right value accordingly. State $q_3$ is a halting state where no transitions are enabled.

We conclude this example by illustrating the definitions associated with transitions. The transition $t$ from $q_0$ to $q_1$ can be denoted by the quintuple $(q_0,\true,q_1,\bot,(\true,\false))$. For $t$, we have $\src(t) = q_0$, $\trg(t) = q_1$, $\otpt(t) = \bot$, $\grd(t) = \true$, $\smvars(t) = \lgvars(t) = \used(t) = \emptyset$, $\avars(t) = \set{\rvar_1}$, and $\navars(t) = \set{\rvar_2}$. In this case $t$ is a non-input, assignment transition for variable $\rvar_1$. In contrast, the transition $t'$ from $q_2$ to itself, is an input transition that is not an assignment transition for any variable. Here we have $\smvars(t') = \set{\rvar_1}$, $\lgvars(t') = \set{\rvar_2}$, and $\used(t') = \set{\rvar_1,\rvar_2}$.
\end{example}

\subsection{Semantics}

An \emph{execution/run} of a {\dipa} $\cA = \defaut$, $\rn = t_0t_1\cdots t_{n-1}$, is a sequence of transitions $t_i$ such that for every $0 < i < n$, $\trg(t_{i-1}) = \src(t_i)$ (i.e., the sequence $\rn$ corresponds to a path in the ``graph'' of $\cA$). We extend the notation of length, the $i$th transition, sub-sequence and suffix from (general) sequences: thus, $\len{\rn} = n$, $\ith{\rn} = t_i$, $\subseq{\rn}{i}{j} = t_i\cdots t_{j-1}$ and $\suffix{\rn} = t_jt_{j+1}\cdots t_{n-1}$. We also extend the notion for source and target from transitions to a run --- $\src(\rn) = \src(t_0)$ and $\trg(\rn) = \trg(t_{n-1})$. Using the notation developed for transitions, $\grd(\ith{\rn})$ is the guard of the $i$th transition $t_i$ of $\rn$. A run $\rn$ is a \emph{cycle} if $\src(\rn) = \trg(\rn)$, i.e., the run begins and ends in the same state. Finally, given two runs $\rn_1$ and $\rn_2$ such that $\trg(\rn_1) = \src(\rn_2)$, $\rn_1\rn_2$ is the run which is the concatenation of $\rn_1$ followed by $\rn_2$.

Recall that transitions of {\dipa} $\cA$ compare values stored in the variables $\rvarx_i$ ($1 \leq i \leq k$) and $\svar$. Thus, to define the semantics of the {\dipa}, we need to make sure that the value of variable $\rvarx_i$ is defined before it is used in a comparison in the guard of a transition. Therefore, we make the technical assumption that on every run starting from the initial state $\qinit$, a variable is assigned a value before it is referenced in a guard. We assume that all {\dipa} $\cA$ considered in this paper are \emph{initialized} as defined formally below.

\vspace*{0.05in}
\noindent
{\bf \Initcond:}
We say that a {\dipa} $\cA = \defaut$ is {\em initialized} if for any run $\rn$ starting from the initial state $\qinit$ (i.e., $\src(\rn) = \qinit$), if $\grd(\ith{\rn})$ references variable $\rvarx_\ell$ (i.e., $\rvarx_\ell \in \used(\ith{\rn})$) then there is $j < i$ such that $\ith[j]{\rn}$ is an assignment transition for $\rvarx_\ell$ (i.e., $\rvarx_\ell \in \avars(\ith[j]{\rn})$).

We need to define one more concept associated with a run $\rho$. For any storage variable $\rvarx$ and position $j \in \set{0,1, \ldots \len{\rn}}$, the \emph{last position} when $\rvarx$ was assigned before $j$ is the maximum index $i < j$ such that $\rvarx$ was assigned on transition $\ith{\rn}$. More precisely,
\[
\lavert_\rn(\rvarx,j) = \max \setpred{i}{i < j,\ \rvarx \in \avars(\ith{\rn})}.~\footnote{As always $\max \emptyset = -\infty$ and $\min \emptyset = \infty$.}
\]
When the run $\rn$ is clear from the context, we will drop the subscript and simply refer to the last assigned position before $j$ for $\rvarx$ as $\lavert(\rvarx,j).$

To define the semantics of a {\dipa} $\cA$, we need to define the probability of ``executions''. But runs, as defined above, do not have all the information we need. For example, the real numbers read as input determine the values of $\svar$ and $\svar'$, which in turn determine whether a transition is enabled and what is stored in the variables. Next, on transitions where either $\svar$ or $\svar'$ are output, to define a meaningful measure space, we need to associate an interval $(v,w)$ in which the output value lies. Thus, we define when a run corresponds to a certain sequence of inputs and outputs.
\begin{definition}[{\Uruninpout}]
Consider {\dipa} $\cA = \defaut$ and a run $\rn$ of $\cA$. Let $\inpsq \in (\Reals \cup \set{\noinp})^*$ be an \emph{input sequence} and $\outsq \in (\outalph \cup (\RealsInf \times \RealsInf))^*$ be an \emph{output sequence}. We say that $\rn$ is a \emph{run on $\inpsq$ producing output $\outsq$} if the following conditions hold.
\begin{enumerate}
\item $\len{\rn} = \len{\inpsq} = \len{\outsq}$.
\item For any $i$, $\ith{\inpsq} = \noinp$ iff $\src(\ith{\rn}) \in \epsstates$. That is, symbol $\noinp$ is only read in non-input states.
\item For any $i$, $\ith{\outsq} \in \outalph$ iff $\otpt(\ith{\rn}) \in \outalph$. Further for such $i$, $\otpt(\ith{\rn}) =\ith{\outsq}$. That is, outputs in $\rn$ ``match'' outputs in $\outsq$, with the only difference being that when $\svar$ or $\svar'$ is output in $\rn$, the corresponding position in $\outsq$ is an interval $(v,w) \in \RealsInf^2$.
\end{enumerate}
When $\rn$ is a run on $\inpsq$ producing $\outsq$, the tuple $\pth = (\rn, \inpsq, \outsq)$ will be called a \emph{\runinpout}.
\end{definition}

For a {\runinpout} $\pth = (\rn,\inpsq,\outsq)$ of {\dipa} $\cA$, the suffix starting at position $j$ is $\suffix{\pth} = (\suffix{\rn}, \suffix{\inpsq}, \suffix{\outsq})$. Notice that $\suffix{\pth}$ (for any $j$) is also a {\runinpout} of $\cA$ since $\suffix{\rn}$ is a run on $\suffix{\inpsq}$ producing $\suffix{\outsq}$. 
{Also, we use length of $\pth$, $\len{\pth}$ to be  $\len{\rn}$ ($=\len{\inpsq} = \len{\outsq}$), the length of the run $\rn$.}

\subsubsection*{Probability of Computations}

We will now define what the probability of each {\runinpout} is. Recall that in each step, the automaton samples two values from Laplace distributions, and if the transition is from an input state, it adds the read input value to the sampled values and compares the result with the values stored in the variables $\rvarx_i$, $1\leq i\leq k$. The step also outputs a value, and if the value output is one of the two sampled values, the {\runinpout} requires it to belong to the interval that appears in the output sequence. The probability of such a transition thus is the probability of drawing a sample that satisfies the guard of the transition and (if the output is a real value) producing a number that lies in the interval in the output label. This intuition is formalized in a precise definition. 

Let us fix a {\runinpout} $\pth = (\rn,\inpsq,\outsq)$ of {\dipa} $\cA = \defaut$. Recall that $\Rvars = \setpred{\rvarx_i}{1\leq i\leq k}$. Since the parameters of the Laplace distribution that is used to sample $\svar$ and $\svar'$ depend on the privacy budget $\epsilon$, the probability of $\pth$ will also depend on $\epsilon$. In addition, the values stored in the variables $\rvarx_i \in \Rvars$ at the start of the computation also influence the behavior of $\cA$. Let $\eta: \Rvars \to \Reals$ be the \emph{evaluation} that defines the values of $\rvarx_i$, $1 \leq i \leq k$, initially. The probability of $\pth$ depends on both $\epsilon$ and $\eta$ and is denoted as $\pathprob{\epsilon,\eta,\pth}$. We define this inductively on $\len{\pth}$. For any $\epsilon$ and any {\runinpout} $\pth$ with $\len{\pth} = 0$, $\pathprob{\epsilon,\eta,\pth} = 1$. 

Let us now consider the case when $\len{\pth} > 0$. Before defining the probability in this case, we define the parameters that we will need. Let $\parf(\src(\ith[0]{\pth})) = (d,\mu,d',\mu')$. Define the value $a_0$ as follows --- if $\ith[0]{\inpsq} \in \Reals$ then $a_0 = \ith[0]{\inpsq}$, and if $\ith[0]{\inpsq} = \noinp$ then $a_0 = 0$. Next, let us define the values $\ell$ and $u$. If $\ith[0]{\outsq} \in \outalph$ then $\ell = -\infty$ and $u = \infty$. Otherwise, if $\ith[0]{\outsq} = (v,w)$ then $\ell = v$ and $u = w$. Finally, for a parameter $z$, let $\eta_z$ be the evaluation that modifies $\eta$ by setting all the variables assigned by $\ith[0]{\rn}$ to $z$. In other words,
\[
\eta_z(\rvarx) = \begin{cases}
                     \eta(\rvarx) & \mbox{if } \rvarx \in \navars(\ith[0]{\rn}) \\
                     z & \mbox{if } \rvarx \in \avars(\ith[0]{\rn})
                  \end{cases}
\]
We are now ready to define $\pathprob{\epsilon,\eta,\pth}$ based on whether $\otpt(\ith[0]{\rn}) = \svar'$ or not.

\vspace*{0.05in}
\noindent
{\bf Case $\otpt(\ith[0]{\rn}) = \svar'$:} Set $\ell' = \max \setpred{\eta(\rvarx)}{\rvarx \in \smvars(\ith[0]{\rn})}$ and $u' = \min \setpred{\eta(\rvarx)}{\rvarx \in \lgvars(\ith[0]{\rn})}$. Also define $p$ to be the probability that $\svar' \in (\ell,u) = (v,w) = \ith[0]{\outsq}$, i.e.,
\[
p = \int_\ell^u \frac{d'\epsilon}{2}\eulerv{-d'\epsilon\card{z-\mu'-a_0}}dz
\]
Then,
\[
\pathprob{\epsilon,\eta,\pth} = p\int_{\ell'}^{u'} \left(\frac{d\epsilon}{2}\eulerv{-d\epsilon\card{z-\mu-a_0}}\right) \pathprob{\epsilon,\eta_z,\suffix[1]{\pth}} dz.
\]

\vspace*{0.05in}
\noindent
{\bf Case $\otpt(\ith[0]{\rn}) \neq \svar'$:} In other words, either $\otpt(\ith[0]{\rn}) \in \outalph$ or $\otpt(\ith[0]{\rn}) = \svar$. In this case set $\ell' = \max (\setpred{\eta(\rvarx)}{\rvarx \in \smvars(\ith[0]{\rn})} \cup \set{\ell})$ and $u' = \min (\setpred{\eta(\rvarx)}{\rvarx \in \lgvars(\ith[0]{\rn})} \cup \set{u})$.
\[
\pathprob{\epsilon,\eta,\pth} = \int_{\ell'}^{u'} \left(\frac{d\epsilon}{2}\eulerv{-d\epsilon\card{z-\mu-a_0}}\right) \pathprob{\epsilon,\eta_z,\suffix[1]{\pth}} dz.
\]

In the special case when $\avars(\ith[0]{\rn}) = \emptyset$ (i.e., the first transition of the run does not change the assignment to any variable), observe that $\eta_z = \eta$. Hence, $\pathprob{\epsilon,\eta_z,\suffix[1]{\pth}}$-term on the right hand side of both equations can be pulled out of the integral, and the expression can be simplified.
We will abuse notation and use $\pathprob{\cdot}$ to also 
{refer to the function $\pathprob{\eta,\pth} \coloneqq \epsilon\mapsto \pathprob{\epsilon,\eta,\pth}$}. Notice that when $\rn$ starts from $\qinit$, because of the {\initcond} condition of {\dipa}, the value of $\pathprob{\cdot}$ does not depend on the valuation $\eta$. For such {\runinpout}s, we may drop the valuation $\eta$ from the argument list of $\pathprob{\cdot}$ to reduce notational overhead. Even though we plan to use the same function name, the number of arguments to $\pathprob{\cdot}$ will disambiguate what we mean.


In this paper we study the computational problem of checking differential privacy for {\dipa}s. We conclude with a precise definition of this problem. We start by specializing the definition of differential privacy to the setting of {\dipa}. For a {\dipa} $\cA$, an input sequence $\inpsq \in (\Reals \cup \set{\noinp})^*$ and an output sequence $\outsq \in (\outalph \cup (\RealsInf \times \RealsInf))^*$, let $\runs(\inpsq,\outsq)$ be the set of all runs $\rn$ of $\cA$ starting from the initial state $\qinit$ such that $\rn$ is a run on $\inpsq$ producing $\outsq$. 
\begin{definition}
\label{def:diff-priv-auto}
A {\dipa} $\cA$ is  $\newd\epsilon$-differentially private (for $\newd > 0$, $\epsilon > 0$) iff for every $\inpsq_1,\inpsq_2 \in (\Reals \cup \set{\noinp})^*$ and $\outsq \in (\outalph \cup (\RealsInf \times \RealsInf))^*$ such that $\inpsq_1$ and $\inpsq_2$ are \emph{adjacent}~\footnote{See Definition~\ref{def:adjacency} on Page~\pageref{def:adjacency}.}, 
\[
\sum_{\rn \in \runs(\inpsq_1,\outsq)} \pathprob{\epsilon,(\rn,\inpsq_1,\outsq)} \leq e^{\newd\epsilon} \;
\sum_{\rn \in \runs(\inpsq_2,\outsq)}\pathprob{\epsilon,(\rn,\inpsq_2,\outsq)}.
\]
\end{definition}

\vspace*{0.05in}
\noindent
{\bf Differential Privacy Problem:} A {\dipa} $\cA$ is said to be \emph{differentially private} if there exists a constant $\newd>0$ (independent of $\epsilon$) such that $\cA$ is $\newd\epsilon$-differentially private, $\forall \epsilon>0.$
The differential privacy problem 
is the problem of determining if a given {\dipa} $\cA$ is differentially private. 

\begin{remark}
    {A {\dipa} $\cA$ is a parametric automaton (with parameter $\epsilon$), and the probability of each of its executions on a sequence of input varies with  $\epsilon.$ Thus,  considering its semantics, using $\cA(\epsilon)$ to refer to the automaton may be more appropriate. However, we shall use $\cA$ to reduce the notational overhead.}
\end{remark}

\section{Well Formed {\dipa}}
\label{sec:decidability}

The main goal of the paper is to solve the differential privacy problem described in Section~\ref{sec:dipauto}: Given a {\dipa} $\cA$ determine if there is a $\newd > 0$ such that for all $\epsilon > 0$, $\cA$ is $\newd\epsilon$-differentially private. In this section, we define the sub-class of \emph{well-formed} {\dipautop} that help characterize precisely the class of {\dipa} that are differentially private. Well-formed {\dipa} are automata that don't have four properties that lead to the violation of privacy: (a) \emph{{\criticalcycle}s}, (b) \emph{{\criticalpair}s}, (c) \emph{{\violatingc}s}, and (d) \emph{{\violatingp}s}. We will define what these types of cycles and paths are in this section.

\subsubsection*{Dependency Graph of a Run}
Consider a run $\rn$ of a {\dipa} $\cA$. Guards on transitions and decisions to store $\svar$ in storage variables, demand that if $\cA$ follows the run $\rn$, then the values sampled as $\svar$ at different steps must be ordered in a certain way to ensure that guards are satisfied. This partial order on the sampled values demanded by a run is conveniently captured as a directed graph that we call the \emph{dependency graph}.
\begin{definition}[Dependency Graph]
\label{def:depend-graph}
Let $\cA = \defaut$ be a {\dipa} and let $\rn = t_0t_1\cdots t_{n-1}$ be a run of $\cA$. The \emph{dependency graph} of $\rn$ is the directed graph $G_\rn = (V,E)$ where
\begin{itemize}
\item $V = \setpred{i}{0\leq i < n}$, and
\item $E$ is defined as $E' \cap (V \times V)$ where
\begin{align*}
E' = &\setpred{(j,\lavert_\rn(\rvarx,j))}{j \in V,\ \rvarx \in \lgvars(t_j)}\\ 
 &\cup \setpred{(\lavert_\rn(\rvarx,j), j)}{j \in V,\ \rvarx \in \smvars(t_j)}.
\end{align*}
\end{itemize}
\end{definition}
Notice that $E = E' \cap (V \times V)$ ensures that an edge $(j,\lavert_\rn(\rvarx,j))$ (or $(\lavert_\rn(\rvarx,j), j)$) is present only when $\lavert_\rn(\rvarx,j) \neq -\infty$ (i.e., when $\rvarx$ is assigned before position $j$). Also observe that an edge $(i,j)$ in $G_\rn$ means that, to satisfy the guards, $\svar$ at position $i$ in the run $\rn$ must be less than $\svar$ at position $j$.

Given the intuition that the dependency graph $G_\rn$ captures the ordering constraints imposed by the guards in $\rn$, one can conclude that a cycle in $G_\rn$ means that $\rn$ places contradictory demands on the values sampled and is therefore not a valid execution of the {\dipa}. We define a run $\rn$ of {\dipa} $\cA$ to be \emph{feasible} iff $G_\rn$ is acyclic. Feasibility is consistent with our semantic intuitions --- if $\rn$ is feasible then there is some evaluation $\eta$ such that for any $\epsilon > 0$, any input sequence $\inpsq$ and any output sequence $\outsq$ in which all output intervals are given by the interval $(-\infty,\infty)$,  for which $\rn$ is a run on $\inpsq$ that produces $\outsq$, $\pathprob{\epsilon,\eta,(\rn,\inpsq,\outsq)} > 0$. 

Let us consider a feasible run $\rn = t_0t_1\cdots t_{n-1}$ of {\dipa} $\cA$. Let $q_i = \src(t_i)$ and let $\parf(\src(t_i)) = (d_i,\mu_i,d'_i,\mu'_i)$. We say that $\rn$ is \emph{strongly feasible} if in addition whenever there is a path from $i$ to $j$ in $G_\rn$ and $q_i,q_j \in \epsstates$ then $\mu_i < \mu_j$. Thus, $\rn$ is strongly feasible if whenever guards require two $\svar$ values on non-input transitions to be ordered, the corresponding means of the Laplace distributions are ordered in the same way. We only consider {\dipa} that satisfy the following \emph{strong feasibility assumption}.

\vspace{0.05in}
\noindent
{\bf Strong Feasibility:} All feasible runs from the initial state $\qinit$ are strongly feasible.

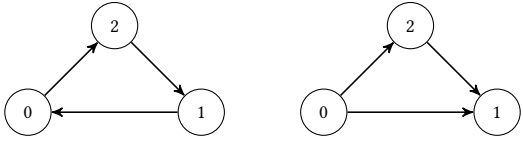
\begin{figure}
\begin{center}
{\footnotesize
\begin{tikzpicture}[node distance=1cm,auto]
\node[state](q0) {$0$};
\node[state](q2) [above right=of q0] {$2$};
\node[state](q1) [below right=of q2] {$1$};
\path[->] (q1) edge (q0)
          (q0) edge (q2)
          (q2) edge (q1);
\node[state](p0) [right=of q1] {$0$};
\node[state](p2) [above right=of p0] {$2$};
\node[state](p1) [below right=of p2] {$1$};
\path[->] (p0) edge (p2)
               edge (p1)
          (p2) edge (p1);
\end{tikzpicture}
}
\end{center}
\captionsetup{font=footnotesize,labelfont=footnotesize}
\caption{Dependency graphs for runs $\rn_1$ and $\rn_2$ from Example~\ref{ex:feasibility}. $G_{\rn_1}$ is on the left and $G_{\rn_2}$ is on the right.}
\label{fig:dependency}
\end{figure}

\begin{example}
\label{ex:feasibility}
Let us look at two example runs of length 3.
{\footnotesize
\begin{align*}\
\rn_1 = & (q_0,\true,q_1,\bot,(\true,\false))(q_1,\svar < \rvar_1,q_2,\bot,(\false,\true))\\
 &(q_2,\svar\geq\rvar_1 \wedge \svar < \rvar_2,q_3,\bot,(\false,\false))\\
 \rn_2 = & (q_0,\true,q_1,\bot,(\true,\false))(q_1,\svar \geq \rvar_1,q_2,\bot,(\false,\true))\\
 &(q_2,\svar\geq\rvar_1 \wedge \svar < \rvar_2,q_3,\bot,(\false,\false))
\end{align*}
}
The only difference between $\rn_1$ and $\rn_2$ is the guard on the second transition, which goes from state $q_1$ to $q_2$. Their dependency graphs are shown in Figure~\ref{fig:dependency}. $G_{\rn_1}$ is on the left and can be explained as follows. Transition $0$ sets variable $\rvar_1$ and transition $1$ sets variable $\rvar_2$. The guard $\svar<\rvar_1$ in transition $1$ results in the edge from $1$ to $0$. The conjunct $\svar \geq \rvar_1$ in transition $2$ results in an edge from \rcomment{$0$ to $2$}, and the conjunct $\svar < \rvar_2$ results in the edge from $2$ to $1$. $G_{\rn_1}$ is cyclic which means that $\rn_1$ is not feasible. Graph $G_{\rn_2}$ on the right in Figure~\ref{fig:dependency} is similar but the guard $\svar \geq \rvar_1$ in transition $1$ results in an edge from $0$ to $1$ (instead of from $1$ to $0$ in $G_{\rn_2}$) which removes the cycle. Thus, $\rn_2$ is feasible.
\end{example}

\subsubsection*{\Criticalcycle}
We are now ready to present the first graph theoretic condition on {\dipa} that demonstrates a violation of differential privacy.
\begin{definition}[\Criticalcycle]
\label{def:leaky-cycle}
A run $\rn$ of $\cA = \defaut$ from the initial state $\qinit$ (i.e., $\src(\rn) = \qinit$) is said to be a \emph{\criticalcycle} if there is an index $0 \leq j < \len{\rn}$ and a storage variable $\rvarx \in \Rvars$ such that the following conditions hold.
\begin{description}
\item[Cycle:] $C = \suffix{\rn}$ is a cycle.
\item[Leak:] There are indices $i_1$ and $i_2$ in $C$ (i.e., $j \leq i_1,i_2$) such that $\rvarx \in \avars(\ith[i_1]{\rn})$ and $\rvarx \in \used(\ith[i_2]{\rn})$.
\item[Repeatability:] $C$ can be repeated arbitrarily many times. That is, for every $m \geq 0$, the run $\rn C^m$ is feasible.~\footnote{$C^m$ denotes the $m$-fold concatenation of $C$ with $C^0 = \emptystr$.}
\end{description}
\end{definition}
Intuitively, the condition Leak in Definition~\ref{def:leaky-cycle} is to ensure that variable $\rvarx$ is assigned a value in the cycle $C$ that is later tested against in a guard.~\footnote{Definition~\ref{def:leaky-cycle} does not require $i_1 < i_2$. Therefore, strictly speaking the assignment in $i_1$ may not be before the test in $i_2$. But this can be easily addressed by taking $C^2$ instead of $C$ as the cycle.} The main effect of the 3 conditions in Definition~\ref{def:leaky-cycle}, is to identify two transitions (namely, those corresponding to assignment and test) that can be taken arbitrarily many times (since they are on a repeatable cycle) such that the $\svar$ values sampled in the two transitions are ordered in the same way each time the transitions are taken. This property leads to a ``leaking'' of the privacy budget, as shall be explained when we sketch the proof.

A cycle $C$ that does not satisfy the condition Leak will be said to be \emph{non-leaking}. 
\begin{definition}[Non-{\criticalcycle}]
\label{def:non-leak}
A run $C$ is a \emph{non-\criticalcycle} if $C$ is a cycle and for every $\rvarx \in \Rvars$ and $i$, if $\rvarx \in \used(\ith{C^2})$ then $\lavert_{C^2}(\rvarx,i) = -\infty$, i.e, $\rvarx$ is not assigned a value in $C$. Here $C^2$ is the concatenation of $C$ with itself.
\end{definition}
In Definition~\ref{def:non-leak}, we use the run $C^2$ to ensure that we also account for the case when $\rvarx$ is assigned \emph{after} it is used in $C$. One important property about non-{\criticalcycle} is that it is always repeatable; this is the content of the next proposition. Thus repeatability is a non-trivial requirement only for cycles that have a leak.
\begin{proposition}
\label{prop:non-leaky-repeat}
Let $\rn$ be a feasible run of $\cA = \defaut$ from the initial state $\qinit$ such that $C = \subseq{\rn}{i}{j}$ (for some $0 \leq i < j \leq \len{\rn}$) is a non-{\criticalcycle}. Then for every $m > 0$, $\subseq{\rn}{0}{i}(\subseq{\rn}{i}{j})^m\suffix{\rn}$ is feasible.
\end{proposition}

\subsubsection*{\Criticalpair}
Recall that the key property of a {\criticalcycle} that leads to the violation of differential privacy is finding two transitions that can be repeated arbitrarily many times such that the $\svar$ value sampled in the two transitions is ordered every time they are taken. {\Criticalcycle}s achieve this by finding both transitions on a cycle that can be repeated. However, that is not the only way such a pair of transitions can arise --- the two transitions could be on two different cycles that can each be repeated. This leads to the definition of a \emph{\criticalpair}. The definition of a {\criticalpair} is subtle and we will discuss its details after presenting it formally.
\begin{definition}[\Criticalpair]
\label{def:leakingpair}
A feasible run $\rn$ of $\cA = \defaut$ from the initial state $\qinit$ is a \emph{\criticalpair} if there are indices $0 \leq i_1 < j_1 \leq \len{\rn}$ and $0 \leq i_2 < j_2 \leq \len{\rn}$ such that the following conditions hold.
\begin{description}
\item[Cycles:] $C_1 = \subseq{\rn}{i_1}{j_1}$ and $C_2 = \subseq{\rn}{i_2}{j_2}$ are both non-{\criticalcycle}s.
\item[Disjointness:] Either $j_1 \leq i_2$ or $j_2 \leq i_1$. That is, $C_1$ and $C_2$ are non-overlapping subsequences of $\rn$.
\item[Order:] There is a path $k_1,k_2,\ldots k_m$ in the dependency graph $G_\rn$ such that $i_1 \leq k_1 < j_1$ ($k_1$ is on $C_1$), $i_2 \leq k_m< j_2$ ($k_m$ is on $C_2$), $k_2 < k_1$ and $k_{m-1} < k_m$.
\end{description}
\end{definition}
\begin{figure*}
\begin{minipage}[b]{0.44\linewidth}
\begin{center}
\begin{tikzpicture}
\footnotesize
\node[mynonstate, initial] (q0) {\mystrut $q_0$ \nodepart{two} \mystrut $\frac{1}{4},\ 1$};
\node[mynonstate, right of=q0] (q1) {\mystrut $q_1$ \nodepart{two} \mystrut $\frac{1}{4},\ 0$};
\node[myinstate, above of=q1] (q2) {$q_2$ \nodepart{lower} $\frac{1}{4},\ 0$};
\node[myinstate, left of=q2] (q3) {$q_3$ \nodepart{lower} $\frac{1}{4},\ 0$};
\node[myinstate, left of=q3] (q4) {$q_4$ \nodepart{lower} $\frac{1}{4},\ 0$};
\draw (q0) edge node[myedgelabel] {$\true,\: \bot$ \nodepart{two} $(\true,\false)$} (q1);
\draw (q1) edge node[myedgelabel, right] {$\true,\: \bot$ \nodepart{two} $(\false,\true)$} (q2);
 \draw (q2) edge[out=240,in=210,looseness=8] node[myedgelabel, below] {$g_1,\: \top$ \nodepart{two} $(\false,\false)$} (q2);
 \draw (q2) edge node[myedgelabel, above] {$g_2,\: \bot$ \nodepart{two} $(\false,\false)$} (q3);
  \draw (q3) edge[loop below] node[myedgelabel] {$g_3,\: \top$ \nodepart{two} $(\false,\false)$} (q3);
  \draw (q3) edge node[myedgelabel, above] {$g_4,\: \bot$ \nodepart{two} $(\false,\false)$} (q4);
\end{tikzpicture}
\end{center}
\captionsetup{font=footnotesize,labelfont=footnotesize}
\caption{{\dipa} $\cA_{\mathsf{leakp}}$ from Example~\ref{ex:leakingpair}. $\cA_{\mathsf{leakp}}$  has two variables, $\rvar_1$ and $\rvar_2$, assigned in the first and the second transition, respectively.  
The guards $g_1 = (\svar \geq \rvar_1)$, $g_2 = (\svar < \rvar_1)$, $g_3 = (\svar < \rvar_2)$, $g_4 = (\svar < \rvar_1) \wedge (\svar \geq \rvar_2)$.}
\label{fig:leakingpairone}
\end{minipage}
\hfill
\begin{minipage}[b]{0.45\linewidth}

\begin{center}
{\footnotesize
\begin{tikzpicture}[node distance=0.8cm,auto]
\node[state](q0) {$0$};
\node[state](q1) [right=of q0] {$1$};
\node[state](q2) [right=of q1] {$2$};
\node[state](q3) [right=of q2] {$3$};
\node[state](q4) [right=of q3] {$4$};
\path[->] (q0) edge[bend right=330] (q2)
          (q3) edge[bend left] (q0)
          (q4) edge[bend left] (q1);
%

\node[state](q7) [below of = q2, node distance=2.2cm] {$2$};
\node[state](q6) [left=of q7] {$1$};
\node[state](q5) [left=of q6] {$0$};
\node[state](q8) [right=of q7] {$3$};
\node[state](q9) [right=of q8] {$4$};
\node[state](q10) [right=of q9] {$5$};
\path[->] (q5) edge[bend right=330] (q7)
          (q8) edge[bend left] (q5)
          (q9) edge[bend left] (q6)
          (q6) edge[bend right=330] (q10)
          (q10) edge[bend left=40] (q5);

\end{tikzpicture}
}
\end{center}
\captionsetup{font=footnotesize,labelfont=footnotesize}
\caption{Dependency graphs for runs $\rn_1$ and $\rn_2$ from Example~\ref{ex:leakingpair}. $G_{\rn_1}$ is on the top and $G_{\rn_2}$ is on the bottom. The nodes are numbered according to the order in which the corresponding transition appears in the run.}
\label{fig:dependency:leakp}
\end{minipage}
\end{figure*}

As mentioned before Definition~\ref{def:leakingpair}, the motivation behind {\criticalpair}s is to identify a pair of transitions $t$ and $t'$ that can be executed multiple times and such that the $\svar$ value each time $t$ is taken is smaller than the $\svar$ value each time $t'$ is taken. Such a pair of transitions represents a ``leak'' of the privacy budget that can be exploited to prove that {\dipa} is not differentially private. Definition~\ref{def:leakingpair} achieves this goal in the following manner. The desired transitions $t$ and $t'$ are $\ith[k_1]{\rn}$ and $\ith[k_m]{\rn}$, respectively. The fact that $t$ and $t'$ are on cycles $C_1$ and $C_2$ which are disjoint (in $\rn$) and non-{\critical}, ensures that they can be repeated thanks to Proposition~\ref{prop:non-leaky-repeat}. The condition Order in Definition~\ref{def:leakingpair} is the most subtle. The fact that $k_2 < k_1$ and $(k_1,k_2)$ is an edge in $G_\rn$ means that there is a storage variable $\rvarx \in \Rvars$ such that $\rvarx$ is assigned in $\ith[k_2]{\rn}$ and $\lttestx$ is one of the conjuncts in $\grd(\ith[k_1]{\rn})$. Further since $C_1$ is non-{\critical}, $\rvarx$ is not updated within $C_1$ and hence $\ith[k_2]{\rn}$ is taken \emph{before} $C_1$. Similar conclusions can be drawn about $k_{m-1}$ and $k_m$ --- there is a variable $\rvary \in \Rvars$ that is assigned in $\ith[k_{m-1}]{\rn}$ which is taken before $C_2$, and $\getesty$ is a conjunct in $\grd(\ith[k_m]{\rn})$. Finally, the path from $k_1$ to $k_m$ means that the $\svar$ value sampled in $\ith[k_1]{\rn}$ is less than the value assigned to $\rvarx$ in $\ith[k_2]{\rn}$, which in turn is less than the value assigned to $\rvary$ in $\ith[k_{m-1}]{\rn}$ and that is less than the $\svar$ value sampled in $\ith[k_m]{\rn}$. $\ith[k_2]{\rn}$ is before $C_1$ which means that the value assigned to $\rvarx$ in $\ith[k_2]{\rn}$ does not change no matter how many times $C_1$ and $C_2$ are repeated. 
{Next, $\ith[k_{m-1}]{\rn}$ is before $C_2$. It is possible that $\ith[k_{m-1}]{\rn}$ is on $C_1$, in which case the value assigned to $\rvary$ changes when $C_1$ is repeated. However, one can show by induction, that the presence of a path in the dependency graph from $\ith[k_2]{\rn}$ to $\ith[k_{m-1}]{\rn}$ and an edge from $\ith[k_{m-1}]{\rn}$ to $\ith[k_m]{\rn}$ means that when $C_1$ and $C_2$ are repeated, there will be a path from $\ith[k_2]{\rn}$ and the last instance of $\ith[k_{m-1}]{\rn}$ and the last value assigned to $\rvary$ in $\ith[k_{m-1}]{\rn}$ will be less than every $\svar$ value sampled in $\ith[k_m]{\rn}$. Thus, every $\svar$ value sampled in $\ith[k_1]{\rn}$ will be less than every $\svar$ value sampled in $\ith[k_m]{\rn}$, no matter how many times $C_1$ and $C_2$ are repeated.}


\begin{example}
    \label{ex:leakingpair}
    Consider the automaton $\cA_{\mathsf{leakp}}$  in Figure~\ref{fig:leakingpairone}. The automaton is drawn following the convention outlined in Example \ref{ex:range-auto}. The automaton has two real variables $\rvar_1$ and $\rvar_2$, assigned in the first and the second transition, respectively. For states $q_i,q_j$ of $\cA_{\mathsf{leakp}}$, let $t_{ij}$ denote the unique transition of $\cA_{\mathsf{leakp}}$ from state $q_i$ to $q_j.$ Observe that $t_{22}$ and $t_{33}$ are cycles. Consider the run $\rn_1=t_{01} t_{12} t_{22}t_{23} t_{33}$ that visits both the cycles $t_{22}$ and $t_{33}$ and its extension $\rn_2=\rn_1t_{34}.$ Their dependency graphs for these runs are shown in Figure~\ref{fig:dependency:leakp}. The nodes $2$ and $4$ correspond to the cycle transitions $t_{22}$ and $t_{33}$ respectively. Considering just the run $\rn_1$, these cycles do not constitute a leaking pair. However, when we consider the extended run, $\rn_2$, we see that these cycles form a leaking pair via the path $4\to 1 \to 5\to 0 \to 2.$
    \end{example}

    Before moving onto the other two properties needed to define well-formed {\dipa}, it is useful to remark that the cycles $C_1$ and $C_2$ in Definition~\ref{def:leakingpair} may be the ``same cycle'', i.e., $C_1$ and $C_2$ could, respectively, be the first and second iterations of the same sequence of $\cA$ transitions.
\subsubsection*{\Violatingc}
Real valued outputs present another avenue through which private information in the input can be leaked. The condition identified by {\criticalcycle}s and {\criticalpair}s do not account for such violations because they are agnostic to the type of output produced by the {\dipa}. Our next condition \emph{\violatingc}, identifies a transition that can be executed repeatedly, and which outputs a pertubed input. 
\begin{definition}[\Violatingc]
\label{def:disclosingcycle}
A feasible run $\rn$ of $\cA = \defaut$ from the initial state $\qinit$ is a \emph{\violatingc} if there are indices $0 \leq j \leq i < \len{\rn}$ such that the following conditions hold.
\begin{description}
\item[Cycle:] $C = \suffix{\rn}$ is a non-{\criticalcycle}.
\item[Disclosing:] $\ith{\rn}$ is an input transition that outputs a real value, i.e., $\src(\ith{\rn}) \in \instates$ with $\otpt(\ith{\rn}) \in \set{\svar,\svar'}$.
\end{description}
\end{definition}
Observe that in Definition~\ref{def:disclosingcycle}, $\ith{\rn}$ is a transition that is on cycle $C$. Moreover, since $C$ is non-{\criticalcycle}, by Proposition~\ref{prop:non-leaky-repeat}, the run $\rn C^m$ is feasible for every $m \geq 0$. Thus, the transition $\ith{\rn}$ can be executed repeatedly. Since $\ith{\rn}$ is an input transition that outputs a real-value, each time it is executed it reveals some information about the input which results in a loss of privacy. 

\subsubsection*{\Violatingp}
We now present the last property needed to define well formed {\dipa}. This last property also concerns privacy violations that arise from real valued outputs. {\Criticalcycle}s and {\criticalpair}s identify a transition that is executed arbitrarily many times where the sampled $\svar$ value is bounded by values sampled in another transition (that is also executed many times) on the same run. However, with real valued outputs, we could have a situation where this bound is revealed once, explicitly in an output. This is captured in our next definition.
\begin{definition}[\Violatingp]
\label{def:violating}
A feasible run $\rn$ of $\cA = \defaut$ from the initial state $\qinit$ is a \emph{\violatingp} if there are indices $0 \leq i \leq j \leq \len{\rn}$ such that the following conditions hold.
\begin{description}
\item[Cycle:] $C = \subseq{\rn}{i}{j}$ is a non-{\criticalcycle}.
\item[Privacy Violation:] There is a path $k_1,k_2,\ldots k_m$ in the dependency graph $G_\rn$ such that either
(a) $\otpt(\ith[k_1]{\rn}) = \svar$, $k_{m-1} < k_m$, and $i \leq k_m < j$, i.e., $\ith[k_m]{\rn}$ is on cycle $C$, or
(b) $i \leq k_1 < j$ ($\ith[k_1]{\rn}$ is on cycle $C$), $k_2 < k_1$, and $\otpt(\ith[k_m]{\rn}) = \svar$.
\end{description}
\end{definition}

It is useful to see how Definition~\ref{def:violating} captures the intuitions laid out before. The path from $k_1$ to $k_m$ in $G_\rn$ ensures that the $\svar$ value sampled in $\ith[k_1]{\rn}$ is less than the $\svar$ value sampled in $\ith[k_m]{\rn}$. Moreover, since $C$ is non-{\critical}, by Proposition~\ref{prop:non-leaky-repeat}, it is repeatable. Condition (a) in (Privacy Violation) says that $\ith[k_m]{\rn}$ is a transition on $C$, and the edge $(k_{m-1},k_m)$ in $G_{\rn}$ along with $k_{m-1} < k_m$ means that there is a variable $\rvarx \in \Rvars$ that is set in $\ith[k_{m-1}]{\rn}$ and $\getestx$ is in $\grd(\ith[k_m]{\rn})$. Moreover, since $C$ is non-{\critical}, $\rvarx$ is not updated in $C$ and hence $k_{m-1}$ is before $C$. Thus, the presence of the path means that the value output in $\ith[k_1]{\rn}$ is less than the $\svar$ value sampled in $\ith[k_{m-1}]{\rn}$ which in turn is less than the $\svar$ value sampled in $\ith[k_m]{\rn}$ every time $C$ is repeated. Therefore, there is a lower bound, which is output in $\ith[k_1]{\rn}$, for arbitrary many $\svar$ values that are generated in $\ith[k_m]{\rn}$. Condition (b) in (Privacy Violation) is similar but \emph{dual}. Here $\ith[k_1]{\rn}$ is on $C$, $\ith[k_2]{\rn}$ is before $C$ and sets a variable $\rvarx$ that is an upper bound on the values sampled in $\ith[k_1]{\rn}$, and finally, $\ith[k_m]{\rn}$ outputs a value that upper bounds all these values, no matter how many times $\ith[k_1]{\rn}$ is executed by repeating $C$.

\subsubsection*{Well-formed {\dipautop}}
The properties defined in this section identify witnesses for the violation of privacy. The class of \emph{well-formed} automata are those that do not suffer from these deficiencies.
\begin{definition}[Well-formed {\dipa}]
\label{def:well-formed}
A {\dipa} $\cA$ is said to be {\em well-formed} if $\cA$ does not have any {\criticalcycle}s, {\criticalpair}s, {\violatingc}s, and {\violatingp}s. 
\end{definition}

Our main results are: (i) a well-formed {\dipa}  is differentially private; (ii) if a {\dipa} satisfying  the {\outcond} property  (see Definition \ref{def:outcond}) is differentially private then it must be well-formed.  We will also show that there is an effective procedure for checking if a {\dipa} is well-formed. These observations together will provide a decidability result for solving the differential privacy problem for {\dipa} that satisfy {\outcond} property. 

\section{Well-formed {\dipa} are Differentially Private}
\label{sec:sufficiency}

One of our main results, which we call the {\em sufficiency theorem}, is that well-formed {\dipa}s  are differentially private.
\ifdefined\short
\else
The proof of this Theorem is involved and carried out in Appendix~\ref{app:sufficient}.
\fi
\begin{theorem}
\label{thm:main}
Let $\cA$ be a {\dipa}. If $\cA$ is well-formed then there is a $\newd > 0$ such that for every $\epsilon > 0$, $\cA$ is $\newd\epsilon$-differentially private. Further, such a $\newd$ can be computed in time exponential in the size of the automaton $\cA.$
\end{theorem}
\begin{proof}[Proof Sketch]
Let $\cA$ be a well-formed {\dipa}. Given a feasible run $\rn=t_0\cdots t_n$ of $\cA$ from the initial state, fix computations $\pth_i=(\rn,\inpsq_i,\outsq)$ for $i=1,2$ such that $\inpsq_1$ and $\inpsq_2$ are adjacent.  For each $j,$ let $\lt_j$ be the {\lq\lq}less than{\rq\rq} relation  on $\Rvars$ imposed by the prefix $\rn[0:j-1]$ --- $(\rvar,\rvar') \in \lt_j$
if there is a path of non-zero length from $\lavert_\rn(\rvar,j)$ to $\lavert_\rn(\rvar',j).$ Similarly,
$\eqs_j$ is the  {\lq\lq}equality{\rq\rq} relation on $\Rvars$ imposed by the prefix $\rn[0:j-1]$ --- $(\rvar,\rvar') \in \eqs_j$
if $\lavert_\rn(\rvar,j)= \lavert_\rn(\rvar',j).$

We can show that there are numbers $\wt_j$ and functions $m_j:\Rvars \rightarrow \set{-1,0,1}$
such that 
\begin{enumerate}
\item For any valuations $\val_1,\val_2$ such that 
$\val_2=\val_1+m_j,$~\footnote{For functions $f,g: A \to \Reals$, $f+g$ is the function that adds the result of $f$ and $g$ for each argument, i.e., $(f+g)(a) = f(a) + g(a)$.}
$$\Prob[\epsilon,\val_2,\pth_2[j:]] \leq \eulerv{\sum_{\ell=j}^{n-1} \wt_\ell }\Prob[\epsilon,\val_1,\pth_1[j:]].$$
\item If $t_{j_1}=t_{j_2},$ $\lt_{j_1}=\lt_{j_2}$ and $\eqs_{j_1}=\eqs_{j_2}$ for $j_1\leq j_2$ then $\wt_{j_1}=0$
\item $\wt_j\leq 2d_j+ d_j'$ where $d_j$ and $d_j'$ are such that $\parf(\src(t_j))=(d_j,\mu_j,d_j',\,\mu_j').$

\end{enumerate}
Observe that the last two conditions imply that there is a number $\newd$ independent of $\rn$ such that $ {{\sum_{\ell=j}^{n-1} \wt_\ell }}<\newd.$ Note that as $\rn$ is a run from initial state then $\Prob[\epsilon,\val_i,\pth_i]$ is independent of $\val_i.$ The above observations imply that
$$ \Prob[\epsilon,\pth_2] \leq \eulerv{\newd }\Prob[\epsilon,\pth_1].$$ This shows that $\cA$ is $\newd\epsilon$-differentially private. To carry out the formal proof, we construct an augmented automaton $\aug\cA$, whose  states are triples of the form $(q,\lt,\eqs)$ where $q$ is a state of $\cA$,  $\lt$, and $\eqs$ are strict partial orders and equivalence relations on $\Rvars.$ The value for $\newd$ is also computed using the augmented automaton. \end{proof}

The problem of checking well-formedness can be shown to be in {\pspace}. 
\ifdefined\short
\else 
The proof is in Appendix~\ref{app:sufficient}. 
\fi 
\begin{theorem}
\label{thm:pspace}
The problem of checking whether a {\dipa} is well-formed is in {\pspace}. When the number of variables is taken to be a constant $k$, then the problem of checking whether a {\dipa} is well-formed is decidable in polynomial time. 
\end{theorem}

\section{Differentially Private {\dipa} are well-formed}
\label{sec:necessity}

While well-formedness is sufficient for ensuring differential privacy, it is not a necessary condition for differential privacy
as illustrated by the following example. 

\begin{figure}
\begin{center}
\begin{tikzpicture}
\footnotesize
\node[mynonstate, initial] (q0) {\mystrut $q_0$ \nodepart{two} \mystrut $\frac{1}{4},\ 0$};
\node[myinstate,right of = q0] (q1) {\mystrut $q_1$ \nodepart{lower} \mystrut $\frac{1}{4},\ 1$};

\draw (q0) edge[bend right] node[below,myedgelabel] {$\true,\: \top$ \nodepart{two} $\true$} (q1);
\draw (q1) edge[bend right] node[above,myedgelabel] {$g_1,\: \top$ \nodepart{two} $\false$} (q0);
\draw (q1) edge[loop right] node[right,myedgelabel] {$g_2,\: \top$ \nodepart{two} $\false$} (q1);


\end{tikzpicture}
\end{center}
\captionsetup{font=footnotesize,labelfont=footnotesize}
\caption{{\dipa} $\cA_{\mathsf{nwf}}$ with one variable $\rvar$ is not well-formed but differentially private. 
The guards $g_1 = (\svar \geq \rvar)$ and $g_2 = (\svar < \rvar)$.}
\label{fig:exam-nwf}
\end{figure}




\begin{example}
   Consider the {\dipa} $\cA_{\mathsf{nwf}}$ with one variable $\svar$ given in Figure~\ref{fig:exam-nwf}. The automaton is drawn following the convention outlined in Example \ref{ex:range-auto}. As each transition outputs $\top,$ 
  $\cA_{\mathsf{nwf}}$, on any input of length $n$,  outputs the  string $\top^n$ with probability $1$. Thus, $\cA_{\mathsf{nwf}}$  is trivially differentially private. However, $\cA_{\mathsf{nwf}}$ is not well-formed as it has a leaking cycle, $t_{a}t_{b}$ where $t_{a}$ is the transition from $q_0$ to $q_1$ and $t_{b}$ is the transition from $q_1$ to $q_0.$
\end{example}

We show, however, that differentially private {\dipa} that satisfy an additional technical property of \emph{{\outcond}} are well-formed. Thus, for {\dipa} satisfying this property, well-formedness is a  precise characterization of when they are differentially private. Before presenting this \emph{restricted necessity} theorem and proof sketch, let us define what it means for a {\dipa} to satisfy the condition of {\outcond}.


\begin{definition}[\Outcond]
\label{def:outcond}
A {\dipa} $\cA = \defaut$ satisfies \emph{\outcond} if the following holds: If $t_1$ and $t_2$ are distinct transitions of $\cA$ such that $\src(t_1) = \src(t_2)$ then $\otpt(t_1) \neq \otpt(t_2)$ and $\set{\otpt(t_1),\otpt(t_2)} \cap \outalph \neq \emptyset$.
\end{definition}

{\Outcond} demands that distinct outgoing transitions from a state have different outputs and at most one of the outgoing transitions outputs a real value. In particular, there cannot be two transitions out of a state $q$ that output $\svar$ and $\svar'$. Distinct outputs on transitions ensure that given a starting state $q$ and an output sequence $\outsq$, there is at most one run $\rn$ starting from $q$ that can produce $\outsq$. 
Observe that the automaton of Figure~\ref{fig:exam-nwf} does not satisfy {\outcond}~property. 
 The necessity proof proceeds by showing that if $\cA$ is not well-formed, then given $\newd$, there are computations $(\rn,\inpsq_1,\outsq)$ and  $(\rn,\inpsq_2,\outsq)$ with the same run $\rn$ such that $\rn$  outputs $\outsq$, $\inpsq_1, \inpsq_2$ are adjacent and the ratio of the probability measures of these computations is $> \eulerv{\newd\epsilon}$ for sufficiently large $\epsilon$. Output distinction guarantees that $\rn$ is the \emph{only} run on $\inpsq_1, \inpsq_2$ that outputs $\outsq$, allowing us to conclude that  {$\cA$} is not differentially private for non-well formed $\cA$. Without output distinction, the deficit in probability measures of $\outsq$ can be made up by other paths. The output distinction property is also needed in~\cite{ChadhaSV21} for the case of a single variable.
We are now ready to present the main result of this section.

\begin{theorem}
\label{thm:necessity}
Let $\cA$ be a {\dipa} that satisfies the {\outcond} property. If $\cA$ is not well-formed, then it is not differentially private.
\end{theorem}

\begin{proof}[Proof Sketch]
Let us fix a {\dipa} $\cA = \defaut$ that satisfies the {\outcond} property. Recall that the {\outcond} property ensures that for any input sequence $\inpsq$ and output sequence $\outsq$, $|\runs(\inpsq,\outsq)| \leq 1$. We sketch the main ideas behind the proof; the full details can be found in
\ifdefined\short
\cite{ChadhaSVB23}.
\else
Appendix~\ref{app:necessity}.
\fi
Assume that $\cA$ is not well-formed. Now, for each value of $\newd$ and $\epsilon$, the proof identifies a run $\rn$, an output sequence $\outsq$, and a pair of adjacent input sequences $\alpha$ and $\beta$ such that the computations $(\rn,\alpha,\outsq)$ and $(\rn,\beta,\outsq)$ demonstrate a violation of differential privacy (Definition~\ref{def:diff-priv-auto}). The construction of witnesses is based on the following sequence of observations.
\begin{enumerate}[leftmargin=*,align=left, itemindent=1em]
\item Let us fix a run $\rn$ from $\qinit$ and an output sequence $\outsq$ consistent with $\rn$. Observe that the number read in an input transition determines the mean of the distributions from which $\svar$ and $\svar'$ are drawn in that step. Let us call an input sequence $\inpsq$ \emph{strongly compliant} with $\rn$ and $\outsq$, if the sampling means satisfy the constraints imposed by $\rn$ and $\outsq$. This has two requirements. First, whenever there is a path from $i$ to $j$ in $G_\rn$, the sample mean at step $i$ is less than the sample mean at step $j$. Notice that strong feasibility ensures this when $i$ and $j$ are non-input transitions, and here we are requiring this to hold when either $i$ or $j$ is an input transition in which case the mean is determined by $\inpsq$. Second, if $\otpt(\ith{\rn}) \in \set{\svar,\svar'}$ (real outputs), the sample mean at step $i$ is in the interval $\ith{\outsq}$. Intuitively, for a strongly compliant input sequence $\inpsq$, the probability of computation $(\rn,\inpsq,\outsq)$ is likely to be ``high''. On the flip side, let us call an input sequence $\inpsq$ \emph{non-compliant} at $i$, if the sample mean set by $\inpsq$ at step $i$ either violates an order constraint or an output constraint. Again intuitively, one can imagine that, as the number of non-compliant transitions increase in $\inpsq$, the probability of the computation $(\rn,\inpsq,\outsq)$ decreases. Now one can prove that if we consider two input sequences $\inpsq_1$, which is strongly compliant, and $\inpsq_2$, which has non-compliant transitions, then the ratio of the probabilities of $(\rn,\inpsq_1,\outsq)$ and $(\rn,\inpsq_2,\outsq)$ grows as the number of non-compliant transitions in $(\rn,\inpsq_2,\outsq)$ increases.
\item Observations in (1) above provide a template for how to identify witnesses for differential privacy violation: the presence of a {\criticalcycle}, {\criticalpair}, {\violatingc}, or {\violatingp} help identify a run, and we then construct two input sequences $\alpha$, which is strongly compliant, and $\beta$ which has many non-compliant steps. Observe that each witness to non-well-formedness is a run containing a cycle that can be repeated arbitrarily many times and contains a transition that will be made non-compliant in the input sequence $\beta$. The intuitions laid out in Section~\ref{sec:decidability} for defining well-formed {\dipa} will be used and we spell this out in each case. A {\criticalcycle} has a transition with index $i_1$ (see Definition~\ref{def:leaky-cycle}) that sets a variable which is then used later in the transition indexed $i_2$. Since the guard of $i_2$ is not $\true$, it is an input transition. We will construct the run $\rn$ by repeating the cycle as many times as needed (based on $\newd$ and $\epsilon$), and in $\beta$ the sample mean at step $i_2$ will be in the wrong order with respect to $i_1$ in each repetition, making it non-compliant.
In a leaking pair (Definition~\ref{def:leakingpair}) there is a pair of transitions indexed $k_1$ and $k_m$ on cycles that can be repeated, and whose sampled values need to be ordered each time they are executed. Moreover, transitions $k_1$ and $k_m$ are input transitions because their guards are not $\true$ (see discussion after Definition~\ref{def:leakingpair}). Thus, in $\beta$ we will flip the order of the sample means at these steps to create an arbitrary number of non-compliant steps. 
The transition indexed $i$ in a {\violatingc} (Definition~\ref{def:disclosingcycle}) is an input transition on a cycle that can be repeated. To create non-compliant steps in $\beta$ we will set the mean of these transitions to not be in the output interval given for this step. 
Finally, in a {\violatingp} (Definition~\ref{def:violating}) there is an input transition with index $k_m$ for case (a) (or $k_1$ for case (b)) that is on a repeatable cycle whose sampled value is required to be larger than (smaller than in case (b)) the value output in step $k_1$ (step $k_m$ for case (b)). To construct the input sequence $\beta$, we set the input for each time $k_m$ ($k_1$ in case (b)) is taken to be smaller than the value output in $k_1$, and thereby creating arbitrarily many non-compliant steps.
\item The general principles behind constructing the input sequences $\alpha$ and $\beta$ are laid out in (2). However, one key requirement for $\alpha$ and $\beta$ to constitute a witness to privacy violation is that they be \emph{adjacent} (Definition~\ref{def:adjacency}) which demands that the values in $\alpha$ and $\beta$ be not too far apart. One challenge is carrying this out is the presence of non input transitions, where the sample means are \emph{fixed}. This can be overcome by carefully analyzing the dependency graph $G_\rn$ and the parameters decorating the states appearing in the run $\rn$. \qedhere
\end{enumerate}
\ifdefined\short
\else
The precise proof based on the above ideas is long and deferred to Appendix~\ref{app:necessity}.
\fi
\end{proof}

In Section~\ref{sec:sufficiency}, we showed that there is a {\pspace} algorithm to determine if an output-distinct {\dipa} $\cA$ is well-formed (Theorem~\ref{thm:pspace}). This complexity bound is tight; we show that the problem of determining if a {\dipa} is differentially private is {\pspace}-hard. (See 
\ifdefined\short
\cite{ChadhaSVB23}
\else
Appendix~\ref{app:pspacehard} 
\fi for the proof.)

\begin{theorem}
\label{thm:lower-bound}
Given an output-distinct {\dipa} $\cA$, the problem of determining if there is a $\newd>0$ such that for all $\epsilon$, $\cA$ is $\newd\epsilon$-differentially private, is {\pspace}-hard.
\end{theorem}



\section{Experiments}
\label{sec:experiments}

\sloppy
We implemented the algorithm that checks whether a {\dipa} $\cA$ is well-formed. In case $\cA$ is well-formed, it computes a bound $\newd$, which we call the \emph{weight} of the automaton, such that $\cA$ is $\newd\epsilon$-differentially private for all $\epsilon$.  The software tool, {\toolaut}, is built in Python 3.9.5 and is available for download at ~\cite{DipAut}. It uses the {\PLY} package~\cite{PLY} for parsing the program and the {\igraph} package~\cite{igraph} to store the input automaton as a graph. The {\igraph} package is also used to perform graph-theoretic operations on the input automaton. 




{\toolaut} has three major components. The first component, called \emph{core}, tokenizes and parses the input using {\PLY}. The second component, \emph{builders}, constructs the augmentation of the input automaton. The augmentation is built using a breadth-first-search of the (implicit) graph of the augmentation. The relations $\lt$
and $\eqs$ are stored as dictionaries during augmentation. To prepare for checking of {\criticalpair} and {\violatingp}, the automaton also builds an {\lq\lq}enhanced{\rq\rq} augmentation. For example, it also builds the graphs that include assignments to the variables $V_1$ and $V_2$ in the
algorithm for checking {\criticalpair} (See the proof of Theorem~\ref{thm:pspace}). The third component \emph{DP tests}, implements the finals checks for {\criticalcycle}, {\criticalpair}, {\violatingp}
and {\violatingc} from the augmentations. If the automaton is well-formed, it also computes the weight of the automaton. If it is not well-formed, it further checks if it is output-distinct. In that case, we report that the automaton is not differentially private.

{\toolaut} was evaluated against a suite of examples (See Table~\ref{table:simple-examples}), which we describe briefly.  

\begin{table*}[!ht]\setlength\tabcolsep{6pt}
    \centering
    \footnotesize{
    \begin{tabular}{l|lll|llll|ll}
    \hline

      \multicolumn{4}{c|}{Benchmarks} &  \multicolumn{4}{c|}{\toolaut} & \multicolumn{2}{c}{{\CheckDP} ~\cite{CheckDP} }  \\ \hline 
     Example &  $v$  & $s$ & $trans$ & \vtop{\hbox{\strut wt calc }\hbox{\strut time (s)}} &  \vtop{\hbox{\strut total }\hbox{\strut time (s)}} & \vtop{\hbox{\strut differentially }\hbox{\strut private?}}    & $\newd$ & 
     \hbox{\strut time (s)} & \vtop{\hbox{\strut Counterexample}\hbox{\strut Validated?}} \\ \hline
        {\SVT}  & 1 & 3 & 3 & 0.00046 & 0.238 & \checkmark & 5/4 & 29.92 &  N.A. \\
         
         {\Spse} & 1 & 3 & 3 & 0.00045 & 0.249 & $\checkmark$ & 7/4 & 52.43 & N.A. \\
         

                {\DC} & 2 & 4 & 5 & N.A. & 0.237 &  $\times,$ \resultdc & N.A. & 43.59 & \timeouta \\ 
        
               \NumericRangeOne & 2 & 4 & 4 & N.A. & 0.234 &  $\times,$ \resultpv  & N.A. & 316.05 & \timeouta \\ 
        
       \NumericRangeTwo & 2 & 4 & 4 & 0.00078 & 0.231 & \checkmark & 5/4 & 1909.43 & \timeouta \\
        

               \LC & 2 & 4 & 4 & N.A. & 0.231 & $\times,$ \resultlc & N.A. & \timeouta & ~ \\ 
        

              {\TRangeOne}   & 3 & 6 & 10 & N.A. & 0.239 & $\times,$ 
        \resultlp   & N.A. & \timeouta & ~ \\
        
        {\TRangeTwo} & 3 & 7 & 11 & 0.00258 & 0.277 & \checkmark & 2 & \timeouta & ~ \\ 
        
        2-\minmax & 2 & 4 & 7 & 0.00065 & 0.220 & \checkmark & 1 & \timeouta & ~\\ 
        
        10-\minmax & 2 & 12 & 31 & 0.00221 & 0.230 & \checkmark & 1 & \memoryerror & ~\\ 
        
        20-\minmax & 2 & 22 & 61 & 0.00434 & 0.248 & \checkmark & 1 & \memoryerror & ~ \\

        100-\minmax & 2 & 102 & 301 & 0.0291 & 0.409 & \checkmark & 1 & \memoryerror & ~ \\ 

        200-\minmax & 2 & 202 & 601 & 0.0803 & 0.643 & \checkmark & 1 & \memoryerror & ~ \\ 
        1-\Range & 2 & 4 & 5 & 0.00083 & 0.227 & \checkmark & 1 & \timeouta & ~\\ 
        
        10-\Range & 20 & 31 & 50 & 0.00797 & 0.611 & \checkmark & 1 & \memoryerror & ~  \\ 
        
        20-\Range & 40 & 61 & 100 & 0.0212 & 3.469 & \checkmark & 1 & \memoryerror & ~  \\ 

        40-\Range & 80 & 121 & 200 & 0.06242 & 35.89 & \checkmark & 1 & \memoryerror & ~  \\ 

        80-\Range & 160 & 241 & 400 & 0.25867 & 506.3  & \checkmark & 1 & \memoryerror & ~  \\
    \hline
    \end{tabular}

    \captionsetup{font=footnotesize,labelfont=footnotesize}
    \caption{Summary of experimental results for {\toolaut} and comparison with {\CheckDP}. The columns in the table are as follows. $v$ is the number of variables in the automaton. $s$ is the number of states in the automaton. $trans$ is the number of transitions in the automaton. The weight calculation time and total time taken by {\toolaut} averaged over six executions are reported next, and are measured in seconds.  Differentially private indicates if the automaton is differentially private or not. In case, it is not, we report the reason detected by the tool: DC/PV/LC/LP means that disclosing cycle/privacy-violating path/leaking cycle/leaking pair, respectively is detected. $\newd$ is the weight of the automaton computed by the algorithm in case it is differentially private. 
    For {\CheckDP}, the time column indicates the running time for CheckDP measured in seconds.  The last column indicates the time taken for counterexample validation by {\PSI} in case a counterexample is generated. {\timeouta} denotes that the tool did not finish in $30$ minutes. {\memoryerror} indicates that {\CheckDP} reported a memory error.
    }
    \label{table:simple-examples}}
\end{table*}

\subsection{Description of Examples}
\label{sec:expdesc}
 The first examples we consider are the standard Sparse Vector Technique (\SVT)~\cite{DNRRV09} and the Numeric Sparse (\Spse)~\cite{DR14}. These algorithms use one variable. Detailed discussion of these algorithms can be found in~\cite{DNRRV09,DR14}. Apart from {\SVT} and {\Spse}, all other examples use more than one variable.
 \ifdefined\short
 \else
The details of these algorithms are also located in Appendix~\ref{app:experiments}. 
\fi
 

We also designed new examples, described below.  The first set of examples was designed to ensure that the tests of well-formedness were implemented correctly. 
A second set of examples were designed to evaluate the scalability of our tool. They include $k$-{\minmax} (for each $k>0$) and $m$-{\Range} (for each $m>0$). The {$1$-\Range} is the range query algorithm given in Example~\ref{ex:range-query-pgm}. 

\subsubsection*{Examples {\LC} and {\DC}} The algorithm {\LC} and {\DC} are variants of {$1$-\Range}.
The algorithm {\LC} is designed to have a {\criticalcycle} and {\DC} is designed to have a {\violatingc}. A detailed description of the algorithms can be found in 
\ifdefined\short
\cite{ChadhaSVB23}.
\else
Appendix ~\ref{app:experiments}.
\fi

\subsubsection*{Examples {\NumericRangeOne} and {\NumericRangeTwo}}

The algorithm {\NumericRangeOne} is the variant of {$1$-Range} which outputs $\svar$ (instead of $\top$) when the sampled value $q[i]$ is greater than $\mathsf{high}.$ The algorithm {\NumericRangeTwo} on the other hand outputs $\svar'.$ {\NumericRangeTwo} is well-formed, output-distinct and hence differentially private but {\NumericRangeOne} has a  privacy-violating path. A detailed description of the algorithms can be found in 
\ifdefined\short
\cite{ChadhaSVB23}.
\else
Appendix ~\ref{app:experiments}.
\fi

\subsubsection*{Examples {\TRangeOne} and {\TRangeTwo}}
{\TRangeOne} is a variant of {$1$-\Range}.  In both algorithms, at the beginning, three thresholds, $T_\ell$, $T_m$, and $T_u$, are perturbed by adding noise sampled from the Laplace distribution. The algorithms then proceed to process the queries, checking if the remaining noisy queries are between the noisy $T_\ell$ and $T_m.$ If at some point the input noisy query exceeds the noisy $T_m$, {\TRangeOne} checks that the remaining queries are in between the noisy $T_m$ and the noisy $T_u.$ In contrast, the algorithm {\TRangeTwo} resamples $T_m$ before checking that the remaining queries are in between the noisy $T_m$ and the noisy $T_u.$ 
{\TRangeOne} has a {\criticalpair} and is not differentially privacy. {\TRangeTwo}, on the other hand, is well-formed, output distinct, and hence differentially private {\TRangeOne} and {\TRangeTwo} are described in detail in
\ifdefined\short
\cite{ChadhaSVB23}.
\else
Appendix ~\ref{app:experiments}.
\fi

\subsubsection*{Example $k$-\minmax}
\RestyleAlgo{boxed} 
\begin{algorithm}
\DontPrintSemicolon
\SetAlgoLined

\KwIn{$q[1:N]$}
\KwOut{$out[1:N]$}
\;

$\s{min}, \s{max} \gets \Lap{ \frac{\epsilon}{4k},q[1])}$\;

\For{$i\gets 2$ \KwTo $k$}
{
    
    $\rv\gets \Lap{\frac{\epsilon}{4k} , q[i]}$\;
    
    \uIf{$(\rv > \s{max}) \wedge (\rv > \s{min})$}{
        $\s{max} \gets \rv$\;
    }\ElseIf{$(\rv < \s{min}) \wedge (\rv < \s{max})$ } {
        $\s{min} \gets \rv$\;
        
    }
    $out[i] \gets \mathsf{read}$
}

\For{$i\gets k+1$ \KwTo $N$}
{
    $\rv\gets \Lap{\frac{\epsilon}{4} , q[i]}$\;
    \uIf{$(\rv \geq \s{min}) \wedge (\rv < \s{max})$}{
      $out[i] \gets \bot$}
    \uElseIf{$(\rv \geq \s{min}) \wedge (\rv \geq \s{max})$}{
      $out[i] \gets \top$\;
      exit
    } \ElseIf{$(\rv < \s{min}) \wedge (\rv < \s{max})$}{
      $out[i] \gets \bot$\;
      exit
    }

}

\caption{$k$-{\minmax} algorithm. $k$-{\minmax} is differentially private.}
\label{fig:k-min-max}
\end{algorithm}
One set of examples designed to check scalability of our algorithm is {$k$-\minmax} ($k\geq 2$). Initially, {$k$-\minmax} reads $k$-queries, adds noise from the Laplace distribution at each step, remembering the maximum and minimum amongst the perturbed queries. During this phase, the outputs do not inform the observer whether the noisy query being processed updates the maximum or minimum. 

After reading the first $k$-queries, each subsequent query is perturbed by adding noise, and the algorithm checks if the noisy query is between the maximum and minimum found in the first $k$-noisy queries. It continues processing the queries as long as it is between those two. Otherwise, it quits.  
Observe that  {$k$-\minmax} is a parametric set of examples, one for each value of $k$. For each $k$, the {\dipa} modeling {$k$-\minmax} has $two$ variables, has $k+2$ states and $3k+1$ transitions. Further, {$k$-\minmax} does not satisfy output distinction for any $k$ as the outputs do not distinguish whether maximum or minimum is being updated in the first phase. However, it is well-formed and $\epsilon$-differentially private. Psuedocode for $k$-{\minmax} is shown as Algorithm~\ref{fig:k-min-max}. 

\subsubsection*{Examples $m$-Range} Another set of examples for scalability is {$m$-\Range} (for each $m$). {$m$-\Range} is the $m$-dimensional version of {\Range}. It repeatedly checks whether a sequence of points
in the $m$-dimensional space is contained in a $m$-dimensional rectangle. The rectangle is specified by giving the upper and lower threshold for each coordinate of the rectangle. The algorithm initially adds Laplacian noise to each of these $2m$ thresholds, then processes the points by adding noise to each coordinate and checking that each noisy coordinate is within the noisy thresholds for that coordinate.  Observe that  {$m$-\Range} is a set of examples, one for each $m.$ For each $m$, the {\dipa} modeling {$m$-\Range} has $2m$ variables, has $3m+1$ states and $5m$ transitions. For each $m$, {$m$-\Range} satisfies output distinction,  is well-formed, and is $\epsilon$-differentially private. 
$m$-{\Range} is given in Algorithm~\ref{fig:mRange2}. {Here the arrays $T_1$ and $T_2$ store the $m$-lower and $m$-upper thresholds, respectively.
The arrays $\s{low}$ and $\s{high}$ store the noisy version of the lower and upper thresholds.} 
In the experiments, $T_1$ is taken to be all $0$s, and $T_2$ is taken to be all $1$s.

\RestyleAlgo{boxed} 
\begin{algorithm}
\DontPrintSemicolon
\SetAlgoLined

\KwIn{$q[1:m]$}
\KwOut{$out[1:Nm]$}
\;

\For{$j\gets 1$ \KwTo $m$}{
    $\s{low[j]} \gets \Lap{ \frac{\epsilon}{4m} , T_1[j]}$\;
    $\s{high[j]} \gets \Lap{ \frac{\epsilon}{4m}, T_2[j]}$\;
    $out[j] \gets \s{cont}$\;
}

\For{$i\gets 1$ \KwTo $N$}
{
    \For{$j\gets 1$ \KwTo $m$}{
        $\rv\gets \Lap{\frac{\epsilon}{4} , q[m(i-1) + j]}$\;
        \uIf{$(\rv \geq \s{low[j]}) \wedge (\rv < \s{high[j]})$}{
        $out[m(i-1) + j] \gets \s{cont}$\;  }
        \ElseIf{$((\rv \geq \s{low[j]}) \wedge (\rv > \s{high[j]}))$}{
          $out[m(i-1) + j] \gets \top$\;
          exit
        } 
        \ElseIf{$((\rv < \s{low[j]}) \wedge (\rv < \s{high[j]}))$}{
          $out[m(i-1) + j] \gets \bot $\;
          exit
        }
    }
}
\caption{$m$-{\Range} algorithm. $m$-Range is differentially private.}
\label{fig:mRange2}
\end{algorithm}

\subsection{Summary of experimental results}

\begin{figure}
  \centering
  \includegraphics[width=0.8\columnwidth]{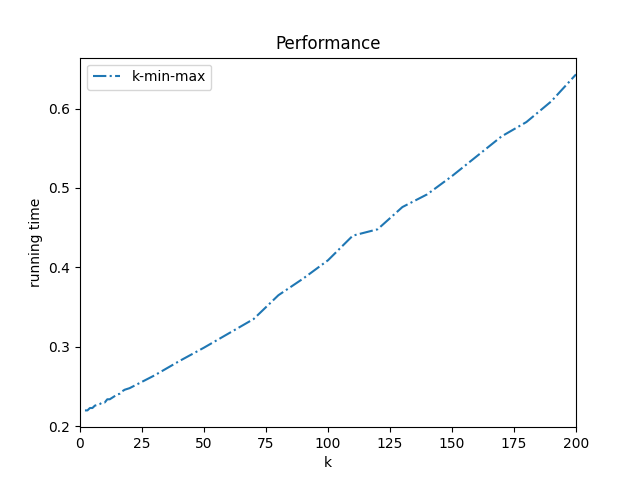}
  \captionsetup{font=footnotesize,labelfont=footnotesize}
  \caption{
  Running time for $k$-\minmax. The $y$-axis gives  the running time measured in seconds, while the $x$-axis gives $k$. The size of the {\dipa} is linear in $k.$  $k$-{\minmax} is differentially private with weight $1.$
  }
  \label{fig:kminmax}
\end{figure}

\begin{figure}
  \centering
  \includegraphics[width=0.8\columnwidth]{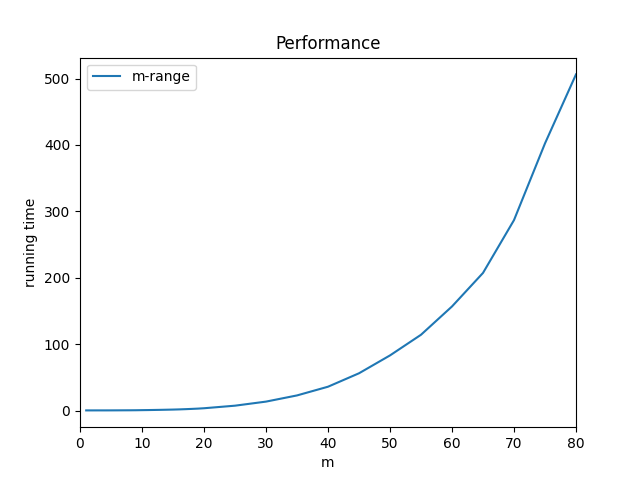}
  \captionsetup{font=footnotesize,labelfont=footnotesize}
  \caption{
  Running time for $m$-\Range. The $y$-axis gives  the running time measured in seconds, while the $x$-axis gives $m$. The size of the {\dipa} is linear in $m.$ $m$-{\Range} is differentially private with weight $1.$}
   \label{fig:mRange}
\end{figure}

The experimental results are summarized in Table~\ref{table:simple-examples}. All experiments were run on a macOS computer with a 1.4 GHz Quad-Core Intel Core i5 CPU processor with 8GB RAM. The running time is benchmarked using {\pyperf}~\cite{pyperf}, which runs each example 6 times and takes the average over the 6 instances. Figure~\ref{fig:kminmax} plots the running time of our implementation for $k$-{\minmax}. As predicted, the tool confirms that $k$-{\minmax} is $\epsilon$-differentially private. A close examination  of the algorithm for checking well-formedness reveals that the algorithm can check the well-formedness of $k$-{\minmax} in time that is linear in $k$. This  observation is confirmed by the experimental results. 
Note that the size of the {\dipa} modeling $k$-{\minmax} is linear in $k$, and hence the running time is also linear in the size of {\dipa}. In contrast,
a careful analysis reveals that the algorithm checking well-formedness takes time that is cubic in $m$ for $m$-{\Range}. This  observation is also confirmed by the experimental results. 
(See Figure~\ref{fig:mRange}).  As predicted, the tool confirms that $m$-{\Range} is $\epsilon$-differentially private.  Note that the number of variables in $m$-{\Range} is $2m$, implying a quartic dependence on the number of variables as well. Data used to generate the graphs is given in
\ifdefined\short
{\cite{ChadhaSVB23}.}
\else {Appendix~\ref{app:experiments}}.
\fi

Salient observations about our tool are as follows:
\begin{enumerate}
\item {\toolaut} is able to check whether the algorithm described by a {\dipa} is well-formed in reasonable time. 
\item In case the automaton $\cA$ is well-formed, it is able to compute a weight $\newd$ that $\cA$ is $\newd\epsilon$-differentially private. The computed values match the theoretical values. Further, the computation of weight has little overhead. 
\item As predicted by the theory, the number of variables plays a crucial role in performance. 
{While the theory predicts that this dependence is exponential (since the augmentation can be of exponential size), 
nevertheless, there are interesting examples in which the dependence is polynomial and not exponential.} 
\item {\toolaut} is not only able to verify differential privacy for examples but also find violations of privacy in a reasonable time, as shown in Table~\ref{table:simple-examples}. 
\end{enumerate}

\subsubsection*{Comparison with {\CheckDP}}
We compare the performance of our tool, {\toolaut} with {\CheckDP}~\cite{CheckDP}. 
CheckDP employs the randomness alignment technique and attempts to  prove differential privacy. If it fails to prove differential privacy, it generates a potential counterexample that must be validated using the {\PSI} probabilistic model checker~\cite{gehr2016psi}. The key differences between {\CheckDP} and {\toolaut} are as follows: \begin{enumerate*}
\item {\CheckDP} supports other arithmetic operations besides comparison operators.
\item However, {\CheckDP} is sound but incomplete and may fail to prove or disprove differential privacy.
\item {\CheckDP} checks if a program is $\newd\epsilon$ differentially private for a given $\newd$. {\toolaut}, on the other hand, computes a $\newd$ for which the program is $\newd\epsilon$ differentially private. \item {\toolaut} operates as a standalone tool, assessing the differential privacy of a given mechanism.  \end{enumerate*} 
The results of the comparison are summarized in Table~\ref{table:simple-examples}. Apart from {\SVT} and {\Spse}, {\CheckDP} times out on all other examples. For those two examples, {\toolaut} significantly outperforms {\CheckDP}.

\section{Related Work}
\label{sec:related}

\subsubsection*{Online Programs and Comparison with~\cite{ChadhaSV21}}
The results in this paper are an extension of those presented in~\cite{ChadhaSV21}. However, the automaton model proposed in~\cite{ChadhaSV21} has only one storage variable, whereas we consider the generalization where the automaton has finitely many real-valued storage variables. Even though we use {the same} name for the automata model and for the conditions characterizing well-formed {\dipa}, the generalization to handle multiple real-valued storage variables is a significant extension. Defining {\criticalcycle}s, {\criticalpair}s, {\violatingp}s and {\violatingc}s, requires a careful analysis of the ordering constraints imposed on values sampled in a run based on what gets stored in variables and the Boolean constraints that guard transitions. These concepts cannot be defined using just the underlying graph of the {\dipa} as in~\cite{ChadhaSV21}; they require introducing the notion of a dependency graph of a run. Even with dependency graphs, the definition of these graph-theoretic conditions is subtle. For example, two cycles contained in a run may not form a {\criticalpair}. However, they may become a leaking pair in an extension of the run as the additional transitions in the extension introduce new dependencies in the dependency graph (see Example \ref{ex:leakingpair} on Page \pageref{ex:leakingpair}). In the case of a single variable~\cite{ChadhaSV21}, such a situation does not arise.

Next, even though the proof showing that well-formedness is necessary for an output-distinct {\dipa} to be differentially private uses a strategy similar to the case for one variable~\cite{ChadhaSV21}, it is significantly more involved. For example, in showing that a {\criticalcycle} is a witness to privacy violation, complications arise due to the need to track the dependency between multiple storage variables and the presence of non-input transitions. When constructing a pair of adjacent inputs that witness the violation of privacy, intervals of real numbers called \emph{bands} need to be carefully identified, where the input of certain transitions is restricted to lie (see \ifdefined\short
\cite{ChadhaSVB23}
\else 
Appendix~\ref{app:necessity}
\fi). The proof that a {\criticalpair} is a witness to privacy violation uses new ideas. In~\cite{ChadhaSV21}, the proof constructs, given $\newd$, two adjacent computations whose ratio is $> \eulerv{\newd \epsilon}$ for each $\epsilon>0$. In this paper, the adjacent computations have a ratio  $> \eulerv{\newd\epsilon}$ only for sufficiently large $\epsilon$.

The proof showing that a well-formed {\dipa} is differentially private is also innovative. In~\cite{ChadhaSV21}, the proof is by induction on the number of assignments to the stored variable in a run. In contrast, here the induction is on the number of transitions in a run, and the induction hypothesis is constructed by classifying the dependency graph nodes as $\gcyclevertex$ or $\lcyclevertex.$ \ifdefined \short
\else (See Appendix~\ref{app:sufficient}). 
\fi

\subsubsection*{Privacy proof construction}
Techniques based on type systems have been proposed in many papers~\cite{GHHNP13, AGHK18,RP10,AmorimAGH15,ZK17,CheckDP} for generating  proofs of differential privacy. Some of these methods such as \cite{GHHNP13, AGHK18,RP10,AmorimAGH15} employ linear dependent types, for which the type-checking and type-inference may be challenging. 
In ~\cite{BKOZ13,BGGHS16,BFGGHS16,AH18} methods based on probabilistic couplings and random alignment arguments have been employed for proving differential privacy.
Shadow execution-based method was introduced in~\cite{WDWKZ19}.
Probabilistic I/O automata are used in~\cite{TschantzKD11} to model interactive differential privacy algorithms and simulation-based methods are used to verify differential privacy, but these methods have not been shown to be complete.

%

\subsubsection*{Counterexample generation} 
Automated techniques to search for privacy violations by generating counter examples have been proposed in~\cite{DingWWZK18,BichselGDTV18,CheckDP}. Techniques include the use of statistical hypothesis testing~\cite{DingWWZK18}, optimization techniques and symbolic differentiation~\cite{BichselGDTV18} and program analysis~\cite{CheckDP}. These methods search over a bounded number of inputs.



\subsubsection*{Model-checking/Markov Chain approaches}
Probabilistic model checking approach for verifying $\epsilon$-differential privacy is employed in~\cite{ChatzGP14,LiuWZ18,ChistikovKMP20}, where it is assumed that the program is given as a finite Markov Chain.
These approaches do not allow for sampling from continuous random variables. 
\ignore{Instead, they assume that the program behavior is given as a finite Markov Chain, and the transition probabilities are specified as inputs. Thus, they also implicitly assume a bounded sequence of inputs and a concrete value of $\epsilon.$
In~\cite{ChistikovKMP20}, the authors use labeled Markov Chains to model differential privacy algorithms. 
They consider discrete probabilities only. They only model inputs taking values from a finite set and implicitly assume a concrete value of $\epsilon.$ Further, they check whether the ratio of probabilities of observations on neighboring inputs is bounded by a constant. If it is bounded, it implies  the algorithm is $\epsilon$-differentially private for sufficiently large $\epsilon$. 
}


\subsubsection*{Decision Procedures}
The decision problem of checking whether a randomized program is differentially private is studied in~\cite{BartheCJS020}, where it is shown to be undecidable for programs with a single input and single output, assuming that the program can sample from Laplacian distributions. A decidable sub-class is identified where the inputs and outputs are constrained to be from a finite domain and have bounded length.   
\ignore{The decision procedure in~\cite{BartheCJS020}
relies on the decision procedure for checking the validity of a sentence in the fragment of the theory of Reals with exponentiation identified in~\cite{mccallum2012deciding}, and has very high complexity.}

\subsubsection*{Complexity} Gaboardi et. al~\cite{GaboardiNP19} study the complexity of deciding differential privacy for randomized Boolean circuits, and show that the problem is $\mathbf{coNP^{\#P}}$-complete. 
The results are extended to Boolean programs~\cite{Gaboardi22} for which the verification problem is {\pspace}-complete. In this line of work, programs have a finite number of inputs, the only probabilistic choices are fair coin tosses, and $\euler^{\epsilon}$ is taken to be a fixed rational number.

\section{Discussion}
\label{sec:refelction}

We discuss the restrictions used in various definitions in this paper.  

\subsubsection*{Strong feasability} 
From the theoretical point of view, strong feasibility is used only to prove the necessity of well-formedness (Theorem~\ref{thm:necessity}). The sufficiency proof (Theorem~\ref{thm:main}) does not require the condition of strong feasibility. Nevertheless, we believe that all differential privacy mechanisms are strongly feasible. We have not encountered examples that violate the strong feasibility condition. Our intuition for this belief is as follows. First, any DiPA that \emph{does not} have any non-input states is, by definition, strongly feasible. For DiPA with non-input states, the condition implies that the mean of the distribution at any two non-input states respects the order given by the dependency graph of a run. Let us consider the {\lq\lq}deterministic{\rq\rq} version of the automaton in which no noise is added. Intuitively, the {\lq\lq}deterministic{\rq\rq} version captures the behavior of the automaton in the limit as the privacy budget $\epsilon$ tends to infinity, i.e., becomes unlimited.  A strongly feasible run implies that we can choose inputs such that the probability of that run tends to 1 as $\epsilon$ tends to $\infty$ and is executable in the {\lq\lq}deterministic{\rq\rq} version. A path that is not strongly feasible implies that the probability of this path tends to $0$ as $\epsilon$ tends to $\infty$, irrespective of the choice of inputs, and will never be executed in the "deterministic version" because the insample values stored at the non-input states do not follow the order given by the dependency graph. The deterministic version of the automaton is relevant as a differentially private algorithm is often the noisy version of a deterministic algorithm (with noise added to make the automaton differentially private). 

\subsubsection*{Output-distinction} Some examples do not meet the condition output distinction. For example, the $k$-{\minmax}  
 (See Section~\ref{sec:expdesc}) and {\Noisy} ~\cite{DR14} are \emph{not} output distinct. However, other examples (m-Range, SVT, NumericSparse) are output distinct. The output distinction condition is only needed to establish necessity but not for sufficiency. In other words, if an automaton is well-formed, it is differentially private, \emph{even if it is not output distinct}. This is true for the $k$-{\minmax} examples. However, the traditional {\Noisy} is neither well-formed nor output distinct, and hence our technique does not establish its differential privacy. Some variants of {\Noisy} (like checking if the $k$th input is maximum) are well-formed and hence can be handled by our techniques.

\subsubsection*{Adjaceny Relations}
For algorithms working on a sequence of answers to queries on a database like {\SVT} and {\Spse} (see ~\cite{DR14}, pages 56 and 57), the assumption that queries are \emph{1-sensitive} is common; here $1$-sensitive means that adding or removing a member from a database can cause a difference of at most $1$ in the output of each query. This assumption is satisfied by all counting queries and can be found in Algorithms 1, 2, 3 in~\cite{DR14} on pages 58, 62, 64, first paragraph on page 5 of~\cite{AH18} and third paragraph of Section 4 in~\cite{DingWWZK18}.

More generally, our results also apply to a sequence of queries each of which is $\Delta$-sensitive. The computation of $\newd$ will change, but the theorems of the sufficiency of well-formedness and necessity for well-formedness for output distinct {\dipautop} remain true.  

\subsubsection*{Boolean Guards on transitions in {\criticalcycle}} 
In the definition of a {\criticalcycle} (see Definition~\ref{def:non-leak}),
it is possible that the constraint involving $\rvar$ in the guard of $\exec[i_2]$ is superfluous. When this happens, there have to be other variables in the guard of $\exec[i_2]$. However, we can show that after removing all superfluous checks from $\exec[i_2]$, either the original cycle will be a leaking cycle for some (possibly different) variable, or the leaking cycle gives rise to a leaking pair when repeated twice. Therefore, in principle, even a superfluous test does leak information (though indirectly).

\subsubsection*{The expressiveness of multi-variable {\dipautop}  vs one-variable {\dipautop}}
We can prove that multi-variable {\dipautop} are strictly more expressive than one-variable {\dipautop}. For example, we can formally show that the {\dipautop} $\mathcal{A}_{\Range}$ (See Figure ~\ref{fig:range-auto}) cannot be modeled using single-variable {\dipautop}.

\section{Conclusions}
\label{sec:conclusions}

\sloppy
We 
extended the {\dipautos} model introduced in~\cite{ChadhaSV21}  for modeling online algorithms that process a stream of unbounded real values representing answers to queries and, in response, produce a sequence of real or discrete output values. In the extended model, a {\dipa} $\cA$ may use \emph{multiple} storage variables to store noisy input values when executing transitions that are used in Boolean conditions that guard transitions. 
Our main contribution is a precise characterization of when {\dipa}s are differentially private using the notion of well-formed automata. The definition of well-formed automata is subtle and complicated, and requires the use of new graph structures associated with the runs of the automata, called dependency graphs. Well-formed {\dipa}s are shown to be differentially private and {\dipa}s satisfying the condition of {\outcond} that are differentially private are necessarily well-formed. The problem of checking well-formedness is {\pspace}-complete. The algorithm for checking differential privacy has been implemented in a tool called {\toolaut}, and our experimental results demonstrate its promise. 

As future work, it will be interesting to identify necessary conditions for classes of automata that do not satisfy the {\outcond} property. Extending {\dipa}s to allow a richer class of comparisons in the guards and a richer class of assignments, like using expressions involving additions of storage variables and/or constants in the guard conditions, is left for future exploration. Computing the optimal weight $\newd$ is another open problem.


\subsubsection*{Acknowledgements}
The authors would like to thank Lipsy Gupta and the anonymous reviewers for their interesting and valuable comments.  Rohit Chadha was partially supported by NSF CNS 1553548 and NSF CCF 1900924. A. Prasad Sistla was partially supported by NSF CCF 1901069, Mahesh Viswanathan was partially supported by NSF CCF 1901069 and NSF CCF 2007428, and Bishnu Bhusal was partially supported by NSF CCF 1900924.


\bibliographystyle{ACM-Reference-Format}
\bibliography{header.bib, main.bib}


\newcommand{\SortNoop}[1]{}
\begin{thebibliography}{31}


\ifx \showCODEN    \undefined \def \showCODEN     #1{\unskip}     \fi
\ifx \showDOI      \undefined \def \showDOI       #1{#1}\fi
\ifx \showISBNx    \undefined \def \showISBNx     #1{\unskip}     \fi
\ifx \showISBNxiii \undefined \def \showISBNxiii  #1{\unskip}     \fi
\ifx \showISSN     \undefined \def \showISSN      #1{\unskip}     \fi
\ifx \showLCCN     \undefined \def \showLCCN      #1{\unskip}     \fi
\ifx \shownote     \undefined \def \shownote      #1{#1}          \fi
\ifx \showarticletitle \undefined \def \showarticletitle #1{#1}   \fi
\ifx \showURL      \undefined \def \showURL       {\relax}        \fi
\providecommand\bibfield[2]{#2}
\providecommand\bibinfo[2]{#2}
\providecommand\natexlab[1]{#1}
\providecommand\showeprint[2][]{arXiv:#2}

\bibitem[\protect\citeauthoryear{Albarghouti and Hsu}{Albarghouti and
  Hsu}{2018}]%
        {AH18}
\bibfield{author}{\bibinfo{person}{Aws Albarghouti} {and}
  \bibinfo{person}{Justin Hsu}.} \bibinfo{year}{2018}\natexlab{}.
\newblock \showarticletitle{Synthesizing coupling proofs of differential
  privacy}. In \bibinfo{booktitle}{\emph{Proceedings of the ACM SIGPLAN
  Symposium on Principles of Programming Languages (POPL)}}.
  \bibinfo{pages}{58:1--58:30}.
\newblock


\bibitem[\protect\citeauthoryear{Barthe, Chadha, Jagannath, Sistla, and
  Viswanathan}{Barthe et~al\mbox{.}}{2020}]%
        {BartheCJS020}
\bibfield{author}{\bibinfo{person}{Gilles Barthe}, \bibinfo{person}{Rohit
  Chadha}, \bibinfo{person}{Vishal Jagannath}, \bibinfo{person}{A.~Prasad
  Sistla}, {and} \bibinfo{person}{Mahesh Viswanathan}.}
  \bibinfo{year}{2020}\natexlab{}.
\newblock \showarticletitle{Deciding Differential Privacy for Programs with
  Finite Inputs and Outputs}. In \bibinfo{booktitle}{\emph{35th Annual
  {ACM/IEEE} Symposium on Logic in Computer Science}}.
  \bibinfo{pages}{141--154}.
\newblock


\bibitem[\protect\citeauthoryear{Barthe, Fong, Gaboardi, Gr{\'{e}}goire, Hsu,
  and Strub}{Barthe et~al\mbox{.}}{2016a}]%
        {BFGGHS16}
\bibfield{author}{\bibinfo{person}{Gilles Barthe},
  \bibinfo{person}{No{\'{e}}mie Fong}, \bibinfo{person}{Marco Gaboardi},
  \bibinfo{person}{Benjamin Gr{\'{e}}goire}, \bibinfo{person}{Justin Hsu},
  {and} \bibinfo{person}{Pierre{-}Yves Strub}.}
  \bibinfo{year}{2016}\natexlab{a}.
\newblock \showarticletitle{Advanced Probabilistic Couplings for Differential
  Privacy}. In \bibinfo{booktitle}{\emph{Proceedings of the 2016 {ACM} {SIGSAC}
  Conference on Computer and Communications Security (CCS)}}.
  \bibinfo{pages}{55--67}.
\newblock


\bibitem[\protect\citeauthoryear{Barthe, Gaboardi, Gr{\'e}goire, Hsu, and
  Strub}{Barthe et~al\mbox{.}}{2016b}]%
        {BGGHS16}
\bibfield{author}{\bibinfo{person}{Gilles Barthe}, \bibinfo{person}{Marco
  Gaboardi}, \bibinfo{person}{Benjamin Gr{\'e}goire}, \bibinfo{person}{Justin
  Hsu}, {and} \bibinfo{person}{{P}ierre-{Y}ves Strub}.}
  \bibinfo{year}{2016}\natexlab{b}.
\newblock \showarticletitle{Proving differential privacy via probabilistic
  couplings}. In \bibinfo{booktitle}{\emph{{IEEE} {S}ymposium on {L}ogic in
  {C}omputer {S}cience ({LICS})}}. \bibinfo{pages}{749--758}.
\newblock


\bibitem[\protect\citeauthoryear{Barthe, K{\"{o}}pf, Olmedo, and
  Zanella{-}B{\'{e}}guelin}{Barthe et~al\mbox{.}}{2013}]%
        {BKOZ13}
\bibfield{author}{\bibinfo{person}{Gilles Barthe}, \bibinfo{person}{Boris
  K{\"{o}}pf}, \bibinfo{person}{Federico Olmedo}, {and}
  \bibinfo{person}{Santiago Zanella{-}B{\'{e}}guelin}.}
  \bibinfo{year}{2013}\natexlab{}.
\newblock \showarticletitle{Probabilistic Relational Reasoning for Differential
  Privacy}.
\newblock \bibinfo{journal}{\emph{ACM Transactions on Programming Languages and
  Systems}} \bibinfo{volume}{35}, \bibinfo{number}{3} (\bibinfo{year}{2013}),
  \bibinfo{pages}{9}.
\newblock


\bibitem[\protect\citeauthoryear{Beazley}{Beazley}{2022}]%
        {PLY}
\bibfield{author}{\bibinfo{person}{David Beazley}.}
  \bibinfo{year}{2022}\natexlab{}.
\newblock \bibinfo{title}{{G}it{H}ub - dabeaz/ply: {P}ython {L}ex-{Y}acc ---
  github.com}.
\newblock \bibinfo{howpublished}{\url{https://github.com/dabeaz/ply}}.
\newblock
\newblock
\shownote{[Accessed 24-Jan-2023].}


\bibitem[\protect\citeauthoryear{Bhusal, Chadha, Sistla, and
  Viswanathan}{Bhusal et~al\mbox{.}}{2023}]%
        {DipAut}
\bibfield{author}{\bibinfo{person}{Bishnu Bhusal}, \bibinfo{person}{Rohit
  Chadha}, \bibinfo{person}{A.~Prasad Sistla}, {and} \bibinfo{person}{Mahesh
  Viswanathan}.} \bibinfo{year}{2023}\natexlab{}.
\newblock \bibinfo{title}{bhusalb/DiPAut: Version 1.0.1}.
\newblock
\newblock
\urldef\tempurl%
\url{https://doi.org/10.5281/zenodo.8332275}
\showDOI{\tempurl}


\bibitem[\protect\citeauthoryear{Bichsel, Gehr, Drachsler{-}Cohen, Tsankov, and
  Vechev}{Bichsel et~al\mbox{.}}{2018}]%
        {BichselGDTV18}
\bibfield{author}{\bibinfo{person}{Benjamin Bichsel}, \bibinfo{person}{Timon
  Gehr}, \bibinfo{person}{Dana Drachsler{-}Cohen}, \bibinfo{person}{Petar
  Tsankov}, {and} \bibinfo{person}{Martin~T. Vechev}.}
  \bibinfo{year}{2018}\natexlab{}.
\newblock \showarticletitle{{DP-Finder}: Finding Differential Privacy
  Violations by Sampling and Optimization}. In
  \bibinfo{booktitle}{\emph{Proceedings of the 2018 {ACM} {SIGSAC} Conference
  on Computer and Communications Security ({CCS})}}. \bibinfo{pages}{508--524}.
\newblock


\bibitem[\protect\citeauthoryear{Bun, Gaboardi, and Glinskih}{Bun
  et~al\mbox{.}}{2022}]%
        {Gaboardi22}
\bibfield{author}{\bibinfo{person}{Mark Bun}, \bibinfo{person}{Marco Gaboardi},
  {and} \bibinfo{person}{Ludmila Glinskih}.} \bibinfo{year}{2022}\natexlab{}.
\newblock \showarticletitle{The Complexity of Verifying Boolean Programs as
  Differentially Private}. In \bibinfo{booktitle}{\emph{2022 IEEE 35th Computer
  Security Foundations Symposium (CSF)}}. \bibinfo{pages}{396--411}.
\newblock
\urldef\tempurl%
\url{https://doi.org/10.1109/CSF54842.2022.9919653}
\showDOI{\tempurl}


\bibitem[\protect\citeauthoryear{Chadha, Sistla, and Viswanathan}{Chadha
  et~al\mbox{.}}{2021a}]%
        {ChadhaSV21}
\bibfield{author}{\bibinfo{person}{Rohit Chadha}, \bibinfo{person}{A.~Prasad
  Sistla}, {and} \bibinfo{person}{Mahesh Viswanathan}.}
  \bibinfo{year}{2021}\natexlab{a}.
\newblock \showarticletitle{On Linear Time Decidability of Differential Privacy
  for Programs with Unbounded Inputs}. In \bibinfo{booktitle}{\emph{36th Annual
  {IEEE} Symposium on Logic in Computer Science (LICS)}}.
  \bibinfo{pages}{1--13}.
\newblock


\bibitem[\protect\citeauthoryear{Chadha, Sistla, and Viswanathan}{Chadha
  et~al\mbox{.}}{2021b}]%
        {ChadhaSV21b}
\bibfield{author}{\bibinfo{person}{Rohit Chadha}, \bibinfo{person}{A.~Prasad
  Sistla}, {and} \bibinfo{person}{Mahesh Viswanathan}.}
  \bibinfo{year}{2021}\natexlab{b}.
\newblock \showarticletitle{On Linear Time Decidability of Differential Privacy
  for Programs with Unbounded Inputs}.
\newblock \bibinfo{journal}{\emph{CoRR}} (\bibinfo{year}{2021}).
\newblock
\showeprint[arxiv]{2104.14519}
\urldef\tempurl%
\url{https://arxiv.org/abs/2104.14519}
\showURL{%
\tempurl}


\bibitem[\protect\citeauthoryear{Chatzikokolakis, Gebler, Palamidessi, and
  Xu}{Chatzikokolakis et~al\mbox{.}}{2014}]%
        {ChatzGP14}
\bibfield{author}{\bibinfo{person}{Konstantinos Chatzikokolakis},
  \bibinfo{person}{Daniel Gebler}, \bibinfo{person}{Catuscia Palamidessi},
  {and} \bibinfo{person}{Lili Xu}.} \bibinfo{year}{2014}\natexlab{}.
\newblock \showarticletitle{Generalized Bisimulation Metrics}. In
  \bibinfo{booktitle}{\emph{35th International Conference on Concurrency Theory
  ({CONCUR})}}. \bibinfo{pages}{32--46}.
\newblock


\bibitem[\protect\citeauthoryear{Chistikov, Kiefer, Murawski, and
  Purser}{Chistikov et~al\mbox{.}}{2020}]%
        {ChistikovKMP20}
\bibfield{author}{\bibinfo{person}{Dmitry Chistikov}, \bibinfo{person}{Stefan
  Kiefer}, \bibinfo{person}{Andrzej~S. Murawski}, {and} \bibinfo{person}{David
  Purser}.} \bibinfo{year}{2020}\natexlab{}.
\newblock \showarticletitle{The Big-O Problem for Labelled Markov Chains and
  Weighted Automata}. In \bibinfo{booktitle}{\emph{31st International
  Conference on Concurrency Theory (CONCUR)}}
  \emph{(\bibinfo{series}{LIPIcs})}, Vol.~\bibinfo{volume}{171}.
  \bibinfo{pages}{41:1--41:19}.
\newblock


\bibitem[\protect\citeauthoryear{Csardi and Nepusz}{Csardi and Nepusz}{2006}]%
        {igraph}
\bibfield{author}{\bibinfo{person}{Gabor Csardi} {and} \bibinfo{person}{Tamas
  Nepusz}.} \bibinfo{year}{2006}\natexlab{}.
\newblock \showarticletitle{The igraph software package for complex network
  research}.
\newblock \bibinfo{journal}{\emph{InterJournal}}  \bibinfo{volume}{Complex
  Systems} (\bibinfo{year}{2006}), \bibinfo{pages}{1695}.
\newblock
\urldef\tempurl%
\url{https://igraph.org}
\showURL{%
\tempurl}


\bibitem[\protect\citeauthoryear{de~Amorim, Gaboardi, Arias, and Hsu}{de~Amorim
  et~al\mbox{.}}{2014}]%
        {AmorimAGH15}
\bibfield{author}{\bibinfo{person}{Arthur~Azevedo de Amorim},
  \bibinfo{person}{Marco Gaboardi}, \bibinfo{person}{Emilio Jes{\'{u}}s~Gallego
  Arias}, {and} \bibinfo{person}{Justin Hsu}.} \bibinfo{year}{2014}\natexlab{}.
\newblock \showarticletitle{Really Natural Linear Indexed Type Checking}. In
  \bibinfo{booktitle}{\emph{26th 2014 International Symposium on Implementation
  and Application of Functional Languages ({IFL})}}.
  \bibinfo{pages}{5:1--5:12}.
\newblock


\bibitem[\protect\citeauthoryear{de~Amorim, Gaboardi, Hsu, and
  Katsumata}{de~Amorim et~al\mbox{.}}{2019}]%
        {AGHK18}
\bibfield{author}{\bibinfo{person}{Arthur~Azevedo de Amorim},
  \bibinfo{person}{Marco Gaboardi}, \bibinfo{person}{Justin Hsu}, {and}
  \bibinfo{person}{Shin{-}ya Katsumata}.} \bibinfo{year}{2019}\natexlab{}.
\newblock \showarticletitle{Probabilistic Relational Reasoning via Metrics}. In
  \bibinfo{booktitle}{\emph{34th Annual {ACM/IEEE} Symposium on Logic in
  Computer Science ({LICS})}}. \bibinfo{pages}{1--19}.
\newblock


\bibitem[\protect\citeauthoryear{Ding, Wang, Wang, Zhang, and Kifer}{Ding
  et~al\mbox{.}}{2018}]%
        {DingWWZK18}
\bibfield{author}{\bibinfo{person}{Zeyu Ding}, \bibinfo{person}{Yuxin Wang},
  \bibinfo{person}{Guanhong Wang}, \bibinfo{person}{Danfeng Zhang}, {and}
  \bibinfo{person}{Daniel Kifer}.} \bibinfo{year}{2018}\natexlab{}.
\newblock \showarticletitle{Detecting Violations of Differential Privacy}. In
  \bibinfo{booktitle}{\emph{Proceedings of the 2018 {ACM} {SIGSAC} Conference
  on Computer and Communications Security (CCS)}}. \bibinfo{pages}{475--489}.
\newblock


\bibitem[\protect\citeauthoryear{Dwork, McSherry, Nissim, and Smith}{Dwork
  et~al\mbox{.}}{2006}]%
        {dmns06}
\bibfield{author}{\bibinfo{person}{Cynthia Dwork}, \bibinfo{person}{Frank
  McSherry}, \bibinfo{person}{Kobbi Nissim}, {and} \bibinfo{person}{Adam
  Smith}.} \bibinfo{year}{2006}\natexlab{}.
\newblock \showarticletitle{Calibrating noise to sensitivity in private data
  analysis}. In \bibinfo{booktitle}{\emph{{IACR} {T}heory of {C}ryptography
  {C}onference (TCC)}}. \bibinfo{pages}{265--284}.
\newblock


\bibitem[\protect\citeauthoryear{Dwork, Naor, Reingold, Rothblum, and
  Vadhan}{Dwork et~al\mbox{.}}{2009}]%
        {DNRRV09}
\bibfield{author}{\bibinfo{person}{Cynthia Dwork}, \bibinfo{person}{Moni Naor},
  \bibinfo{person}{Omer Reingold}, \bibinfo{person}{Guy~N. Rothblum}, {and}
  \bibinfo{person}{Salil~P. Vadhan}.} \bibinfo{year}{2009}\natexlab{}.
\newblock \showarticletitle{On the complexity of differentially private data
  release: efficient algorithms and hardness results}. In
  \bibinfo{booktitle}{\emph{{ACM} {SIGACT} {S}ymposium on {T}heory of
  {C}omputing (STOC)}}. \bibinfo{pages}{381--390}.
\newblock


\bibitem[\protect\citeauthoryear{Dwork and Roth}{Dwork and Roth}{2014}]%
        {DR14}
\bibfield{author}{\bibinfo{person}{Cynthia Dwork} {and} \bibinfo{person}{Aaron
  Roth}.} \bibinfo{year}{2014}\natexlab{}.
\newblock \showarticletitle{The Algorithmic Foundations of Differential
  Privacy}.
\newblock \bibinfo{journal}{\emph{Foundations and Trends in Theoretical
  Computer Science}} \bibinfo{volume}{9}, \bibinfo{number}{3--4}
  (\bibinfo{year}{2014}), \bibinfo{pages}{211--407}.
\newblock


\bibitem[\protect\citeauthoryear{Foundation}{Foundation}{2023}]%
        {pyperf}
\bibfield{author}{\bibinfo{person}{Python~Software Foundation}.}
  \bibinfo{year}{2023}\natexlab{}.
\newblock \bibinfo{title}{pyperf: A toolkit to write, run and analyze
  benchmarks}.
\newblock \bibinfo{howpublished}{\url{https://github.com/psf/pyperf}}.
\newblock
\newblock
\shownote{Accessed on 24-Jan-2023.}


\bibitem[\protect\citeauthoryear{Gaboardi, Haeberlen, Hsu, Narayan, and
  Pierce}{Gaboardi et~al\mbox{.}}{2013}]%
        {GHHNP13}
\bibfield{author}{\bibinfo{person}{Marco Gaboardi}, \bibinfo{person}{Andreas
  Haeberlen}, \bibinfo{person}{Justin Hsu}, \bibinfo{person}{Arjun Narayan},
  {and} \bibinfo{person}{Benjamin~C Pierce}.} \bibinfo{year}{2013}\natexlab{}.
\newblock \showarticletitle{Linear dependent types for differential privacy}.
  In \bibinfo{booktitle}{\emph{{ACM} {SIGPLAN--SIGACT} {S}ymposium on
  {P}rinciples of {P}rogramming {L}anguages ({POPL})}}.
  \bibinfo{pages}{357--370}.
\newblock


\bibitem[\protect\citeauthoryear{Gaboardi, Nissim, and Purser}{Gaboardi
  et~al\mbox{.}}{2020}]%
        {GaboardiNP19}
\bibfield{author}{\bibinfo{person}{Marco Gaboardi}, \bibinfo{person}{Kobbi
  Nissim}, {and} \bibinfo{person}{David Purser}.}
  \bibinfo{year}{2020}\natexlab{}.
\newblock \showarticletitle{The Complexity of Verifying Loop-Free Programs as
  Differentially Private}. In \bibinfo{booktitle}{\emph{47th International
  Colloquium on Automata, Languages, and Programming, ({ICALP})}}
  \emph{(\bibinfo{series}{LIPIcs})}, Vol.~\bibinfo{volume}{168}.
  \bibinfo{pages}{129:1--129:17}.
\newblock


\bibitem[\protect\citeauthoryear{Gehr, Misailovic, and Vechev}{Gehr
  et~al\mbox{.}}{2016}]%
        {gehr2016psi}
\bibfield{author}{\bibinfo{person}{Timon Gehr}, \bibinfo{person}{Sasa
  Misailovic}, {and} \bibinfo{person}{Martin Vechev}.}
  \bibinfo{year}{2016}\natexlab{}.
\newblock \showarticletitle{{PSI}: Exact Symbolic Inference for Probabilistic
  Programs}. In \bibinfo{booktitle}{\emph{International Conference on Computer
  Aided Verification}}. Springer, \bibinfo{pages}{62--83}.
\newblock


\bibitem[\protect\citeauthoryear{Liu, Wang, and Zhang}{Liu
  et~al\mbox{.}}{2018}]%
        {LiuWZ18}
\bibfield{author}{\bibinfo{person}{Depeng Liu}, \bibinfo{person}{Bow{-}Yaw
  Wang}, {and} \bibinfo{person}{Lijun Zhang}.} \bibinfo{year}{2018}\natexlab{}.
\newblock \showarticletitle{Model Checking Differentially Private Properties}.
  In \bibinfo{booktitle}{\emph{Programming Languages and Systems - 16th Asian
  Symposium, ({APLAS})}} \emph{(\bibinfo{series}{Lecture Notes in Computer
  Science})}, Vol.~\bibinfo{volume}{11275}. \bibinfo{pages}{394--414}.
\newblock


\bibitem[\protect\citeauthoryear{Lyu, Su, and Li}{Lyu et~al\mbox{.}}{2017}]%
        {lyu2016understanding}
\bibfield{author}{\bibinfo{person}{Min Lyu}, \bibinfo{person}{Dong Su}, {and}
  \bibinfo{person}{Ninghui Li}.} \bibinfo{year}{2017}\natexlab{}.
\newblock \showarticletitle{Understanding the Sparse Vector Technique for
  Differential Privacy}.
\newblock \bibinfo{journal}{\emph{{Proceedings of VLDB}}} \bibinfo{volume}{10},
  \bibinfo{number}{6} (\bibinfo{year}{2017}), \bibinfo{pages}{637--648}.
\newblock


\bibitem[\protect\citeauthoryear{Reed and Pierce}{Reed and Pierce}{2010}]%
        {RP10}
\bibfield{author}{\bibinfo{person}{Jason Reed} {and}
  \bibinfo{person}{Benjamin~C. Pierce}.} \bibinfo{year}{2010}\natexlab{}.
\newblock \showarticletitle{Distance Makes the Types Grow Stronger: A Calculus
  for Differential Privacy}. In \bibinfo{booktitle}{\emph{Proceedings of the
  15th ACM SIGPLAN International Conference on Functional Programming (ICFP)}}.
  \bibinfo{pages}{157–168}.
\newblock


\bibitem[\protect\citeauthoryear{Tschantz, Kaynar, and Datta}{Tschantz
  et~al\mbox{.}}{2011}]%
        {TschantzKD11}
\bibfield{author}{\bibinfo{person}{Michael~Carl Tschantz},
  \bibinfo{person}{Dilsun~Kirli Kaynar}, {and} \bibinfo{person}{Anupam Datta}.}
  \bibinfo{year}{2011}\natexlab{}.
\newblock \showarticletitle{Formal Verification of Differential Privacy for
  Interactive Systems (Extended Abstract)}. In \bibinfo{booktitle}{\emph{27th
  Conference on the Mathematical Foundations of Programming Semantics
  ({MFPS})}} \emph{(\bibinfo{series}{Electronic Notes in Theoretical Computer
  Science})}, Vol.~\bibinfo{volume}{276}. \bibinfo{pages}{61--79}.
\newblock


\bibitem[\protect\citeauthoryear{Wang, Ding, Kifer, and Zhang}{Wang
  et~al\mbox{.}}{2020}]%
        {CheckDP}
\bibfield{author}{\bibinfo{person}{Yuxin Wang}, \bibinfo{person}{Zeyu Ding},
  \bibinfo{person}{Daniel Kifer}, {and} \bibinfo{person}{Danfeng Zhang}.}
  \bibinfo{year}{2020}\natexlab{}.
\newblock \showarticletitle{{CheckDP}: An Automated and Integrated Approach for
  Proving Differential Privacy or Finding Precise Counterexamples}. In
  \bibinfo{booktitle}{\emph{2020 {ACM} {SIGSAC} Conference on Computer and
  Communications Security (CCS)}}. \bibinfo{pages}{919--938}.
\newblock


\bibitem[\protect\citeauthoryear{Wang, Ding, Wang, Kifer, and Zhang}{Wang
  et~al\mbox{.}}{2019}]%
        {WDWKZ19}
\bibfield{author}{\bibinfo{person}{Yuxin Wang}, \bibinfo{person}{Zeyu Ding},
  \bibinfo{person}{Guanhong Wang}, \bibinfo{person}{Daniel Kifer}, {and}
  \bibinfo{person}{Danfeng Zhang}.} \bibinfo{year}{2019}\natexlab{}.
\newblock \showarticletitle{Proving differential privacy with shadow
  execution}. In \bibinfo{booktitle}{\emph{Proceedings of the 40th {ACM}
  {SIGPLAN} Conference on Programming Language Design and Implementation,
  {(PLDI)}}}. \bibinfo{pages}{655--669}.
\newblock


\bibitem[\protect\citeauthoryear{Zhang and Kifer}{Zhang and Kifer}{2017}]%
        {ZK17}
\bibfield{author}{\bibinfo{person}{Danfeng Zhang} {and} \bibinfo{person}{Daniel
  Kifer}.} \bibinfo{year}{2017}\natexlab{}.
\newblock \showarticletitle{{LightDP}: towards automating differential privacy
  proofs}. In \bibinfo{booktitle}{\emph{Proceedings of the 44th {ACM} {SIGPLAN}
  Symposium on Principles of Programming Languages ({POPL})}}.
  \bibinfo{pages}{888--901}.
\newblock


\end{thebibliography}
\ifdefined\short
\else
\appendix

\section{Sufficiency of well-formedness}
\label{app:sufficient}
\subsection*{Well-formed automata are differentially private}
We shall now show that a well-formed {\dipa} is also differentially private. 
We start by constructing for each {\dipautos} $\cA$, another automaton $\aug \cA$, which we shall call the {\em augmentation} of $\cA.$ Intuitively, $\aug \cA$ captures all the paths of $\cA$ that occur with non-zero probability (See Proposition~\ref{prop:augmentation}). 


In the augmented automaton, $\aug \cA$, each state will carry additional information regarding the relationships that must hold amongst the values stored in the real variables $\rvar_i, 1\leq i \leq k.$ In particular, each state will carry two binary relations over the set of real variables $\Rvars =\set{\rvar_i,1\leq i \leq k}$; the first relation will capture the {\lq\lq}less-than{\rq\rq} relation  and the second relation shall  capture the {\lq\lq}equals-to{\rq\rq} relation.

\begin{definition}[Augmentation of a {\dipa}]
\label{def:augdipa}
The augmentation of a  \emph{\dipautos} $\cA = \defaut$ is the automaton 
$\aug \cA=\defautaug$ defined as follows. Let $\Rvars=\vars\setminus \set{\svar,\svar'}.$
\begin{itemize}
\item  The set $\aug \states $ is the set of states $(q,\lt,\eqs)$ such that $q\in \states,$ $\lt \subseteq \Rvars \times \Rvars$ is a strict partial order, $\eqs \subseteq \Rvars \times \Rvars$ is an equivalence relation,  $\lt \intersect \eqs=\emptyset$ and $\lt \union \eqs$ is transitive. The state $(q,\lt,\eqs)$ is an input state if and only if $q\in \instates.$
 
\item $\aug \qinit =(\qinit, \emptyset,\id_\Rvars).$ where $\id_\Rvars=\set{(x,x) \st x\in \Rvars}.$ 
\item $\aug{\parf}((q, \lt, \eqs)) = \parf(q)$ for each $(q, \lt, \eqs)  \in \aug{\states}.$

\item  $\aug{\transf}: (\aug{\states} \times \cnds) \pto (\aug{\states} \times (\outalph \cup \set{\svar,\svar'}) \times \set{\true,\false}^k)$ is defined as follows.
   \begin{itemize}
       
       \item If $\transf((q,c))$ is undefined, then so is $\aug{\transf} ((q, \lt, \eqs),c)$ for each  possible $\lt$ and $\eqs$. 
       \item Otherwise, assume that $\transf(q,c) = (q_1,o,\vec b).$ Let $\lt$ be a strict partial order, and let $\eqs$ be an equivalence relation such that 
       $\lt \intersect \eqs =\emptyset$ 
       and $\lt \union \eqs$ is transitive. 
       Consider the following definitions:
        $$\begin{array}{l}
            \svars  = \set{\rvar_i \in \Rvars \st  \exists \rvar_j.  (\rvar_i,\rvar_j) \in \lt\union \eqs,  \\
                   \hspace*{2cm}\svar \geq \rvar_j \mbox{ is a conjunct of }c}\\
            \lvars  = \set{\rvar_i \in \Rvars \st \exists \rvar_j.  (\rvar_j,\rvar_i) \in \lt\union \eqs, \\
           \hspace*{2cm} \svar< \rvar_j  \mbox{ is a conjunct of }c}\\
            \ltb   = \lt \union \set{(\rvar_i,\rvar_j)\st \rvar_i\in \svars, \rvar_j\in \lvars}.
        \end{array}$$
        
        Now, $\transf((q,c))$ is defined only if $\ltb  \intersect \eqs = \emptyset.$ In this case, $\aug{\transf}((q,\lt,\eqs),c)= ((q_1,\lta,\eqs_{\mathsf{after}}), o, \vec b )$ where
        $\lta$ and $\eqs_{\mathsf{after}}$ are defined as follows:
        $$\begin{array}{l}
            \avars  = \set{\rvar_i \st b[i]=\true}\\
            \navars = \set{\rvar_i \st b[i]=\false}\\
            \lta = (\ltb \intersect (\navars \times \navars)) \\
                \hspace*{0.13cm}    \union \set{(\rvar_i,\rvar_j)\st \rvar_i\in\svars\intersect \navars, \rvar_j\in\avars} \\
               \hspace*{0.26cm} \union \set{(\rvar_i,\rvar_j)\st \rvar_i\in\avars, \rvar_j\in\lvars \intersect \navars}\\
            \eqs_{\mathsf{after}} = ({\eqs}\intersect (\navars \times \navars)) \, \union\, \\
            \hspace*{0.8cm} \set{(\rvar_i,\rvar_j)\st \rvar_i,\rvar_j\in \avars}. 
        \end{array}$$


        \end{itemize}
       
       
       
          


\end{itemize}
\end{definition}

\begin{definition}
We shall say that a valuation $\val$ is compatible with an augmented state $(q,\lt,\eqs)$ 
if $\val(\rvar_i)<\val(\rvar_j)$ whenever $(\rvar_i, \rvar_j)\in \lt$ and $\val(\rvar_i)=\val(\rvar_j)$ whenever $(\rvar_i,\rvar_j)\in \eqs.$


For each transition $t=((q,\lt,\eqs),c,(q',\lt',\eqs'),o,b),$ of $\aug{\cA},$ there is a unique transition $\proj t=(q,c,q',o,b)$ of $\cA.$  
If $ \rn=t_0t_1\cdots t_{n-1}$ is an execution of $\aug{\cA}$, the execution
$\proj {\rn} = \proj{t_0}\proj{t_1}\cdots \proj{t_{n-1}}$  is said to be the projection of $\rn.$ Observe that if $\rn$ is a run of $\aug \cA$ on $\inpsq$ outputting $\outsq$ then $\proj \rn$ is a run of $\cA$ on $\inpsq$ outputting $\outsq.$ For a computation $\pth=(\rn,\inpsq,\outsq)$ of $\aug \cA$ we write 
$\proj{\pth}=(\proj \rn, \inpsq,\outsq).$ As always, 
a run $\rn$ is said to be a run from the initial state if $\src(\rn)$ is the initial state of $\aug \cA.$ 
 
\end{definition}

We have the following proposition:
\begin{proposition}
\label{prop:augmentation}
Let $\cA$ be a {\dipautos} and $\aug \cA$ be its augmentation. 
\begin{enumerate}
   \item Let $\rn =t_0\cdots t_{n-1}$ be a run of  $\aug \cA$ from the initial state of $\cA$. For each $0< i\leq n,$ let $q_i,\lt_i,\eqs_i$ be such that $\trg(t_{i-1})=(q_i,\lt_i,\eqs_i).$ 
    For each $0< i\leq n$, $\rvar_1,\rvar_2\in\Rvars,$
      \begin{itemize}
      \item 
      $\rvar_1 \lt_i \rvar_2$ if and only if there is a path of non-zero length from $\lavert_\rn(\rvar_1,i)$ to  $\lavert_\rn(\rvar_2,i)$
      in $G_{\rn[0:i]}.$
      \item $\rvar_1 \eqs_i \rvar_2$ if and only $\lavert_\rn(\rvar_1,\rcomment{i})=  \lavert_\rn(\rvar_2,\rcomment{i}).$
      \item $G_\rn$ is acyclic. Hence, every path of $\aug \cA$ from the initial state is feasible. 
      \end{itemize}
    \item 
    If $\pth=(\rn,\inpsq,\outsq)$ is a computation of $\aug \cA$ and $\val$ a valuation compatible with $\fstst (\rn),$ then 
    \begin{itemize}
    \item $\pathprob{\epsilon,\val,\pth} = \pathprob{\epsilon,\val,\proj{\pth}}.$
    \item The dependency graph of $G_\rn$ is the same as the dependency graph of $ G_{\proj{\rn}}.$
    \end{itemize}
    \item If $\rn$ is a feasible run of $\cA$ from the initial state of $\cA$, then there is a unique run $\rn^\dagger$ of
    $\aug \cA$ from the initial state of $\aug \cA$ such that $\proj {\rn^\dagger} =\rn$.

   \item  $\aug{\cA}$ is well-formed if and only if $\cA$ is well-formed.
   \item $\aug{\cA}$ is $(d,\epsilon)$-differentially private if and only if $\cA$ is $(d,\epsilon)$-differentially private.
 \end{enumerate}

\end{proposition}

Thanks to Proposition~\ref{prop:augmentation}, if we can show that if the augmentation of a well-formed automaton $\cA$ is $(d,\epsilon)$-differentially private, then so is  $\cA.$ 
For the rest of the section, without loss of generality, we shall assume that all states of an augmented automaton, $\aug{\cA}$ are reachable from the initial state of the $\aug \cA$.
We start with some useful definitions. 
\begin{definition}

Let $\aug{\cA}$ be the augmented automaton of $\cA.$
\begin{itemize}
\item A transition $t$ of $\aug \cA$ is said to be a cycle transition if there is a cycle $C=t_0\cdots t_{n-1}$ of $\aug \cA$ such that $t=t_i$ for some $0\leq i<n.$  
\item Given a transition $t$ of $\aug {\cA}$,
Let $P(\src(t)) = (d,\mu,d',\mu').$ We define
$$          d(t)= d\quad 
            \mu(t)=\mu \quad
            d'(t)=d'\quad \mu'(t) =\mu'.$$
\end{itemize}            
\end{definition}

\begin{definition}
\label{def:weight}
Let $\cA$ be a {\dipautos} and $\aug{\cA} $ be its augmentation. Let $\rn=t_0\cdots t_{n-1}$ be a run from the initial state of $\aug{\cA}.$
Let $G_\rn$ be the dependency graph of $\rn.$ For $0\leq j\leq n$, let $\rn_j= \rn[j:].$\footnote{Recall, by convention (See Section~\ref{sec:prelims}),  $\rn_n$ is the empty string.}
 
 \begin{enumerate}
    
\item    The vertex $j$ of $G_\rn$ is said to be a $\gcyclevertex$ $(\lcyclevertex)$ if there
is a path $i_1,\ldots,i_k$ in 
 $G_\rn$ such that $i_1=j$ ($i_k=j$ resp.), $i_{k-1}<i_{k}$ ($i_2<i_1$ resp.) and $t_{i_k}$ ($t_{i_1}$ resp.) is a cycle transition.
 \item 
 The sets \rcomment{$\used(\rn_j)$ for $0\leq j< n$}  are defined by backward induction on $j$ as follows: 
 $$\begin{array}{lcl}
 \used(\rn_{n})&=& \emptyset\\
 \used(\rn_j)&=& \smvars(t_j) \union \lgvars(t_j) \\
     &&\hspace*{0.4cm}\union  (\navars(t_j) \intersect \used(\rn_{j+1}))\end{array}$$
 \item The transition $t_j$ is said to be a \emph{quasi-cycle} transition if  the following hold:
 \begin{enumerate}
   \item $t_j$ does not output $\svar,$ ie, $\svar \notin\otpt(t_j).$
   \item $\avars(t_j)\intersect \used(\rn_{j+1})=\emptyset,$ 
    \item if $\rvar\in \smvars(t_j)$ then $\lavert_\rn(j,\rvar)$ is a $\gcyclevertex$ and
     \item if $\rvar\in \lgvars(t_j)$ then $\lavert_\rn(j,\rvar)$ is a $\lcyclevertex.$ 
 \end{enumerate}
 
 \item The weight of transition $t_j$ at position j, $\weight{t_j}\stackrel{\mbox{def}}=f_j(a_j + b_j) d(t_j) + c_j d'(t_j)$ 
 where 
 $$ \begin{array}{lcl}
  f_j &=&\begin{cases}
 1 & \mbox{if } t_j \mbox{ is \emph{not} quasi-cyclic } \\
 0 & \mbox{otherwise}\\
 \end{cases}\\
 a_j &=&\begin{cases}
 1 & \mbox{if } j \mbox{ is a } \lcyclevertex \mbox{ or a } \\
  & \gcyclevertex \mbox{ of } G_\rn\\
 0 & \mbox{otherwise}\\
 \end{cases}\\
  b_j &=&\begin{cases}
 1 & \mbox{if } t_j \mbox{ is an input transition}  \\
 0 & \mbox{otherwise}

 \end{cases}\\
 

 c_j &=&\begin{cases}
 1 & \mbox{if } \svar'\in \otpt(t_j)\\
 & \mbox { and } t_j \mbox{ is an input transition} \\
 0 & \mbox{otherwise}
 \end{cases}\\
 \end{array}$$

 \item The weight the run $\rn$, $\weight{\rn}=\sum_{j<n} \weight {t_j} .$
 \end{enumerate}
 \end{definition}
 
 \begin{proposition}
 \label{prop:consequences}
 Let $\cA$ be a well-formed {\dipautos} and $\aug{\cA} $ be its augmentation. Let $\rn=t_0\cdots t_{n-1}$
be a run  
from the initial state of $\aug{\cA}$
and $G_\rn$ be the dependency graph of $\rn.$
For each $j,j'$ We consider each part of the Proposition. 
 \begin{enumerate}
     
   \item If $\rvar \in \smvars(t_j)$ ($\rvar \in \lgvars(t_j)$ resp.) is such that $\lavert_\rn(j,\rvar)$ is a $\lcyclevertex$ ($\gcyclevertex$ resp.) then $j$ is a $\lcyclevertex$ ($\gcyclevertex$ resp.) also. 
     \item A $\gcyclevertex$ of $G_\rn$ cannot be a $\lcyclevertex.$
     \item If $j$ is a $\gcyclevertex$ or a $\lcyclevertex$ then the transition $t_j$ cannot output $\svar.$
      
       \item If $j<j'$ and $t_j=t_{j'},$ then $\weight{t_j}=0.$
       
 \end{enumerate}

 \end{proposition}
\begin{proof}
Thanks to Proposition~\ref{prop:augmentation} $\aug \cA$ is well-formed as $\cA$ is. 
\begin{enumerate}
   
    \item Immediate from the definitions of $\lcyclevertex$ and $\gcyclevertex.$
    \item Immediate from the fact that $\aug \cA$ is a well-formed {\dipautos}, and hence has no leaking pairs.
  \item Immediate from the fact that $\aug \cA$ is a well-formed {\dipautos}, and hence has no privacy-violating path.
  \item Follows from well-formedness:
  If $t_j=t_{j'}$, then 
  $t_jt_{j+1}\ldots t_{j'}$
  is a cycle. As $\aug \cA$ is well-formed, $t_j$ does not output $\svar'$ and is easily seen as a quasi-cycle transition.\qedhere
    
\end{enumerate}
\end{proof} 
 \begin{proposition}
\label{prop:obs}
For $i=1,2$, let \rcomment{$\pth =(\rn,\inpsq,\outsq)$
be a computation of $\aug{\cA}.$}
 Let $\val^1,\val^2$ be valuations compatible with $\src(\rn).$ If 
        $\val^1(\rvar)=\val^2(\rvar)$
        for each $\rvar\in \used(\rn),$ then 
        \rcomment{$$\pathprob{\epsilon,\val^1,\pth} =  \pathprob{\epsilon,\val^2,\pth}.$$}
\end{proposition}
\begin{proof}
By induction on $\abs{\exec}.$ 
\end{proof}
  
\begin{theorem}
\label{thm:suffapp}
Let $\cA$ be a well-formed {\dipa}.
For $i=1,2$, let $\pth_i =(\rn,\inpsq_i,\outsq)$
be computations of $\aug{\cA}$ such that $\inpsq_1$ and $\inpsq_2$ are adjacent and
$\rn$ starts from the initial state of $\aug \cA.$
Then
$$\pathprob{\epsilon,\pth_2} \leq \eulerv{\weight{\rn} \epsilon}\, \pathprob{\epsilon,\pth_1}.$$

\end{theorem}

\begin{proof}

Let $\rn=t_0\cdots t_{n-1}.$ 
Let $G_\rn$ be the dependency graph of $\rn.$
 For each $0\leq j< n,$ define $m_j: \Rvars \to \set{-1,0,1}$ as follows:
 $$ m_j(\rvar) =\begin{cases}
                   1 & \mbox{if } \lavert(\rvar,j) \mbox{ is  a }\gcyclevertex\\
                    -1 & \mbox{if } \lavert(\rvar,j) \mbox{ is  a }\lcyclevertex\\
                    0 & \mbox{otherwise}
 \end{cases}$$

For $0\leq j\leq n,$ $\rn_j=\rn[j:].$
For $i=1,2$, $0\leq j\leq n$ let $\pth_{i,j}=\rcomment{\pth_i}[j:].$ 
\footnote{By convention, (See Section~\ref{sec:prelims}), $\rn_n$  and $\pth_n$ is the empty string $\emptystr.$} 
For $0\leq j\leq n,$ we shall say that a valuation $\val$ is compatible with $\rn_j$ if $\val$ is compatible $\src(\rn_j)$ if $j<n$ and with $\trg(\rn)$ otherwise.
The theorem follows from Proposition~\ref{prop:obs} and the following claim.

\begin{claim}
Let  $0\leq j\leq n,$ and
let $\val^1,\val^2$ be valuations such that  $\val^1,\val^2$ are compatible with $\rn$.    If  
$$ \restr {\val^2}{\used(\rcomment{\rn})}=\restr {(\val^1+m_j)}{\used(\rcomment{\rn})},$$
 then 
$$\pathprob{\epsilon,\val^1,\pathtwo  1 j} \geq \eulerv{-\wt_j \epsilon}\, \pathprob{\epsilon,\val^2,\pathtwo 2 j}$$
where 
$$\wt_j = \sum^{n-1}_{u=j} \weight{t_u} .$$
\end{claim}

The proof is by induction on $n-j.$
\paragraph*{Base Case: $j=n$} The claim follows from the definitions.
\paragraph*{Induction Hypothesis: $j<n$}  Assume that the claim is true for $j+1.$ Fix $\val^1,\val^2.$


 For $i=1,2$, $\epsilon>0,$ and $z\in \Reals$   let 
    $$\begin{array}{lcl}
        \rcomment{\inp_i}&=&\fstst(\inpsq_i([j]))\\
        \nu_i&=&\begin{cases}
          \mu(t_j)+\inp_i & \mbox{if } \inp_i\ne \tau\\
           \mu(t_j) & \mbox{otherwise}

         \end{cases}\\
         \nu'_i&=&\begin{cases}
          \mu'(t_j)+\inp_i & \mbox{if } \inp_i\ne \tau\\
           \mu'(t_j) & \mbox{otherwise}

         \end{cases}\\
         \mathsf{low}_i&=&\max_{\rvar \in \smvars(t_j)}{\val^i(\rvar)}\\
         \low_i & =& \begin{cases}
                   \max(\mathsf{low}_i,r) & \mbox{if } \outsq[j]=(\svar,r,s)\\
                    \mathsf{low}_i & \mbox{otherwise} 
         \end{cases}\\
           \rcomment{\mathsf{up}_i}&=&\min_{\rvar \in \lgvars(t_j)}{\val^i(\rvar)}\\
         \up_i & =& \begin{cases}
                   \min(\mathsf{up}_i,s) & \mbox{if } \outsq[j]=(\svar,r,s)\\
                    \mathsf{up}_i & \mbox{otherwise} 
                    \end{cases}\\
                        
        \rcomment{\val^i_z(\rvarx)} &=&   \begin{cases}
                     \eta(\rvarx) & \mbox{if } \rvarx \in \navars(t_j) \\
                     z & \mbox{if } \rvarx \in \avars(t_j)
                  \end{cases}
        \end{array}
    $$

Now if $\low_2 \geq \up_2$ then $\pathprob{\epsilon,\val^2,\exec_{2,j}}=0$ and the claim is trivially true. Hence, without loss of generality, we assume that $\low_2 < \up_2.$ We will shortly argue that if $\low_2 < \up_2$ then $\low_1<\up_1$ also.  Assuming that this is case, for $i=1,2$ and $\epsilon>0,$ let
   \begin{align*}
      q_i(\epsilon) &=
         \begin{cases}
              \frac {d'(t_j)\epsilon} 2 \int_{r}^{s} \eulerv{-d'(t_j)\epsilon\; \abs{z-\nu'_i}} dz
                    & \mbox{if } \outsq[j]=(\svar',r,s)\\
                    1 & \mbox{otherwise}                
                    \end{cases}\\
          p_i(\epsilon)&=  \frac {d(t_j)\epsilon} 2 \int_{\low_i}^{\up_i} \eulerv{-d(t_j)\epsilon\; \abs{z-\nu_i}} \pathprob{\epsilon,\val^i_z,\pth_{i,j+1}} dz   \end{align*}

  
   By definition,  we will then have 
   $$ \pathprob{\epsilon,\val^i,\pth_{i,j}} =q_{\rcomment{i}}(\epsilon) p_i(\epsilon).$$

   
    Let 
    $$\Delta=\nu_2-\nu_1=\nu'_2-\nu'_1.$$ We have that $-1\leq\Delta \leq 1$ and $\Delta=0$ if $t_j$ is a non-input transition.  Let $f_j,a_j,b_j,c_j$ be as in the Definition~\ref{def:weight}. It is easy to see that $$q_1(\epsilon) \geq \eulerv{-c_j d'(t_j)\epsilon}q_2(\epsilon).$$
    Thus, we shall be done if we can show that $$\low_1< \up_1 \mbox{ and } p_1(\epsilon)\geq \eulerv{\epsilon  (-f_j(a_j+b_j)d(t_j)-\wt_{j+1})}p_2(\epsilon).$$
    We consider three mutually exclusive but exhaustive cases, depending on the values of $f_j$ and $a_j.$
    \begin{enumerate}[label=(\alph*)]
    
    \item \label{case} Let us  consider the case when $f_j=0.$
    Thus, the transition $t_j$ is a quasi-cycle transition. 
    
     By definition of a quasi-cycle transition,
if $\rvar\in \smvars(t_j)$ then $\lavert(\rvar)$ is a $\gcyclevertex$ and
      if $\rvar\in \lgvars(t_j)$ then $\lavert(\rvar)$ is a $\lcyclevertex.$  Thus, we must have 
    $m_j(\rvar) = 1$ for each $\rvar \in \smvars(t_j)$ and $m_j(\rvar)=-1$ for each $\rvar\in \lgvars(t_j).$ Note that $t_j$ does not output $\svar$. 
    Thus, $$\low_2=\low_1+1 \qquad \up_2=\up_1-1.$$
    Thus, from the assumption that 
    $\low_2<\up_2$, it is easy to see that $\low_1<\up_1.$

      Now, by definition of a quasi-cycle transition,
        $$\avars(t_j)\intersect \used(\rcomment{\rn_{j+1}})=\emptyset.$$

        Fix $b$  such that 
        $\low_2<  b < \up_2$ be some number.
        In this case, we can write using Proposition~\ref{prop:obs} 
       \begin{align*}
        p_i(\epsilon)  &= \frac {d(t_j)\epsilon} 2 \int_{\low_i}^{\up_i} \eulerv{-d(t_j)\epsilon\; \abs{z-\nu_i}} \pathprob{\epsilon,\val^i_{\rcomment{z}},
        \pth_{i,j+1}} dz  \\
             &= 
             \pathprob{\epsilon,\val^i_b,
             \pth_{i,j+1}}\,
             (\frac {d(t_j)\epsilon} 2 \int_{\low_i}^{\up_i} \eulerv{-d(t_j)\epsilon\; \abs{z-\nu_i}}  dz) 
             \end{align*}

       

        We have by construction, for all $\rvar \in \used(\rcomment{\rn_{j+1}}),$
      $$\val^2_b
       (\rvar) 
      = \val^1_b
      (\rvar)  + m_{j+1}(\rvar). $$
      It is also easy to see that 
      $\val^i_b 
      $ is compatible with $\rn_{j+1}$ for each $i=1,2.$
      Thus, by the induction hypothesis, we have that       
       $$  \pathprob{\epsilon,
       \val^1_b 
       ,\pth_{1,j+1}} \geq \eulerv{-\epsilon \wt_{j+1}} \pathprob{\epsilon,
       \val^2_b 
       ,\pth_{2,j+1}}.$$

       The lemma, now follows in this case from the following observation:
      \begin{align*}
        \int_{\low_1}^{\up_1}   \eulerv{-d(t_j)\epsilon\; \abs{z-\nu_1}}  dz   &= \int_{\low_1}^{\up_1}   \eulerv{-d(t_j)\epsilon\; \abs{z-\nu_2+\Delta}}  dz\\
             &=  \int_{\low_1+\Delta}^{\up_1+\Delta}   \eulerv{-d(t_j)\epsilon\; \abs{z-\nu_2}} 
             dz\\
             &\geq \int_{\low_1+1}^{\up_1-1}   \eulerv{-d(t_j)\epsilon\; \abs{z-\nu_2}} 
             dz\\
             &=
             \int_{\low_2}^{\up_2}   \eulerv{-d(t_j)\epsilon\; \abs{z-\nu_2}} 
             dz
         \end{align*}    

    \item\label{caseb} Let us  consider the case when $f_j=1$ and $a_j=0.$
     
    Please note that if $a_j=0$ then $j$ is neither a $\lcyclevertex$ nor a $\gcyclevertex.$ Thanks to Proposition~\ref{prop:consequences}, it follows that 
      $m_j(\rvar)\ne -1$ for each $\rvar\in \smvars(t_j)$ and $m_j(\rvar)\ne 1$ for 
      each $\rvar\in \lgvars(t_j).$
      Thus $\low_1\leq \low_2$ and $\up_2\leq \up_1$, and hence $\low_1<\up_1.$
      
      Also, observe that we have by definition $m_{j+1}(\rvar)=0$ for each $\rvar\in \avars(t_j)$ and $m_{j+1}(\rvar)=m_j(\rvar)$ for each $\rvar\in \navars(t_j).$
      From this it is easy to see that for each $\rvar\in \used(\rcomment{\rn_{j+1}}), $
      $$\val^{2}_z(\rvar)= \val^{1}_z(\rvar)+m_{j+1}(\rvar).$$
      Furthermore, it is easy to see that for each $\low_i< z< \up_i,$
      $\val^{k}_z$ is compatible with $\absexec_{j+1}.$
      As $\low_1\leq \low_2$ and $\up_2\leq \up_1$, we get that for each $\low_2< z< \up_2,$  and $k=1,2$ $\val^{k}_z$ is compatible with $\absexec_{j+1}.$
      By induction hypothesis, we get that for each $\low_2< z <\up_2,$ 
$$\pathprob{\epsilon,\val^1
_z
,\pth_{1,j+1}} \geq 
      \eulerv{-\epsilon \wt_{j+1}} \pathprob{\epsilon,\val^2_z,\pth_{2,j+1}}$$
      
      Thus, we have 
      \begin{align*}
          p_1(\epsilon) &=  \frac {d(t_j)\epsilon} 2 \int_{\low_1}^{\up_1} \eulerv{-d(t_j)\epsilon\; \abs{z-\nu_1}} \pathprob{\epsilon,\val^1_z,\pth_{1,j+1}} dz    \\
           & \geq  \frac {d(t_j)\epsilon} 2 \int_{\low_2}^{\up_2} \eulerv{-d(t_j)\epsilon\; \abs{z-\nu_1}} \pathprob{\epsilon,\val^1_z,\pth_{1,j+1}} dz    \\
           & \geq  \eulerv{-\epsilon \wt_{j+1}}\frac {d(t_j)\epsilon} 2 
           \\
           & \hspace*{1cm }
           \int_{\low_2}^{\up_2} \eulerv{-d(t_j)\epsilon\; \abs{z-\nu_1}}\pathprob{\epsilon,\val^2_z,\pth_{2,j+1}} dz    \\
             & =  \eulerv{-\epsilon \wt_{j+1}}\frac {d(t_j)\epsilon} 2 
             \\ & \hspace*{1cm }\int_{\low_2}^{\up_2} \eulerv{-d(t_j)\epsilon\; \abs{z-\nu_2+\Delta}} \pathprob{\epsilon,\val^2_z,\pth_{2,j+1}} dz    \\
         \end{align*}
         
           Now, in case $t_j$ is a non-input transition, $\Delta=0$ and $b_j=0.$ Hence we get 
          $$
          p_1(\epsilon)\geq \eulerv{-\epsilon \wt_{j+1}} p_2(\epsilon)=\eulerv{\epsilon (-f_j(a_j+b_j)-\wt_{j+1})}p_2(\epsilon)$$
          as required.
          
          Otherwise, $b_j=1$. Since $\Delta\in[-1,1]$, we have  that
          $$\eulerv{-d(t_j)\epsilon\; \abs{z-\nu_2+\Delta}} \geq \eulerv{-d(t_j)\epsilon} \eulerv{-d(t_j)\epsilon\; \abs{z-\nu_2}}$$
          Thus, 
          \begin{align*}
          p_1(\epsilon)
          & \geq  \eulerv{-\epsilon \wt_{j+1}}\frac {d(t_j)\epsilon} 2 \eulerv{-d(t_j)\epsilon}
          \\ & \hspace*{0.8cm} \int_{\low_2}^{\up_2} \eulerv{-d(t_j)\epsilon\; \abs{z-\nu_2}} \pathprob{\epsilon,\val^2_z,\pth_{2,j+1}} dz   \\
              & =  \eulerv{\epsilon(-f_j(a_j+b_j)d(t_j)-\wt_{j+1})} p_2(\epsilon)
              \end{align*}
      as required.

    
          
    \item Let us consider the case when $f_j=1$ and $a_j=1.$
    Thus $j$ is a $\lcyclevertex$ or a $\gcyclevertex.$ We consider the case when $j$ is a $\lcyclevertex.$ The case when $j$ is a $\gcyclevertex$ is similar and left out. 
    
    Since $j$ is a $\lcyclevertex$, $t_j$ does not output $\svar.$ Thus, we have that $\low_i=\mathsf{low}^{\rcomment{i}}_k$    and $\up_i=\mathsf{up}^{\rcomment{i}}_k$ for each $i=1,2.$ Further, we must have by definition for each $\rvar\in\lgvars(t_j),$ $\lavert(\rvar,j)$ is also a ${\lcyclevertex}.$ Thus,
    $m_j(\rvar)=-1$ for each $\rvar\in \lgvars(t_j).$ From these observations, we get that 
    $$\up_2=\up_1-1 \qquad \low_2-1 \leq \low_1 \leq \low_2+1.$$
    Now, as we have assumed that $\low_2 < \up_2,$ we get 
    $$\low_1<\up_2+1=\up_1$$
    as desired.

      It is also easy to see that $\val^1_z$ and $\val^2_{z-1}$  is compatible with $\rn_{j+1}$
       for each  $\low_2 < z-1 < \up_2$, ie,   for each $\low_2 +1 <z<\up_1.$ 

       Also, by definition as $j$ is a $\lcyclevertex$, $m_{j+1}(\rvar)=-1$ for each $\rvar \in \avars(t_j).$
       Thus, we have $\rvar\in \used(\rcomment{\rn_{j+1}}), $
      $$\rcomment{\val^2_{z-1}(\rvar)= \val^1_{z}(\rvar)+m_{j+1}(\rvar)}.$$
        By induction hypothesis, we get that for each  $\low_2+1 <z<\up_1,$ 
      $$ \pathprob{\epsilon,\val^1_z,\pth_{1,j+1}} \geq \eulerv{-\epsilon \wt_{j+1}} \pathprob{\epsilon,\val^2_{z-1},\pth_{2,j+1}} $$
      Observe that $\low_2+1 \geq \low_1.$
      We thus have 
      
      \begin{align*}
           p_1(\epsilon)  &= \frac {d(t_j)\epsilon} 2 \int_{\low_1}^{\up_1} \eulerv{-d(t_j)\epsilon\; \abs{z-\nu_1}} \pathprob{\epsilon,\val^1_z,\pth_{1,j+1}} dz \\
          & \geq  \frac {d(t_j)\epsilon} 2 \int_{\low_2+1}^{\up_1} \eulerv{-d(t_j)\epsilon\; \abs{z-\nu_1}} \pathprob{\epsilon,\val^1_z,\pth_{1,j+1}} dz \\
            & \geq  \frac {d(t_j)\epsilon \eulerv{-\epsilon \wt_{j+1}} } 2 
            \\ & \hspace*{0.8cm}
            \int_{\low_2+1}^{\up_1} \eulerv{-d(t_j)\epsilon\; \abs{z-\nu_1}} \pathprob{\epsilon,\val^2_{z-1},\pth_{2,j+1}} dz\\
            &= \frac {d(t_j)\epsilon \eulerv{-\epsilon \wt_{j+1}} } 2 \\ & \hspace*{0.8cm} \int_{\low_2+1}^{\up_1} \eulerv{-d(t_j)\epsilon\; \abs{z-\nu_1}} \pathprob{\epsilon,\val^2_{z-1},
            \pth_{2,j+1}} dz\\
            &=  \frac {d(t_j)\epsilon \eulerv{-\epsilon \wt_{j+1}} } 2 \\ & \hspace*{0.8cm} \int_{\low_2}^{\up_1-1} \eulerv{-d(t_j)\epsilon\; \abs{z+1-\nu_1}} \pathprob{\epsilon,\val^2_z
            ,\pth_{2,j+1}} dz\\ 
          &= \frac {d(t_j)\epsilon \eulerv{-\epsilon \wt_{j+1}} } 2 \\ & \hspace*{0.8cm} \int_{\low_2}^{\up_2} \eulerv{-d(t_j)\epsilon\; \abs{z+1-\nu_2 +\Delta}} \pathprob{\epsilon,\val^2_z
          ,\pth_{2,j+1}} dz\\
            &\geq  \frac {d(t_j)\epsilon \eulerv{-\epsilon \wt_{j+1}-\epsilon d(t_j)(1+ {\Delta})} } 2 \\ & \hspace*{0.8cm} \int_{\low_2}^{\up_2} \eulerv{-d(t_j)\epsilon\; \abs{z-\nu_2}} \pathprob{\epsilon,\val^2_z
            ,\pth_{2,j+1}} dz\\
            &=  \eulerv{-\epsilon \wt_{j+1}-\epsilon d(t_j)(a_j+ {\Delta})} p_2(\epsilon)
       \end{align*}
    The result follows by observing that ${\Delta}\leq b_j$ as 
    $\Delta=0$ if $b_j=0$ and in the interval $[-1,1]$ otherwise. \qedhere
    
    \end{enumerate}

\end{proof}

\begin{corollary}
If the {\dipa} $\cA$ is well-formed then there is a number $\wt(\cA)$ such that $\cA$ is $\wt(\cA)\epsilon$-differentially private. Further, the number $\wt(\cA)$ can be computed from $\aug \cA$ in polynomial time, and hence from $\cA$ is exponential time.  
\end{corollary}
\begin{proof}
Thanks to Proposition~\ref{prop:augmentation} and \rcomment{Theorem~\ref{thm:suffapp}}, it suffices to show that there is a $\newd$ such that $\weight\rn\leq \newd$ for every run $\rn$ of $\aug\cA.$ Now, from that fact that if a transition of the automaton can contribute to the weight of a run at most once (See Proposition~\ref{prop:consequences}), it is immediate to see such a $\newd$ exists. 
We detail below a better bound on $\newd.$

Consider the underlying labeled graph $G$ constructed from the augmented automaton $\aug \cA$ as follows. Its vertices are states of $\aug \cA$ and there is an edge from $q_1$ to $q_2$ if and only if there is a transition $t$ from $q_1$ and $q_2$. The label of the edge is $t.$ We shall also assign weights to the edge as follows. We  assign the edge weight $e w_1+  w2$ where $e$ is $2$ if the transition  is an input transition and $1$ otherwise. $w_1$ is $d(t)$ if either $t$ is not a cycle transition or if there is a variable $\rvar\in\avars(t)$ and a run $\rn$ from starting from $q_2$ in which $\rvar$ is accessed without being assigned. Otherwise, $w_1$ is $0.$  $w_2$ is $d'(t)$ if $\svar'$ is output in $t$ and $0$ otherwise.  

Once the graph $G$ has been constructed, we can construct its component graph $G'$ and assign weights to each node and transition of this graph. We shall take the weight of a component in $C$ to be the sum of the weights of all transitions in $C.$ The weight of the edge from component $C_1$ to $C_2$ labeled $t$ is taken to be the weight of the edge $t.$ Note that $G'$ is a DAG and can be computed in time polynomial in the size of $\aug \cA.$ Now, each path in $G'$ has a weight which is the sum of weights of transitions and nodes. Let $\newd$ be the maximum value amongst the weights of  paths in $G'$ starting from the component containing the initial state of $\aug \cA.$ It is easy to see that weight of any run of $\cA$ is bounded by $\newd$ and that $\newd$ can be computed from $G'$ in linear time. Thus, a desired $\newd$ can be constructed from $\cA$ in polynomial time.  

A better approximation to $\newd$ can be constructed by taking the bisimulation quotient of $\aug\cA$ before running the above algorithm.
\end{proof}

\subsection*{Checking Well-formedness is in {\pspace}.}

\begin{theorem}
The problem of checking whether a {\dipa} is well-formed is decidable in \pspace. When the number of variables is taken to be a constant $k$, then the problem of checking whether a {\dipa} is well-formed is decidable in polynomial time. 
\end{theorem}
\begin{proof}
Recall that $\cA$ is well-formed iff $\aug \cA$ is well-formed. 


Our {\pspace} algorithm will first non-deterministically check if $\aug \cA$ has a 
leaking cycle without needing to construct the whole automaton. This will allow us to conclude that the problem of checking whether $\aug \cA$ has a leaking cycle is in {\pspace},  thanks to Savitch's theorem. 

The non-deterministic algorithm $Alg$ for checking whether $\aug \cA$ has a {\criticalcycle} guesses a variable $\rvar\in \Rvars$ and  a run $\rn\,C$
of $\aug \cA$ incrementally such that \begin{enumerate*}[label=(\roman*)]
\item $C$ is a cycle, and \item there are indices $i_1$ and $i_2$ such that  $\rvar$ is assigned in the transition $t_{i_1}$ and used in transition $t_{i_2}.$
\end{enumerate*}
Note that as all runs of $\aug \cA$ are feasible, the algorithm does not need to check the repeatability of the cycle $C.$

The algorithm $\Alg$ performs the above by guessing the variable $\rvar$ and the run $\rn\,C=t_0\cdots t_{n-1}$  one-by-one from the initial state of $\aug \cA,$ and  at each steo
    \begin{itemize}
    \item checks that the source of the current guessed transition is exactly the target of the last guessed transition,
    \item checks that the current guesses transition is a valid transition of $\aug \cA$,
     \item if has not guessed as yet, $\Alg$ guesses if the current guessed transition is the first transition of $C$; if it guesses that it indeed is, then it remembers $\src(t_i)$ in the memory, and that fact that it guessed cycle $C$ has begun,
     \item if the cycle $C$ begins at position $i$ or has already begun then it additionally checks if 
        \begin{enumerate}
        \item $\rvar$ is assigned in the current guessed transition
        \item $\rvar$ is used in the current guessed transition. 
        \end{enumerate}
        \item $\Alg$ declares that $\aug \cA$ has a leaking cycle if the target of the last transition is exactly the source of the cycle $C$ it guessed, and if $\rvar$ was assigned and used in its guessed cycle $C.$
     \end{itemize} 

It is easy to see that the path $\rn\,C$ can be guessed without explicitly constructing $\aug \cA$ and that the above checks require only space polynomial in the size of $\cA.$ If $\aug \cA$ has a leaking cycle; then we can declare that $\cA$ is not well-formed. Otherwise, we check if $\aug \cA$ has a {\criticalpair}.       

To check for a {\criticalpair} of $\aug \cA$,
we have to search for a run $\rn$ of $\aug \cA$ from the initial state, such that 
 there are indices $0 \leq i_1 < j_1 \leq \len{\rn}$ and $0 \leq i_2 < j_2 \leq \len{\rn}$ such that following conditions hold.
 \begin{enumerate} 
  \item $C_1 = \subseq{\rn}{i_1}{j_1}$ and $C_2 = \subseq{\rn}{i_2}{j_2}$ are cycles. (Note that since $\aug \cA$ does not have leaking cycles by assumption, all cycles of $\aug \cA$ are non-{\criticalcycle}s).
  \item $C_1$ and $C_2$ are non-overlapping.
  \item There is a path $k_1,k_2,\ldots k_m$ in the dependency graph $G_\rn$ such that $i_1 \leq k_1 < j_1$ ($k_1$ is on $C_1$), $i_2 \leq k_m< j_2$ ($k_m$ is on $C_2$), $k_2 < k_1$ and $k_{m-1} < k_m$.
  \end{enumerate}

Now, it is easy to see that a non-deterministic algorithm that runs in  space polynomial in the size of $\cA$ 
can check for a run $\rn$ that satisfies the first two conditions above, as in the case of a {\criticalcycle}. The challenge is to check for the third condition, as maintaining the dependency graph for the entire run may not be possible in polynomial space. However, we will exploit the relations $\lt$ and $\eqs$ in an augmented state. Let $\trg(\rn[i])=(q,\lt,\eqs).$ Recall that 
\begin{enumerate}
\item $(\rvar_1,\rvar_2) \in \lt$ if and only if there is a path from $\lavert_\rn(i,\rvar_1)$ to $\lavert_\rn(i,\rvar_2)$ in the graph $G_{\subseq {\rn} 0 i}.$  
\item and  $(\rvar_1\, \rvar_2)\in \eqs$
if and only if $\lavert_\rn(i,\rvar_1)=\lavert_\rn(i,\rvar_2).$ 
\end{enumerate} 
To exploit the relations $\lt$ and $\eqs$, the algorithm  shall pretend that there are two additional real variables, $V_1$ and $V_2$ that are assigned exactly once each during the run $\rn$. The variable $V_1$ is assigned when the algorithm guesses that the current index is the index ${k_2}$ and the $V_2$ is assigned when the algorithm guesses that the current index is the index ${k_{m-1}}.$ The non-deterministic algorithm, $\Alg_1,$ for checking the existence of a {\criticalcycle} proceeds as follows. It guesses the run $\rn=t_0\cdots t_{n-1}$ incrementally. 
 At each step,
    \begin{itemize}
    \item checks that the source of the current guessed transition is exactly the target of the last guessed transition,
    \item checks that the current guessed transition is a valid transition of $\aug \cA$,
     \item if $\Alg_1$ has not guessed as yet that the index $k_2$ has been encountered, $\Alg_1$ guesses if 
     the current transition is the desired transition $t_{k_2}$ or not. If it guesses that the current transition is the desired transition $t_{k_2}$, then it treats the variable $V_1$ as being assigned in the current transition. 
     \item if $\Alg_1$ has not guessed as yet that the index $k_{m-1}$ has been encountered, $\Alg_1$ guesses if 
     the current transition is the desired transition $t_{k_{m-1}}$ or not, if it guesses that the current transition is the desired transition $t_{k_{m-1}}$, then it treats the variable $V_2$ as being assigned in the current transition, 
     \item if $\Alg_1$ has yet to guess the cycle $C_1$, then it guesses if the current transition is the first transition of cycle $C_1$; if it guesses that it indeed does, then it remembers the source of the current transition in the memory, and the fact that it guessed cycle $C_1$ has begun, 
     \item it $\Alg_1$ has yet to guess the cycle $C_2$,  then it guesses if the cycle $C_2$ begins at the current transition; if it guesses that it indeed does, then it remembers the source of the current transition in the memory, and that fact that it guessed cycle $C_2$ has begun,
     \item makes sure that it is not guessing that it is in cycle $C_1$ and $C_2$ simultaneously,
     \item if $\Alg_1$ is guessing that  the cycle $C_1$ is being processed, then it guesses if the current transition is the transition $t_{k_1}$; and checks the guess by checking if there is a variable $\rvar \in \lgvars(t)$  such that $ (V_1,\rvar)\in \eqs$ where $t$ is the current transition and $\eqs$ is such that $\src(t)=(q,\lt,\eqs)$;
     \item if $\Alg_1$ is guessing that  the cycle $C_2$ is being processed, then it guesses if the current transition is the transition $t_{k_{m}}$; and checks the guess by checking if there is a variable $\rvar \in \smvars(t)$  such that $ (V_2,\rvar)\in \eqs$ where $t$ is the current transition and $\eqs$ is such that $\src(t)=(q,\lt,\eqs)$;
     
    
     \item if it is guessing that the current transition is in the cycle $C_i,$ for $i=1,2$, it guesses if the current guessed transition is the last transition of $C_i$; if that is the case, then it checks that the target of the current transition is exactly the source of the cycle $C_i$ it has stored in its memory, and
     \item once the algorithm guesses that both cycles $C_1$ and $C_2$ are completed, it guesses if the current transition is the final transition of $\rn.$
     If it guesses that the current transition is indeed the final transition and the target of the transition is the triple $(q,\lt,\eqs)$, then it declares that $\aug \cA$ has a {\criticalpair} if all the above checks passed, and if either $(V_1,V_2)\in \lt$ or $(V_1, V_2)\in \eqs.$
     
     \end{itemize} 
     It is easy to show that the above algorithm runs in space polynomial in the size of $\aug \cA$ and that the algorithm declares that $\aug \cA$ has a {\criticalpair} iff $\aug \cA$ has a {\criticalpair} thanks to the properties of $\lt$ and $\eqs$. 

     Now, if $\aug \cA$ does not have a {\criticalcycle} or a {\criticalpair}, then the algorithm for well-formedness will check for {\violatingc} and {\violatingp} next. The {\pspace} algorithm for checking {\violatingc} can be designed along the same lines as the algorithm for checking  for {\criticalcycle}, and the {\pspace} algorithm for checking for
     {\violatingp} can be designed along the same lines as the algorithm for check for {\criticalpair}. 
\end{proof}

\section{Necessity of well-formedness for output-distinct {\dipautop}}
\label{app:necessity}


In this section, we give the proof showing that if {\dipa} $\cA$ satisfying {\Outcond} property is not well-formed then $\cA$ is not differentially private. The proof will be broken into four Lemmas \ref{lem:criticalcnec}, \ref{lem:criticalpnec}, \ref{lem:violatingcnec} and \ref{lem:violatingpnec}, given in this section. 




\subsection*{\textbf{{\Criticalcycle}s implies no privacy}}

\begin{lemma}
\label{lem:criticalcnec}
A  {\dipa} $\cA $, satisfying {\Outcond} property, is 
not differentially private if  it has a  {\criticalcycle}.
\end{lemma}

Let $\cA= \defaut.$ Assume that $\cA$ satisfies {\Outcond} property and has a {\criticalcycle}.

Let $\absexec'=t_0,t_1,...,t_{m'-1}$  be a run of $\cA$, starting from the initial state $\qinit$, that is a {\criticalcycle}. For $0\leq u<m'$, let $c_u$ be the guard of transition $t_u.$ From the definition of a {\criticalcycle}, we see that there exists an integer $m\leq m'-1$ such that the suffix  $C'=t_m,...,t_{m'-1}$ is a cycle that is repeatable, and  
there exist distinct integers $i,j$ such that $m\leq i,j<m'$ and a variable $\rvarx_{i'}$ ($1\leq i'\leq k$) such that $t_i$ is an assignment transition for the variable $\rvarx_{i'}$ and $c_j$ references $\rvarx_{i'}$. Let $n'=m'-m.$
 Now, we extend $\absexec'$ by repeating the cycle $C'$ to get the run $\absexec=\absexec'C'$. We let  $\absexec=t_0,...,t_{m+n-1}$ 
where $n=2n'$. For $0\leq u<m+n$, let $q_u\:=\src(t_u)$ and $\gamma_u=\otpt(t_u).$ Also, let $q_{m+n}=\trg(t_{m+n-1}).$ Note that $q_u=q_{u+n'},\:\gamma_u=\gamma_{u+n'}$ for $m\leq u<m+n'.$ Now, let $C=t_m,...,t_{m+n-1}.$ Note that $C$ is a cycle.  It is not difficult to see the following property is satisfied by $\absexec$: the variable $\rvarx_{i'}$ is referenced in the condition $c_{j+n'}$ and the transition $t_{i}$ is an assignment transition for $\rvarx_{i'}$.



 As before, let $t_u$ be the {$u$-th transition} of $\absexec$ and $c_u$ be the {guard} of the $u$-th transition. 
  Further, let $d_u$ and $\mu_u$ be such that $\parf(q_u) = (d_u,\mu_u)$ for each $u.$

From our discussion above, we see that there exist at least one triple $(u',v',w')$ of integers 
 such that $m\leq
u'<v'<m+n$, $1\leq w'\leq k$ and the following properties are satisfied: (i) $c_{v'}$ references $\rvarx_{w'}$ and (ii) $u'=\lavert_{\absexec}(\rvarx_{w'},v').$
We call such a triple as an {\em assign\_refer} triple.
Now, we give the definitions and proof assuming that there exists at least one triple $(u',v',w')$, as given above, such that the condition
$\svar\geq \rvarx_{w'}$ is a conjunct in the guard $c_{v'}$ (the case when for all assign\_refer triples $(u',v',w')$, the condition 
$\svar< \rvarx_{w'}$ is a conjunct in the guard $c_{v'}$, is handled similarly in a symmetric fashion as outlined later). Now we fix a triple of integers $(i,j,i')$ as follows. If there exists at least one assign\_refer triple $(u',v',w')$ such that $q_{u'}\in\epsstates$ then we take $(i,j,i')$ to be any such triple so that 
$\mu_i$ is the maximum among all such triples; otherwise, we take $(i,j,i')$ be any assign\_refer triple. 
In the remainder of our proof, we fix the triple of integers $i,j,i'$ as specified above.


Consider any integer $\ell>0$. We define a run $\absexec_\ell$ starting
from $\qinit$ by repeating the cycle $C=t_m,\ldots t_{m+n-1}$, $\ell$
times. Formally, $\absexec_\ell=t_0,t_1,...,t_{m+\ell n-1}$ such that 
$q_u=q_{u-n}$ and $\gamma_u=\gamma_{u-n}$ for $m+n\leq
u< m+\ell n.$ Let
$\outgammaseq(\ell)=o_0\cdots o_{m+\ell n-1}$ be the output sequence of length $m+\ell n$ such that $o_u=\gamma_u$ if $\sigma_u\in \outalph,$ otherwise $o_u=(\gamma_u,-\infty,\infty).$
Once again, we let  $c_u$ be the {guard} of the $u$-th transition $t_u.$ 
Now, for the given $\ell>0$, we define two neighboring input sequences
$\inalphaseq(\ell)=\inalpha_0\cdots \inalpha_{m+\ell n-1}$ and $\inbetaseq(\ell)=\inbeta_0\cdots \inbeta_{m+\ell n-1}$ each of length $m+\ell n.$

Let $Z'\:= \{\frac{1}{2}\}\cup\{|\mu_u-\mu_{u'}|\::0\leq u,u'<m+n\ell,\:q_u,q_{u'}\in \epsstates,\:\mu_u\neq \mu_{u'}\},$ and  $\Delta\:=\frac{\min(Z')}{m+n\ell}.$ Observe that $\Delta>0.$
Let $Z\:=\{\mu_u\::m\leq u<m+n,\: q_u\in \epsstates\}.$ Now, we define a constant $z$ as follows. If $Z\neq \emptyset$ then $z\:=\min(Z)-\frac{1}{2}$, otherwise $z\:=-\frac{1}{2}.$
Let $U\:=\{u\::q_u\in\epsstates\:,m\leq u<m+n\ell\}$ and $U'\:=\{u\::q_u\in\epsstates\:,0\leq u<m\}.$


Recall that $G_{\absexec_\ell}\:=(V,E)$ is the dependence graph of the run $\absexec_\ell.$ Note that $V\:=\{u\::0\leq u<m+n\ell\}.$ A {\em source node} in $G_{\absexec_\ell}$ is a node that has no incoming edges and a {\em sink node} is a node that has no outgoing edges. The length of a path in $G_{\absexec_\ell}$ is the number of edges on the path. Note that if the path is a single node, then it's length is zero. Observe that the length of any path is less than $m+n\ell.$
We say that a path $p\:=(u_0,...,u_{r})$ in $G_{\absexec_\ell}$,  is a {\em maximal path} iff  either $u_0$ is a source node or $u_0\in U$, and $\forall k_1,\:0<k_1\leq r$, $u_{k_{1}}\notin U.$ For a maximal path $p$, as given above, we define $weight(p)$ to be the value $z'+r\Delta$ where $z'=\mu_{u_{0}}$ if $u_0\in U$, otherwise $z'=z.$ Now, we define a function $\psi$ that associates a real value with each node in $V$ as follows. For $u\in V$, $\psi(u)$ is as given below: if $u\in U,\:\psi(u)=\mu_u$; if $u\notin U$ and is a source node then $\psi(u)=z$; in all other cases, $\psi(u)$ is the maximum weight of a maximal path ending in $u.$ 


From our assumption about $\cA$, we observe that $\absexec_\ell$ is a strongly feasible run. Using this fact, we establish that, if $(u,u')\in E$ then $\psi(u')\geq \psi(u)+\Delta.$ This is shown as follows. If $u'\notin U$ then $\psi(u')$ is the maximum weight of a maximal path ending in $u'.$ If $u''$ is the node just before $u'$ on such a maximal path, then by definition $\psi(u')=\psi(u'')+\Delta$, and further more $\psi(u'')\geq \psi(u)$, and the desired result follows. Now, consider the case, when $u'\in U.$ Now, consider a predecessor $u''$ of $u'$, i.e., $(u'',u')\in E$, such that $\psi(u'')$ is maximum. Clearly $u''\notin U.$ Consider the maximal path ending in $u''$ whose weight is maximum. If this path starts from a node which is a source node, then by definition, we see that the weight of the path is less than $\min(\{\mu_u\::u\in U\})$ and the result follows from this. On the other, if the above maximal path  ending $u''$ starts from a node $w\in U$, we see that $\mu_w < \mu_{u'}$ (because every feasible execution in $\cA$ is strongly feasible); the required result follows from this observation and the fact the  length of the above maximal path is less than $m+n\ell.$  


For each $u,\:0\leq u<m+n\ell$, if $u\in U'\cup U$ then let $a_u:=\tau$, otherwise let $a_u\:=\psi(u)$. For any such $u$, let $X_u$ be the random variable with distribution $\Lap{d_{u}\epsilon,\mu_{u}}$ or $\Lap{d_{u}\epsilon,a_{u}}$, respectively, depending on whether $u\in U'\cup U$ or not. Consider any $u, 0\leq u<m+n\ell,$ such that
$q_u\notin \epsstates.$ (Note that if $q_u\in \epsstates$ then $c_u$ is the condition $\true$.)
The guard $c_u$ is a conjunction of atomic conditions of the form $\svar \geq \rvarx_{k_1}$ or of the form $\svar <\rvarx_{k_1}$ for some $k_1,\: 1\leq k_1\leq k$. Let $u_1<u$ be the maximum integer such that the transition $t_{u_{1}}$ is an assignment transition for the variable $\rvarx_{k_1}.$ Now, in $c_u$, we replace $\svar$ by the random variable  $X_u$ and replace $\rvarx_{k_1}$ by 
 $X_{u_{1}}.$ Let $c'_u$ be the condition obtained by modifying every atomic condition in $c_u$ as specified above. 
Now, let $\cX(\ell)\:=\{X_u\::0\leq u<m+n\ell\}$, $\cC(\ell)\:=\{c'_u\::0\leq u<m+n\ell\}.$  Let $\exec_{\inalphaseq}(\ell)$  denote the computation given by the triple $(\absexec_\ell,\inalphaseq(\ell),\outgammaseq(\ell)).$ 
Now, $\pathprob{\epsilon,\exec_{\inalphaseq}(\ell)}$ is the probability that the random variables in $\cX(\ell)$ satisfy all the guard conditions in $\cC(\ell).$ Let $R\pathprob{\epsilon,\exec_{\inalphaseq}(\ell)}$ be the probability that the random variables in $\cX(\ell)$ satisfy all the guard conditions in $\cC(\ell)$ and $\forall u\in U',\:X_u\in [\psi(u)-\frac{\Delta}{2},\:\psi(u)+\frac{\Delta}{2}].$ For all $u\in U'$, we call the intervals $[\psi(u)-\frac{\Delta}{2},\:\psi(u)+\frac{\Delta}{2}]$ as \emph{bands}. Clearly, 
$\pathprob{\epsilon,\exec_{\inalphaseq}(\ell)}\geq R\pathprob{\epsilon,\exec_{\inalphaseq}(\ell)}.$
Let $C\pathprob{\epsilon,\exec_{\inalphaseq}(\ell)}$ be the conditional probability that the random variables in  $\cX(\ell)$ satisfy all the guard conditions in $\cC(\ell)$ given that
$\forall u\in U',\:X_u\in [\psi(u)-\frac{\Delta}{2},\:\psi(u)+\frac{\Delta}{2}].$
Now, we see that $R\pathprob{\epsilon,\exec_{\inalphaseq}(\ell)}\:=(C\pathprob{\epsilon,\exec_{\inalphaseq}(\ell)}\cdot\prbfn{\forall u\in U',\:X_u\in [\psi(u)-\frac{\Delta}{2},\:\psi(u)+\frac{\Delta}{2}]}).$ It can be easily shown that there exists $\epsilon' >0$, such that $\forall \epsilon >\epsilon'$,
for every $u\in U'$, $\prbfn{X_u\in [\psi(u)-\frac{\Delta}{2},\:\psi(u)+\frac{\Delta}{2}]} \geq \frac{1}{4}\eulerv{-\epsilon d_{u}|\mu_u-\psi(u)-\frac{\Delta}{2}|}.$ From this, we see that, there exists constants $c_1,c_2,\epsilon'>0$, such that $\forall \epsilon >\epsilon'$,
$\prbfn{\forall u\in U',\:X_u\in [\psi(u)-\frac{\Delta}{2},\:\psi(u)+\frac{\Delta}{2}]}\:\geq c_1 \eulerv{-c_2\epsilon}.$ 

Now, we give a lower bound for 
$C\pathprob{\epsilon,\exec_{\inalphaseq}(\ell)}$, for large values of $\epsilon.$ Recall that, each conjunct in $c'_u\in \cC(\ell)$, for $u\notin U\cup U'$, involves two random variables, say $X_{u},X_{u'} \in \cX(\ell).$ We replace each such conjunct in $c'_u$ as follows, if $u'\in U'$; if the conjunct is the atomic condition $X_{u'}\geq X_{u}$ we replace it by $\psi(u')-\frac{\Delta}{2} \geq X_{u}$, otherwise the conjunct is 
 the atomic condition $X_{u'}< X_{u}$ and we replace it by $\psi(u')+\frac{\Delta}{2} < X_{u}.$ The condition $c'_u$ is unchanged if $u'\notin U'.$ Let the resulting set of conditions be denoted by $c''_u.$  Now, let  
$\cC''(\ell)\:=\{c''_u\::0\leq u<m+n\ell\}.$ Now, it is easily seen that $C\pathprob{\epsilon,\exec_{\inalphaseq}(\ell)}$ is greater than or equal to $p_{\ell}$, where $p_{\ell}$ is the probability that the random variables in $\cX(\ell)$ satisfy all the conditions in $\cC''(\ell).$ Now, using similar proof technique as given in \cite{ChadhaSV21,ChadhaSV21b}, we can show that there exists a constant $\epsilon'_\ell$, such that $\forall \epsilon > \epsilon'_\ell$, $p_{\ell}>\frac{1}{2}.$(This is because, for every conjunct of $c''_u$ of the form $X_{u_{1}}\geq X_{u_{2}} $ or of the form $X_{u_{2}}< X_{u_{1}} $, it is the case that $a_{u_{1}}>a_{u_{2}}$. For every conjunct of the form $X_{u_{1}}\leq c' $, we have $a_{u_1} < c'$ and for conjuncts of the form $c'< X_{u_{1}} $, it is the case that $a_{u_{1}}>c'$, where $c'$ is a constant).
Now putting all the above observations together, by taking $\epsilon_\ell = \max(\epsilon',\epsilon'_{\ell})$, we see that $\forall \epsilon >\epsilon_{\ell}$,
$\pathprob{\epsilon,\exec_{\inalphaseq}(\ell)}>\frac{c_{1}}{2}\eulerv{-c_2\epsilon}.$

Now, we define $\inbetaseq(\ell)=\inbeta_0\cdots \inbeta_{m+\ell n-1}.$ To do this, we prove some properties of $\inalphaseq(\ell).$ For each $u,\:0\leq u<m+n\ell$, we define the real value $\ba_u$ as follows: if $u\in U'\cup U$ then $\ba_u\:=\mu_u$, otherwise  $\ba_u\:=a_u$.
from the way we chose the integers $i,j,i'$  and our assumption that the condition 
$\svar \geq \rvarx_{i'}$ is a conjunct of the guard $c_j$, we see that, for every $\ell'$ such that $0\leq \ell' <\ell$ the following properties are satisfied: $t_{i+n\ell'}$ is an assignment transition for $\rvarx_{i'}$;  the condition $\svar \geq \rvarx_{i'}$ is a conjunct in the guard $c_{j+n\ell'}$; for every
$k_1$ such that $i+n\ell'<k_1<j+n\ell'$, $t_{k_{1}}$ is not an assignment transition for $\rvarx_{i'}.$ Now, we fix $\ell'$ to be any integer such that $0\leq \ell' <\ell.$ 
We show below that $\ba_{i+n\ell'} <\ba_{j+n\ell'}\leq \ba_{i+n\ell'}+\frac{1}{2}.$ 



We proceed as follows. First, observe that $j+n\ell' \notin U$, since $c_{j+n\ell'}$ is not the condition $\true.$ 
From our definition, we see that $\ba_{j+n\ell'}$ is the maximum of the weights of maximal paths 
in $G_{\absexec_\ell}$ that end at the node $j+n\ell'.$ Let $(i_0,i_1,...,i_{\ell_{2}})$ be the maximal path in $G_{\absexec_\ell}$ that ends in the node $j+n\ell'$ (i.e., $i_{\ell_{2}}\:=j+n\ell'$) having maximum weight among all such paths. Clearly the length of this path is $\ell_{2}$ and  $\ell_{2}< m+n\ell.$ If the node $i+n\ell'$ lies on the above path then the desired result follows from the definition of $\psi(j+n\ell').$ Now assume that the node $i+n\ell'$ does not lie on the above path. Now, we have two cases. In the
first case, $i\in U$. Clearly $(i,j)\in E$. From the way we chose $i,j,i'$, it is the case that $\mu_i$ is the maximum of all $\mu_u$ such that there is an assign\_refer triple $(u,v,w)$ where $u\in U$ and $\svar\geq \rvarx_{w}$ is a conjunct of $c_v.$ It should be easy to see that $\mu_{i_{0}}=\mu_i$ and 
$\ba_{i+n\ell'}< \ba_{j+n\ell'}<\ba_{i+n\ell'}+\frac{1}{2}.$ Now, consider the case when $i\notin U$ and 
hence $i_0\notin U$. In this case also, we see, from the definition of $\psi$, that $\ba_{j+n\ell'}\leq z+\frac{1}{2}$ and $\ba_{i+n\ell'}\geq z$ and the desired result follows. 


Now, we give the values of $b_u$, for $0\leq u<m+n\ell.$ For every $\ell'$, such that $0\leq \ell'<\ell$,
$b_{j+n\ell'}\:=a_{j+n\ell'}-1$ and for all other values of $u$, $b_u\:=a_u.$
For each $u,\:0\leq u<m+n\ell$, let $\bb_u$ be defined as follows: if    $u\in U'\cup U$ then $\bb_u\:=\mu_u$, otherwise $\bb_u=b_u.$
Since $\ba_{i+n\ell'}<\ba_{j+n\ell'}\leq \ba_{i+n\ell'}+\frac{1}{2}$, we see that $\bb_{j+n\ell'}\leq \bb_{i+n\ell'}-\frac{1}{2}.$
Clearly, the input sequence $\inbetaseq(\ell)$ is a neighbor of $\inalphaseq(\ell).$   Now, using the same analysis, as in \cite{ChadhaSV21,ChadhaSV21b}, it is easily shown that the input sequences 
$\inalphaseq(\ell)$ and $\inbetaseq(\ell)$ are witnesses for violation of the differential privacy property by $\cA.$

In the above definition of the input sequences $\inalphaseq(\ell)$ and $\inbetaseq(\ell)$, we assumed that there exists an assign\_refer triple $(u,v,w)$ such that the condition $\svar\geq \rvarx_{w}$
 is a conjunct in the guard $c_v.$
Now, we give the construction for the other case when no such assign\_refer triple exists, that is,
for every assign\_refer triple $(u,v,w)$, the condition $\svar< \rvarx_{w}$ is a conjunct of $c_v.$
Now, we chose an assign\_refer triple $(i,j,i')$ as follows. If there exists at least one assign\_refer triple $(u,v,w)$ such that $q_u\in \epsstates$, then we chose $(i,j,i')$ to be one such triple so that $\mu_i$ is the minimum among all such triples; otherwise, we chose $(i,j,i')$ to be any of the assign\_refer triples. Now, the proof is similar to the earlier case with the following changes.
 Basically, we change the definitions (in a symmetric way) of  the function $\psi$, of maximal paths and their weights in $G_{\absexec_\ell}\:=(V,E).$ 

The constant $\Delta$ is same as before, i.e., $\Delta\:=\min(Z')$ where  $Z'\:=\{\frac{1}{2}\}\cup\{|\mu_u-\mu_{u'}|\::0\leq u,u'<m+n\ell,\:q_u,q_{u'}\in \epsstates,\:\mu_u\neq \mu_{u'}\}.$ However, the constant $z$ is given as follows. If $Z\neq \emptyset$ then $z:=\max(Z)+\frac{1}{2}$, otherwise,  $z\:=\frac{1}{2}$; here 
$Z\:=\{\mu_u\::m\leq u<m+n,\: q_u\in \epsstates\}.$
As before, $U\:=\{u\::q_u\in\epsstates\:,m\leq u<m+n\ell\}$ and $U'\:=\{u\::q_u\in\epsstates\:,0\leq u<m\}.$

We say that a path $p\:=(u_0,...,u_{r})$ in $G_{\absexec_\ell}$,  is a {\em maximal path} iff  either $u_r$ is the sink  node or $u_r\in U$, and $\forall k_1,\:0\leq k_1< r$, $u_{k_{1}}\notin U.$ For a maximal path $p$, as given above, we define $weight(p)$ to be the value $z'-r\Delta$ where $z'=\mu_{u_{r}}$ if $u_r\in U$, otherwise $z'=z.$ Now, we define the function $\psi$ that associates a real value with each node in $V$ as follows. For $u\in V$, $\psi(u)$ is as given below: if $u\in U,\:\psi(u)=\mu_u$; if $u\notin U$ and is a sink node then $\psi(u)=z$; in all other cases, $\psi(u)$ is the minimum weight of a maximal path starting with $u.$ 

The definition of $\inalphaseq(\ell)$ is same as before with the modified definition of $\psi.$ To define $\inbetaseq(\ell)$, we modify the earlier approach as follows. First, for each integer $\ell'$, such that $0\leq \ell' <\ell$, we  show that  $\ba_{j+n\ell'} <\ba_{i+n\ell'}\leq \ba_{j+n\ell'}+\frac{1}{2}.$ 
Now, the values of $b_u$, for $0\leq u<m+n\ell$ are given as follows. 
For every $\ell'$, such that $0\leq \ell'<\ell$, 
$b_{j+n\ell'}\:=a_{j+n\ell'}+1$ and
for all other values of $u$, $b_u\:=a_u.$ Now, for each $u,\:0\leq u<m+n\ell$, we define $\bb_u$ to be as in the previous case.
Since $\ba_{j+n\ell'}<\ba_{i+n\ell'}\leq \ba_{j+n\ell'}+\frac{1}{2}$, we see that $\bb_{j+n\ell'}\geq \bb_{i+n\ell'}+\frac{1}{2}.$
The proof remains the same as in the previous case with the above changes.

\subsection*{\textbf{{\Criticalpair} implies no privacy}}
The following technical lemma states the properties of the dependency graph of a path that has a non-leaking cycle repeated many times.
\begin{lemma}
\label{lem:criticalptech}
Let $C$ be a non-leaking cycle of  {\dipa} $\cA$ of length $m$  and $\rho\:=\rho'C\rho''$ be a run of $\cA$ starting from the initial state and $\rho_L\:=\rho'C^L\rho''$ be the run in which the cycle $C$ is repeated $L$ times, for some $L>0$. If $(i,j)$ is an edge in $G_{\rho}$ (respectively, $(j,i)$ is an edge) then the following properties hold.
\begin{enumerate}
    \item At most one of the two indices $i,j$ is on the cycle $C$ (i.e., corresponds to a position on $C$).
    \item If $i$ is before $C$ and $j$ is on $C$ then, in $G_{\rho_{L}}$, there are edges from $i$ to the node $j+km$ (respectively, an edge from $j+km$ to $i$), for every $k,\:0\leq k<L.$
    \item If $i$ is after $C$ and $j$ is on $C$ then, in $G_{\rho_{L}}$, there is an edge from $i+m(L-1)$ to $j+m(L-1)$ (respectively, from  $j+m(L-1)$ to  $i+m(L-1)$).
    
\end{enumerate}
\end{lemma}


\begin{proof}

  We make the following observations. An edge $(i,j)$ (or $(j,i)$) in $G_{\rho}$ indicates that the transition in $\rho$ at the position given by $\min(i,j)$, is an assignment transition for some variable $\rvarx_{\ell}$ which is referenced in the guard of the transition in $\rho$ at the position given by  $\max(i,j).$ Property (1) of the lemma follows from this observation and the fact that $C$ is non-leaking cycle. Property (2) of the lemma follows from the fact that 
  the transition of $\rho$ at the position $i$ is an assignment transition for some variable $\rvarx_{\ell}$ which is referenced in the guard of the transition at the position $j$ which is on $C$, and none of the transitions on $C$ is an assignment transition for $\rvarx_{\ell}.$ (Note that $j+km$ is the position in the $k^{th}$ iteration of $C$  in $\rho_L$ that corresponds to position $j$ in $\rho.$) Property (3) follows from the observation that the transition in $\rho$ at position $j$ is an assignment transition for some variable $\rvarx_{\ell}$ which is referenced in the guard of the transition at position $i.$ 
\end{proof}

\begin{lemma}
\label{lem:criticalpnec}
A  {\dipa} $\cA$ is 
not differentially private if  it has a {\criticalpair}. 

\end{lemma}
%

\begin{proof}
Let $\cA= \defaut.$
Assume that $\cA$  has a {\criticalpair}.
From the definition \ref{def:leakingpair}, we see that there exists 
a feasible run $\rn$ of $\cA$ from the initial state $\qinit$ and there are indices $\ell_1,\ell_2,\ell_3$ and $\ell_4$ such that $0 \leq \ell_1 < \ell_2 \leq \len{\rn}$ and $0 \leq \ell_3 < \ell_4 \leq \len{\rn}$ such that the three conditions specified in the definition are satisfied. These conditions state the following. The sub-runs $C = \subseq{\rn}{\ell_1}{\ell_2}$ and $C' = \subseq{\rn}{\ell_3}{\ell_4}$ are both non-{\criticalcycle}s. The cycles $C,C'$ are non-overlapping and may appear in either order, i.e., either $\ell_2\leq\ell_3$ or $\ell_4\leq\ell_1.$
 Further more, there exists a path $i_1,i_2,...,i_{m-1},i_m$ in $G_{\rho}$ such that $i_1$ is on $C$ (i.e., $\ell_1\leq i_1<\ell_2$),$t_{i_{m}}$ is on $C'$ (i.e.,$\ell_3\leq i_m<\ell_4$), $i_2<i_1$ and $i_{m-1}<i_m$.

Let the lengths of $C,C'$ be $n_1,n_2$, respectively; that is $n_1\:=\ell_2-\ell_1$ and $n_2\:=\ell_4-\ell_3.$ We give the proof assuming $C$ appears before $C'$, i.e., $\ell_2\leq \ell_3.$ (The proof for the case when $C'$ appears before $C$, i.e., $\ell_4 \leq \ell_1$ is similar and is left out). Now, we can write the run $\rn$ as $\rn\:= \rn_1 C \rn_2 C' \rn_3$ where $\rn_1=\subseq{\rn}{0}{\ell_1}$,$\rn_2=\subseq{\rn}{\ell_2}{\ell_3}$ and $\rn_3=\subseq{\rn}{\ell_4}{\len{\rn}}.$
Now, for any $L>0$, consider the  run $\rn_L$ in $\cA$ starting
from $\qinit$ in which the cycles $C,C'$ are repeated $L$ times
each. That is, $\rn_L=\:\rn_1 (C)^{L}\rn_2(C')^{L}\rn_3.$  
\ignore{
Formally,  $\absexec_L$ is given by 
$$\begin{array}{lcl}
\absexec_{L}&=& q_0\sigma_0 \cdots q_{\ell_{1}} \sigma_{\ell_{1}} \cdots q_{\ell_{2}+n_1(L-1)} \sigma_{\ell_{2}+n_1(L-1)}\cdots \\
                        && \hspace*{1cm} \cdots q_{\ell_{3}+n_1(L-1)}\sigma_{\ell_{3}+n_1(L-1)}\cdots q_{\ell_{4}+(n_1+n_2)(L-1)} \sigma_{\ell_{4}+(n_1+n_2)(L-1)}\cdots q_{n+(n_1+n_2)(L-1)}
\end{array}$$

Observe that $i_1,i_2,...,i_{m-1},i_m$ is a path in $G_{\rn}$ and hence in $G_{\absexec}$, $i_2<i_1$, $i_1$ is on $C$, $i_{m-1}<i_m$ and $i_m$ is on $C'.$ Hence $i_2$ is before $C$, $i_{m-1}$ is before $C'$. 
}
Using Lemma \ref{lem:criticalptech}, the following are easily observed. 
For every $k',\: 0\leq k'<L$, we have the following edges in $G_{\rn_{L}}$: there is an edge from the node $i_1+k'n_1$ to $i_2$ (note that the former node is a copy of node $i_1$ in iteration $k'$ of $C$); if $i_{m-1}$ is before $C$ then there is an edge from $i_{m-1}$ to the node $(i_m+n_1(L-1)+n_2k')$ (note the later node is a copy of node $i_m$ in iteration $k'$ of $C'$); if $i_{m-1}$ is on or after $C$ then there is an edge from the node $(i_{m-1}+(L-1)n_1)$ to $(i_m+n_1(L-1)+n_2k').$ Further more, using Lemma \ref{lem:criticalptech}, the following can be shown. If $i_{m-1}$ is before $C$ in $\rn$ then there is a path in $G_{\rn_{L}}$ from $i_2$ to $i_{m-1}$, otherwise there is a path in $G_{\rn_{L}}$ from $i_2$ to $(i_{m-1}+n_1(L-1)).$
(Roughly speaking, such a path in  $G_{\rn_{L}}$ can be obtained by taking the path from $i_2$ to $i_{m-1}$ in $G_{\rn}$, and replacing every node in the path that is on $C$ or on $C'$ by the copy of the same node in the last iteration of that cycle in $\rn_L$).
Putting all the above observations together, it is easily seen that,
for every $k',k''$ such that $0\leq k',k''<L$, there is a path in $G_{\rn_{L}}$ from the node $(i_1+k'n_1)$ to the node $(i_m+n_1(L-1)+k''n_2)$; essentially this states that from each node which is a copy of node $i_1$ in every iteration of $C$, there is a path to the copy of node $i_m$ in every iteration of $C'.$

 Now, for each $k$, let $t_k$ be the $k$-th transition of $\rn_L$, $c_k$ be the guard of the $k$-th transition, $q_k\:=\src(t_k)$ and $\sigma_k\:=\otpt(t_k).$ Note that $\rn\:=\rn_L$ when $L=1.$


Let
$\outgammaseq(L)=o_0\cdots o_{n-1+(n_1+n_2)(L-1)}$ be the output sequence of length $n+(n_1+n_2)(L-1)$ such that $o_k=\sigma_k$ if $\sigma_k\in \outalph,$ otherwise $o_k=(\sigma_k,-\infty,\infty).$

Now, we define two adjacent input sequences $\inalphaseq(L)=\inalpha_0\cdots\inalpha_{n-1+(n_1+n_2)(L-1)}$ and
$\inbetaseq(L)\:=\inbeta_0\cdots\inbeta_{n-1+(n_1+n_2)(L-1)}$ as follows. 
We define $\inalphaseq(L)$ similar to the way we defined the corresponding input sequence in the proof of the Lemma \ref{lem:criticalcnec}. To do this, we consider the dependence graph $G_{\rn_{L}}\:=(V,E)$ where $V\:=\{u'\:: 0\leq u'<n+(n_1+n_2)(L-1)\}.$ For any $u'\in V$, let $\mu_{u'}$ be the constant as defined in the proof of Lemma  \ref{lem:criticalcnec}.
Let $Z'\:=\{\frac{1}{2}\}\cup \{|\mu_{u'}-\mu_{u''}|\::u',u''\in V,\:q_{u'},q_{u''}\in \epsstates,\:\mu_{u'}\neq \mu_{u''}\}$ and $\Delta\:=\frac{\min(Z')}{n+(n_1+n_2)(L-1)}.$
Let $Z\:=\{\mu_{u'}\::\ell_{1}\leq u'<\ell_{2}\;or\;\ell_{3}\leq u'<\ell_{4},\: q_{u'}\in \epsstates\})$. Now, we define the constant $z$ as follows. If $Z\neq \emptyset$ then $z\:=\min(Z)-\frac{1}{2}$, otherwise $z\:=-\frac{1}{2}.$ 
Let $U\:=\{u'\::q_{u'}\in\epsstates\:,\ell_{1}\leq u'<\ell_{2}+n_1(L-2)\;or\;\ell_{3}+n_1(L-1)\leq u'<\ell_{4}+(n_1+n_2)(L-2)\}$ and $U'\:=\{u'\::q_{u'}\in\epsstates\:,0\leq u'<n+(n_1+n_2)(L-1), u'\notin U\}.$ We define maximal paths in $G_{\absexec_{L}}$ and weights of the maximal paths as in the proof of lemma \ref{lem:criticalcnec}. 

First, observe that, the set $U$ is the set of all indices corresponding to the non-input states in the first $L-1$ iterations of $C$ and $C'$; the corresponding indices in the $L^{th}$ iteration (last iteration) of $C$ and $C'$ are included in the set $U'$. We make the following additional observations using the fact that $C$ and $C'$ are non-leaking cycles. 
Any variable that is assigned a value in $C$ is not referenced by another transition in $C$; the same holds for $C'$. Also, if a variable that is assigned a value in $C$ in the run $\rn$ is referenced in a transition in $C'$, then in $\rn_L$ only the value assigned in the last iteration of $C$ (i.e., $L^{th}$ iteration ) is referenced in a transition of any of the iterations of $C'$. 

Based on all these observations, we see that every maximal path in $G_{\rn_{L}}$ starts with node $u_0\notin U.$ (Maximal paths of $G_{\rn_{L}}$ and their weights are as defined in the proof of lemma \ref{lem:criticalcnec}). As a consequence, weights of any two maximal paths differ by at most $\frac{1}{2}.$

Now, we define the function $\psi$ that associates a value with each $u'\in V$, exactly as in the proof of lemma \ref{lem:criticalcnec} using the above values of $\Delta,z,U$ and $U'.$

From the way we defined the graph $G_{\rn_{L}}$ and the above observations,  the following properties are easily shown to hold.
 \begin{itemize}
     \item For every $u',u'', 0\leq u',u''<L,$ there is a path, in $G_{\rn_{L}}$,
     from the node $i_1+u'n_1$  to the node $i_m+n_1(L-1)+u''n_2.$
     \item For every $u', 0\leq u'<L,$ every maximal path, in $G_{\rn_{L}}$, that ends in the node $i_1+u'n_1$ or in the node $i_m+n_1(L-1)+u'n_2$, starts from a node that is not in $U.$
     
 \end{itemize}
 Putting all the above observations together, we see that, for every $u',\: 0\leq u'<L-1$, $\psi(i_1+u'n_1)\leq \frac{1}{2}$ and  $\psi(i_m+n_1(L-1)+u'n_2)\leq \frac{1}{2}.$ Further more, for every $u',u''$, such that $0\leq u',u''<L$, there is a path, in $G_{\rn_{L}}$, from the node $i_1+u'n_1$ to the node $i_m+n_1(L-1)+u''n_2$, and $\psi(i_1+u'n_1)<\psi(i_m+n_1(L-1)+u''n_2)\leq \psi(i_1+u'n_1)+\frac{1}{2}.$
 
 Now using proof similar to that of lemma \ref{lem:criticalcnec}, it is easily shown that there exist constants $\epsilon_\ell,c_1,c_2>0$ such that   $\forall \epsilon >\epsilon_{\ell}$,
$\pathprob{\epsilon,\exec_{\inalphaseq}(L)}>\frac{c_{1}}{2}\eulerv{-c_2\epsilon}.$

Now, let $\inbetaseq(L)\:=\inbeta_0\cdots\inbeta_{n-1+(n_1+n_2)(L-1)}$ be such that, for $0\leq u_1<n+(n_1+n_2)(L-1)$, $\inbeta_{u_{1}}$ is defined as follows: if $u_1=i_m+n_1(L-1)+u'n_2$ for some $u',\:0\leq u'<L-1,$ $\inbeta_{u_{1}}\:=\inalpha_{u_{1}}-1$, otherwise $\inbeta_{u_{1}}\:=\inalpha_{u_{1}}.$ It is easily seen that the input sequences $\inalphaseq(L),\inbetaseq(L)$ are neighboring sequences. For each $u_1,\:0\leq u_1<n+(n_1+n_2)L$, let $X_{u_{1}}$ be the random variable with  
the distribution $\Lap{d_{u_{1}}\epsilon,\inbeta_{u_{1}}}.$
Now, for each $u',\:0\leq u'<L-1,$ consider the pair of random variables 
$X_{i_{1}+u'n_{1}}$ and $X_{i_{m}+n_1(L-1)+u'n_{2}}.$ The guard conditions of all the transitions of the execution of $\absexec_L$, with input sequence $\inbetaseq(L)$, imply that for each $u',\:0\leq u'<L-1,\:X_{i_{1}+u'n_{1}}< X_{i_{m}+n_1(L-1)+u'n_{2}}.$  However, from the definition of $\inbetaseq(L)$, we see that $\inbeta_{i_{m}+n_1(L-1)+u'n_{2}}\leq \inbeta_{i_{1}+u'n_{1}}-\frac{1}{2}.$  Now, using the same analysis, as in \cite{ChadhaSV21,ChadhaSV21b}, it is easily shown that the input sequences 
$\inalphaseq(\ell)$ and $\inbetaseq(\ell)$ are witnesses for violation of the differential privacy property by $\cA.$
\end{proof}

\subsection*{\textbf{{\Violatingc}s implies no privacy}}

\begin{lemma}
\label{lem:violatingcnec}
A  {\dipa} $\cA $ is 
not differentially private if  it has a  {\violatingc}.
\end{lemma}
\begin{proof}
Thanks to Lemma~\ref{lem:criticalcnec} and Lemma~\ref{lem:criticalpnec}, we can assume $\cA$ does not have  {\criticalcycle}s or {\criticalpair}s. 
Assume that $\cA$ has a {\violatingc} $C.$ By definition, there is a feasible run in $\cA$ starting from the initial state, having a suffix which is a non-leaking cycle, say cycle $C$, such that $C$
has a transition whose output is $\svar$ or $\svar'.$ We consider the case when $C$ has a transition whose output is $\svar.$  The proof for the case
when $C$ has a transition whose output is $\svar'$ is simpler and
is left out. Now, if the transition of $C$ whose output is $\svar$ has the guard $\true,$ 
then it can be shown easily that repeating the cycle $\ell$ times incurs a privacy cost linear in  
$\ell\epsilon,$ and hence $\cA$ cannot be $\newd\epsilon$-differentially private for any $\newd>0.$
Thus, we consider more interesting case when the guard is a condition other than $\true.$


As indicated above, we consider the case when $C$ has a transition with output  $\svar.$ 
Let $\rn$  be a run of $\cA$ such that $\len{\rn}=j+m$ where $m>0$, $j\geq 0$, $\src(\rho)=q_{init}$, and the following conditions are satisfied: $\subseq{\rn}{j}{\len{\rn}}\:=C$ and 
 and for each $\ell>0$, the run $\subseq{\rn}{0}{j}C^{\ell}$
(i.e., the run obtained by repeating the cycle $C$, $\ell$ times) is feasible.  
Fix $0\leq r<m$ be such that the output of the $(j+r)$-th transition of $\rn$ is $\svar.$
Let the guard of the $(j+r)$-th transition of $\rn$ be the condition $c_r.$ Let $hset \:= \{ \rvarx_j | \svar <\rvarx_j\; \mbox{is a conjunct of }c_r \}$ and $lset \:= \{ \rvarx_j | \svar \geq \rvarx_j\; \mbox{is a conjunct of }c_r \}.$
Observe that, since $C$ is non-leaking cycle, it has no assignment transitions for any of the variables in $lset \cup hset.$

Fix $\ell>0.$ We define the run $\rn_\ell$ starting
from $\qinit$ by repeating the cycle $C$ in $\rn$, $\ell$
times. Formally, $\rn_\ell=\subseq{\rn}{0}{j}C^{\ell}.$ Observe that $\len{\rn_{\ell}}=j+\ell m.$ 
For each $k', \:0\leq k'<j+\ell m$, let $t_{k'},c_{k'}$ be the $k'$-th transition and it's guard in $\rn_{\ell}$, $q_{k'}=\src(t_{k'})$, $\sigma_{k'}=\otpt(t_{k'})$ and $\parf(q_{k'})=(d_{k'},\mu_{k'},d'_{k'},\mu'_{k'}).$ Observe that 
$t_{k'}=t_{k'-m}$ and $\sigma_{k'}=\sigma_{k'-m}$ for $j+m\leq
k'< j+\ell m.$ 
Observe that $\sigma_{j+nm+r}\:=\svar$, for all $n$ such that $0\leq
n<\ell.$

Now we construct two input sequences 
$\inalphaseq(\ell)=a_0\cdots a_{j+\ell m-1}$ and $\inbetaseq(\ell)=b_0\cdots b_{j+\ell m-1}$ as follows. We take $a_{k'}=-\mu_{k'}$, for all $k', 0\leq k'< j+\ell m$ such that $t_{k'}$ is an input transition, otherwise we take $a_{k'}=\tau.$
We take $b_{k'}=-\mu_{k'}-1$ if $k'=j+nm+r$ for some $0\leq n<\ell$ and $b_{k'}=a_{k'}$ otherwise. We also construct an output sequence $O(\ell)=o_0\cdots o_{j+\ell m-1}$ as follows: for all $k'$, $0\leq k'<j+\ell m$, i) $o_{k'}=\sigma_{k'}$ if $\sigma_{k'} \in \Gamma$, ii) $o_{k'}=(\sigma_{k'},0,\infty)$ if $k'=j+nm+r$ for some $0\leq n<\ell$, and iii) $o_{k'}=(\sigma_{k'},-\infty,\infty)$ otherwise.
Let $\kappa(\ell)$ and $\kappa'(\ell)$ respectively be the computations given by the triples $(\rn_{\ell},\inalphaseq(\ell),O(\ell))$ and $(\rn_{\ell},\inbetaseq(\ell),O(\ell)).$
\ignore{
\begin{itemize}
\item $\absexec=\abst(\exec(\ell)),$ 
\item $\inseq(\exec(\ell))=\inalphaseq(\ell),$  and  
\item all $k'$, i) $o_{k'}=\sigma_{k'}$ if $\sigma_{k'} \in \Gamma$, ii) $o_{k'}=(\sigma_{k'},0,\infty)$ if $k'=j+nm+r$ for some $0\leq n<\ell$, and iii) $o_{k'}=(\sigma_{k'},-\infty,\infty)$ otherwise.
\end{itemize} 
Let $\exec'(\ell)=\execlb{j+\ell m}$  be the  path that is equivalent to $\exec$ and $\inseq(\exec'(\ell))=\inbetaseq(\ell).$
}

Let $\psuffix{k'}{\kappa(\ell)}\:=(\psuffix{k'}{\rn(\ell)},\psuffix{k'}{\inalphaseq(\ell)},\psuffix{k'}{O(\ell)})$ and $\psuffix{k'}{\kappa'(\ell)}\:=(\psuffix{k'}{\rn(\ell)},\psuffix{k'}{\inbetaseq(\ell)},\psuffix{k'}{O(\ell)})$, respectively, be the suffixes of the computations $\kappa(\ell)$ and $\kappa'(\ell)$ from the position $k'.$
 Using backward induction on $k'$,i.e., in decreasing values of $k'$, we can easily show that for each evaluation $\psi,$ for the set of variables $\{\rvarx_{j'}| 1\leq j'\leq k\}$, either both 
$\pathprob{\epsilon,\psi,\psuffix{k'}{\kappa(\ell)}}, \pathprob{\epsilon,\psi,\psuffix{k'}{\kappa'(\ell)}} $ are zero (intuitively speaking, this happens when the interval $[u,v]\cap (0,\infty)\:=\emptyset$ where $u$ is the maximum value of a variable in $lset$ and $v$ is the minimum value of a variable in $hset$ for a given evaluation) ,  or both the probabilities are non-zero and  
$$ \pathprob{\epsilon,\psi,\psuffix{k'}{\kappa(\ell)}} = \euler^{ {\#(k') d_{j+r}\epsilon} } \pathprob{\epsilon,\psi,\psuffix{k'}{\kappa'(\ell)}}$$
where $\#(k')$ is the number of indices $k_1$ such that $k'\leq k_1 < j+m\ell-1$ and $k_1=j+nm+r$ for some $0\leq n<\ell.$ 
Thus,  $$\pathprob{\epsilon,\kappa(\ell)}=\euler^{ {\ell d_{j+r}\epsilon} } \pathprob{\epsilon,\kappa'(\ell)}.$$

Now, $\ell$ is arbitrary and hence for every $\newd>0$, there is an $\ell$ such that $\pathprob{\epsilon,\kappa(\ell)}>\euler^{ {\newd\epsilon}} \pathprob{\epsilon,\kappa'(\ell)}.$
Hence $\cA$ is not differentially private. 
\end{proof}

\subsection*{\textbf{{\Violatingp}s implies no privacy}}

\begin{lemma}
\label{lem:violatingpnec}
A  {\dipa} $\cA $ is 
not differentially private if  it has a  {\violatingp}.
\end{lemma}
\begin{proof}
Thanks to Lemma~\ref{lem:criticalcnec},  Lemma~\ref{lem:violatingcnec} and Lemma~\ref{lem:criticalpnec}, we can assume $\cA$ does not have  {\criticalcycle}s, {\violatingc}s or {\criticalpair}s. 
 We give the proof for one of the two cases of a 
{\violatingp}. (The proof for the other
case of the privacy violating path is similar and is leftout.) Specifically, we give the proof when condition (b) of the definition \ref{def:violating} is satisfied; that is, the {\violatingp} $\rn$ which starts from the the initial state $\qinit$ has a non-leaking cycle $C$,
and the dependency graph $G_{\rn}$ has a path of the form $i_1,...,i_m$ where the index $i_1$ (i.e. the transition $\trname(\ith[i_{1}]{\rn})$) is on $C$, $i_2<i_1$ and the transition  $\trname(\ith[i_{m}]{\rn})$ outputs $\svar.$ Since $C$ is not a {\criticalcycle} and $i_2<i_1$, it follows that there is a variable $\rvarx_{k'}$ such that $\trname(\ith[i_{2}]{\rn})$ is an assignment transition for $\rvarx_{k'}$ and the condition $\svar < \rvarx_{k'}$ is a conjunct of the guard of $\trname(\ith[i_{1}]{\rn}).$

Fix $\ell>0.$ Consider the run  $\rn(\ell)$ of length $n$ from the initial state $\qinit$ such that $\rn(\ell)$ is obtained from $\rn$ by repeating the cycle $C$, $\ell$ times.
Let $k_1,k_2,\ldots,k_\ell$ be the indices where the transition $\trname(\ith[i_{1}]{\rn})$ of $\rn$ (which is on the cycle $C$) occurs in $\rn(\ell).$ Similarly, let $r$ be the index  where the transition  $\trname(\ith[i_{m}]{\rn})$ of $\rn$ occurs in $\rn(\ell).$
Let $\parf(\src(\trname(\ith[i]{\rn(\ell)}))=(d_{i },\mu_{i },d'_{i},\mu'_{i})$, for all $i ,\:0\leq i \leq n.$
Next, we construct two input sequences $\inalphaseq(\ell)=a_0\cdots a_{n-1}$ and $\inbetaseq(\ell)=b_0\cdots b_{n-1}$ of length $n$ as follows. 
If the $i$-th transition of $\rn(\ell)$ is a non-input transition then $a_{i }=b_{i }=\tau.$
If  $i \in \set{k_1,k_2,\ldots,k_\ell}$  then $a_{i }=-\mu_{i }$ and $b_{i }=-\mu_{i }+1.$
For all other values of $i$, $a_{i }=b_{i }=-\mu_{i }.$ For $0\leq i<n$, let $\sigma_i\:=\otpt({\trname(\ith[i]{\rn(\ell)}}).$
Let $O(\ell)\:=o_0,...,o_{n-1}$ be the output sequence defined as follows: for all $i $, $0\leq i<n$, i) $o_{i }=\sigma_{i }$ if $\sigma_{i } \in \Gamma$, ii) $o_{i }=(\sigma_{i },-\infty,0)$ if $i =r$, and iii) $o_{i }=(\sigma_{i},-\infty,\infty)$ otherwise. 

\ignore{
Let $\exec(\ell)=\execl{n}$ be the path such that
\begin{itemize}
\item $\absexec=\abst(\exec(\ell)),$ 
\item $\inseq(\exec(\ell))=\inalphaseq(\ell),$  and  
\item for all $i $, i) $o_{i }=\sigma_{i }$ if $\sigma_{i } \in \Gamma$, ii) $o_{i }=(\sigma_{i },-\infty,0)$ if $i =r$, and iii) $o_{i }=(\sigma_{i},-\infty,\infty)$ otherwise. 

\end{itemize}
Let $\exec'(\ell)=\execlb{n}$ be the  path that is equivalent to $\exec(\ell)$ and $\inseq(\exec'(\ell))=\inbetaseq(\ell).$
 }
 Let $\kappa(\ell)$ and $\kappa'(\ell)$ be the computations given by the triples $(\rn(\ell),\inalphaseq(\ell),O(\ell))$ and  $(\rn(\ell),\inbetaseq(\ell),O(\ell))$ respectively.
Please note that in $\kappa(\ell),\kappa'(\ell),$ the $r$-th output (i.e., the value output in $o_r$) is a non-positive number. 
From the construction of $\kappa(\ell),\kappa'(\ell)$, it can be shown that   $$\pathprob{\epsilon, \kappa(\ell)}=\euler^{ {\ell d_{k_1}\epsilon}} \pathprob{\epsilon,\kappa'(\ell)}.$$ 
As in the case of {\violatingc} (See Lemma~\ref{lem:violatingcnec}),  we can conclude that $\cA$ is not differentially private. 
\end{proof}

\section{\textbf{PSPACE-hardness of checking well-formedness}}
\label{app:pspacehard}
In this sub-section we show that the problem of checking well-formedness of a {\dipa}.
\begin{lemma}
\label{lem:pspace-hard}
The problem of checking if a given  output-distinct {\dipa}  is well-formed is PSPACE-hard.
 
\end{lemma}

\begin{proof}
We prove the lemma by giving a polynomial time reduction from the problem of checking whether a polynomial space bounded single tape Turing Machine (TM) halts on a given input. More specifically, given a single tape TM $M$ that is polynomial space bounded and an input $u$, we give a polynomial time algorithm that outputs a {\dipa} $\cA$ so that $M$ halts on the input $u$ iff $\cA$ has no {\criticalcycle}. 

Let $M$ be the given TM. Without loss of generality, we assume that the input alphabet of $M$ is given by 
$\Sigma\:=\{0,1\}$ and it's tape alphabet $\Upsilon\:=\Sigma\cup \{B\}$ where $B$ is the blank symbol. Let $M$ be given by the 4-tuple $(R,\transf',r_{init},r_{halt})$ where $R$ is the set of it's control states; $r_{init},r_{halt}\in R$ are the initial and halting states respectively and
$$\transf'\::(R-\{r_{halt}\})\times\Upsilon \to R\times \Sigma \times \{Left,Right\}$$
is the transition function.
Intuitively, if $\transf'(r,a)\:=(r',b,d)$, where $d\in \{Left,Right\}$ then $M$ when in the control state $r$ scanning the symbol $a$ in the cell pointed by it's head, it writes the symbol $b$ into the cell and moves it's head in the direction given by $d.$ We assume that if $M$ tries to move it's head further left of the left most cell then it stays in the same position. Without loss of generality, we assume that in each transition, $M$ always writes a value $0$ or $1$ into the current cell its scans; the value it writes may be the same value it read or is a different value. 
Notice that when in the control state $r_{halt}$, $M$  halts, i.e., no transitions are defined from the state $r_{halt}.$ 
We assume that $M$ uses at most $p(n)$ space on any input of length $n$, where $p(n)$ is a polynomial in $n.$ 

Let $u=u_0,\ldots,u_{n-1}$ be the given input to $M.$ Let $N=p(n).$ Now, we give the construction of the automaton 
$\cA\:=\defaut$ as follows. The set of store variables $S\:=\{x\}\cup \{y_i,z_i\: 0\leq i<N\}.$
The set $\states\:=$
$$\{q_j,q'_j\::0\leq j\leq N\}\cup \\
\{(r,i),\:(r',i),\:(r'',i)\::0\leq i<N, r\in R\}.$$ Further more, $\qinit=q_0$, $\outalph\:=\{\top,\bot\}$,$\parf(s)\:=(1,0,1,0)$ for all $s\in\states$ and $\transf$, which defines the transitions of $\cA$  is defined as follows.
  First we define the intuition into the definition of $\transf$ and the working of $\cA.$
  The variables $y_i,z_i$ together with $x$ are used to denote the contents of the $i$-th tape cell of $M.$
  More specifically, the satisfaction of the conditions $z_i<x$, $z_i\geq x$, denote that the cell $i$ has blank symbol (i.e., symbol B) and non-blank symbol, respectively; similarly, satisfaction of the conditions $y_i<x$, $y_i\geq x$, denote that cell $i$ contains $0$  and $1$, respectively. We have a transition from $q_0$ to $q_1$ with guard $\true$ that assigns $\svar$ to $x$ and  outputs $\bot$; this transition initializes $x.$

  We have the following transitions that capture the fact that the first $n$ cells of the tape contain non-blank input symbols, and the remaining cells contain blank symbols.
  For each $j$,$1\leq j\leq n$, we have two transitions: (i) from $q_j$ to $q'_j$ with guard $\svar\geq x$ and with assignment to variable $z_{j-1}$; (ii)  from $q'_j$ to $q_{j+1}$ with guard $\svar<x$ (resp., with guard $\svar \geq x$) when $u_{j-1}=0$ (resp., when $u_{j-1}=1$)  with assignment to variable $y_{j-1}.$
  For each $j$, $n+1\leq j<N$, we have a transition from $q_j$ to $q_{j+1}$ with guard $\svar<x$ and with assignment to $z_{j-1}.$ We also have a transition from $q_N$ to the state $(r_{init},0)$ with guard $\svar<x$ and with assignment to $z_{N-1}.$ All these transitions output $\bot.$

When the automaton $\cA$ is in the states of the form $(r,i)$ ($r\in R,0\leq i<N$), it simulates $M.$ For $r\in R$ and for each $i,0\leq i<N$, $\cA$ has the following transitions.
\begin{itemize}
\item If $\transf'(r,B)\:=(s,b,d)$ then we have the following three transitions in $\cA$: (i) there is a transition 
from $(r,i)$ to $(r',i)$ with guard $z_i<\svar \wedge \svar\leq x$ that outputs $\bot$; (ii) there is a transition from $(r',i)$ to $(r'',i)$ with guard $\svar\geq x$ that assigns $\svar$ to $z_i$ and outputs $\top$ (iii) there is a transition from $(r'',i)$ to $(s,j)$ that assigns $\svar$ to $y_i$, and where $j=i+1$ if $d=Right$ and $j=i-1$ if $d=Left$; if $b=0$ then the guard of the transition is $\svar< x$, otherwise the guard is $\svar\geq x.$

\item If $\transf'(r,a)\:=(s,b,d)$, where $a\neq B$, then we have the following three transitions in $\cA$:
(i) there is a transition from $(r,i)$ to $(r',i)$, if $a=0$ then the guard of the transition is $\svar \geq y_i \wedge \svar<x$ and it's output is $\bot$, otherwise it's guard is $\svar \geq x \wedge \svar<y_i$ and it outputs $\top$; (ii) there is a transition from $(r',i)$ to $(s,j)$ that assigns $\svar$ to $y_i$, if $b=0$ then the guard of the transition is $\svar <x$ and it outputs $\bot$, otherwise the guard is $\svar\geq x$ and it outputs $\top$; further more,$j=i+1$ if $d=Right$ and $j=i-1$ if $d=Left.$ 
 \end{itemize}

 It can easily be shown  that $M$ halts on the input $u$ iff  $\cA$ is well-formed. \end{proof}
 
 \ignore{Proof sketch to be completed later: A configuration of $M$ is triple of the form $(r,i,v)$ where $r\in R, \: 0\leq i<N$ and $v\in \{0,1,B\}^N.$ The configuration $(r_{init},0,uB^{N-n})$ is the initial configuration of $M.$ The unique computation $M$ , on input $u$, is a sequence of configurations starting from the initial configuration computation such that each successive configuration is the resulting configuration when one step of $M$ is executed from the previous configuration.
 An augmented state of $\cA$ is a pair $(q,C)$ where $q$ is a state of $\cA$ and $C$ is a set of inequalities of the form $x<y_i,x\geq y_i, x<z_i$ and $x\geq z_i$ such that for each $i, 0\leq i<N$, exactly one of the two inequalities $x<y_i,x\geq y_i$ and exactly one of the two inequalities $x<z_i,x\geq z_i$ are in $C.$
 
 There is a unique run of $\cA$ starting from it's initial state $qinit$ and which
 }

\section{Details of Examples used in the Experiments}
\label{app:experiments}

\subsection{Pseuedocode of examples in experiments}
We present the pseudocode of the examples described in Section~\ref{sec:expdesc}.

\RestyleAlgo{boxed} 
\begin{algorithm}
\DontPrintSemicolon
\SetAlgoLined

\KwIn{$q[1:N]$}
\KwOut{$out[1:N]$}
\;
$\s{threshold} \gets \Lap{ \frac{\epsilon}{4} , T}$\;

\For{$i\gets 1$ \KwTo $N$}
{
    $\rv\gets \Lap{\frac{\epsilon}{2} , q[i]}$\;
    \uIf{$(\rv \geq \s{threshold})$}{
      $out[i] \gets \top $\;
      exit}
    \Else{
      $out[i] \gets \bot$\;
    }
}

\caption{{\SVT}. {\SVT} is differentially private. In experiments, the (non-private) $\s{threshold}$ is set to $0$.}
\label{fig:svt-algorithm}
\end{algorithm}


\label{ex:numeric-sparse-algorithm}

\RestyleAlgo{boxed} 
\begin{algorithm}
\DontPrintSemicolon
\SetAlgoLined

\KwIn{$q[1:N]$}
\KwOut{$out[1:N]$}
\;
$\s{threshold} \gets \Lap{ \frac{\epsilon}{4} , T}$\;

\For{$i\gets 1$ \KwTo $N$}
{
    $\rv\gets \Lap{\frac{\epsilon}{2} , q[i]}$\;
    \uIf{$(\rv \geq \s{threshold})$}{
      $out[i] \gets  \Lap{\frac{\epsilon}{2}} $\; exit}
      
    \Else{
      $out[i] \gets \bot$\;
    }
}

\caption{{\Spse}. {\Spse} is differentially private. In experiments, the (non-private) $\s{threshold}$ is set to $0$.}
\label{fig:numeric-sparse-algorithm}
\end{algorithm}





\RestyleAlgo{boxed} 
\begin{algorithm}
\DontPrintSemicolon
\SetAlgoLined

\KwIn{$q[1:N]$}
\KwOut{$out[1:N]$}
\;
$\s{x1} \gets \Lap{ \frac{\epsilon}{2} , T_\ell}$\;
$\s{x2} \gets \Lap{ \frac{\epsilon}{2} , T_u}$\;

\For{$i\gets 1$ \KwTo $N$}
{
    $\rv\gets \Lap{\frac{\epsilon}{2} , q[i]}$\;
    \uIf{$((\rv < \s{x1}) \wedge (\rv > \s{x2}))$}{
      $out[i] \gets \top $}
    \uElseIf{$(\rv < \s{x2})$}{
        $\s{x2} \gets \rv$\;
      $out[i] \gets \bot$\;
    }
}

\caption{The example {\LC}. {\LC} is not differentially private as it has a leaking cycle. In the experiments, $T_\ell$ and $T_u$ are taken to be $0$ and $1$, respectively. }
\label{fig:leakingcycle}
\end{algorithm}



\RestyleAlgo{boxed} 
\begin{algorithm}
\DontPrintSemicolon
\SetAlgoLined

\KwIn{$q[1:N]$}
\KwOut{$out[1:N]$}
\;
$\s{low} \gets \Lap{ \frac{\epsilon}{4} , T_\ell}$\;
$\s{high} \gets \Lap{ \frac{\epsilon}{4}, T_u}$\;
\For{$i\gets 1$ \KwTo $N$}
{
    $\rv\gets \Lap{\frac{\epsilon}{4} , q[i]}$\;
    \uIf{$(\rv \geq \s{low}) \wedge (\rv < \s{high})$}{
      $out[i] \gets \rv $}
    \uElseIf{$(\rv \geq \s{low}) \wedge (\rv \geq \s{high})$}{
      $out[i] \gets \top$\;
      exit
    } \ElseIf{$(\rv < \s{low}) \wedge (\rv < \s{high})$}{
      $out[i] \gets \bot$\;
      exit
    }
}

\caption{The example {\DC}. {\DC} is not differentially private as it has a disclosing cycle. In the experiments, $T_\ell$ and $T_u$ are taken to be $0$ and $1$, respectively. }
\label{fig:range-query-disclosing-cycle}
\end{algorithm}


\RestyleAlgo{boxed} 
\begin{algorithm}
\DontPrintSemicolon
\SetAlgoLined

\KwIn{$q[1:N]$}
\KwOut{$out[1:N]$}
\;
$\s{low} \gets \Lap{ \frac{\epsilon}{4} , T_\ell}$\;
$\s{high} \gets \Lap{ \frac{\epsilon}{4}, T_u}$\;
\For{$i\gets 1$ \KwTo $N$}
{
    $\rv\gets \Lap{\frac{\epsilon}{4} , q[i]}$\;
    \uIf{$(\rv \geq \s{low}) \wedge (\rv < \s{high})$}{
      $out[i] \gets \bot$}
    \uElseIf{$(\rv \geq \s{low}) \wedge (\rv \geq \s{high})$}{
      $out[i] \gets \rv$\;
      exit
    } \ElseIf{$(\rv < \s{low}) \wedge (\rv < \s{high})$}{
      $out[i] \gets \top$\;
      exit
    }
}

\caption{The algorithm {\NumericRangeOne}.
{\NumericRangeOne} is not differentially private as it has a privacy-violating path. In the experiments, $T_\ell$ and $T_u$ are taken to be $0$ and $1$, respectively. }
\label{fig:numeric-range-with-pvp}
\end{algorithm}

\RestyleAlgo{boxed} 
\begin{algorithm}[ht]
\DontPrintSemicolon
\SetAlgoLined

\KwIn{$q[1:N]$}
\KwOut{$out[1:N]$}
\;
$\s{low} \gets \Lap{ \frac{\epsilon}{4} , T_\ell}$\;
$\s{high} \gets \Lap{ \frac{\epsilon}{4}, T_u}$\;
\For{$i\gets 1$ \KwTo $N$}
{
    $\rv\gets \Lap{\frac{\epsilon}{4} , q[i]}$\;
    \uIf{$(\rv \geq \s{low}) \wedge (\rv < \s{high})$}{
      $out[i] \gets \bot$}
    \uElseIf{$(\rv \geq \s{low}) \wedge (\rv \geq \s{high})$}{
      $out[i] \gets \Lap{\frac{\epsilon}{4} , q[i]}$\;
      exit
    } \ElseIf{$(\rv < \s{low}) \wedge (\rv < \s{high})$}{
      $out[i] \gets \top$\;
      exit
    }
}

\caption{{\NumericRangeTwo}. {\NumericRangeTwo} is differentially private. In the experiments, $T_\ell$ and $T_u$ are taken to be $0$ and $1$, respectively. }
\label{fig:numeric-range-fixed}
\end{algorithm}


\begin{algorithm}
\DontPrintSemicolon
\SetAlgoLined

\KwIn{$q[1:N]$}
\KwOut{$out[1:N]$}
\;
$\s{u} \gets \Lap{ \frac{\epsilon}{4}, T_{\ell}}$\;

$\s{v} \gets \Lap{ \frac{\epsilon}{4} , T_m}$\;

$\s{w} \gets \Lap{ \frac{\epsilon}{4}, T_u}$\;

\For{$i\gets 1$ \KwTo $N$}
{
    $\rv\gets \Lap{\frac{\epsilon}{4} , q[i]}$\;
    
    \uIf{$(\rv \geq \s{u}) \wedge (\rv < \s{v})$}{
      $out[i] \gets \s{cont}$\;
      }
    \uElseIf{$(\rv < \s{u})$}{
      $out[i] \gets \bot$\;
      exit\;
    } \ElseIf{$(\rv > \s{v}) \wedge (\rv < \s{w})$}{
      $out[i] \gets \top$\;
      break \;
    }
}

\For{$i\gets i+1$ \KwTo $N$}
{
    $\rv\gets \Lap{\frac{\epsilon}{4} , q[i]}$\;
    
    \uIf{$(\rv \geq \s{v}) \wedge (\rv < \s{w})$}{
      $out[i] \gets \s{cont}$\;
      }
    \uElseIf{$(\rv < \s{v})$}{
      $out[i] \gets \bot$\;
      exit
    } \ElseIf{$(\rv > \s{w})$}{
      $out[i] \gets \top$\;
      exit \;
    }
}

\caption{{\TRangeOne} algorithm. {\TRangeOne} is not differentially private as it has a leaking pair. In the experiments, the thresholds $T_\ell, T_m$ and $T_u$ are chosen as $0,1,$ and $2$, respectively}
\label{fig:TRangeOne}
\end{algorithm}

\begin{algorithm}
\DontPrintSemicolon
\SetAlgoLined

\KwIn{$q[1:N]$}
\KwOut{$out[1:N]$}
\;
$\s{u} \gets \Lap{ \frac{\epsilon}{4} , T_\ell}$\;
$\s{v} \gets \Lap{ \frac{\epsilon}{4}, T_m}$\;
$\s{w} \gets \Lap{ \frac{\epsilon}{4}, T_u}$\;

\For{$i\gets 1$ \KwTo $N$}
{
    $\rv\gets \Lap{\frac{\epsilon}{4} , q[i]}$\;
    
    \uIf{$(\rv \geq \s{u}) \wedge (\rv < \s{v})$}{
      $out[i] \gets \s{cont}$\;
      }
    \uElseIf{$(\rv < \s{u})$}{
      $out[i] \gets \bot$\;
      exit
    } \ElseIf{$(\rv > \s{v}) \wedge (\rv < \s{w})$}{
      $out[i] \gets \top$\;
      break \;
    }
}

$\s{v} \gets \Lap{ \frac{\epsilon}{4} , T_}$\;

\For{$i\gets i+1$ \KwTo $N$}
{
    $\rv\gets \Lap{\frac{\epsilon}{4} , q[i]}$\;
    
    \uIf{$(\rv \geq \s{v}) \wedge (\rv < \s{w})$}{
      $out[i] \gets \s{cont}$\;
      }
    \uElseIf{$(\rv < \s{v})$}{
      $out[i] \gets \bot$\;
      exit
    } \ElseIf{$(\rv > \s{w})$}{
      $out[i] \gets \top$\;
      exit \;
    }
}

\caption{{\TRangeTwo}. {\TRangeTwo} is differentially private. In the experiments, the thresholds $T_\ell, T_m$ and $T_u$ are chosen as $0,1,$ and $2$, respectively.}
\label{fig:TRangeTwo}
\end{algorithm}

\newpage

\subsection{Raw data for graph plots}

Table~\ref{table:kminmax} and Table~\ref{table:m-svt-range} give the running times for $k$-{\minmax} and $m$-{\Range} in our experiments. 
We also report the time taken to compute the weight. As we can see, it is minuscule compared to the total running time.

\begin{table}[hb]\setlength\tabcolsep{3.5pt}
    \caption{Experimental result of $m$-{\Range} examples. The columns vars, states, and transitions give the number of variables  of states and transitions in the example. The column wt calculation time gives the average running time in seconds, and running time gives the average running time in seconds. The running times are averaged over 6 executions. In each case, {\toolaut}, returns that the automaton is differentially private with weight $1.$}
    \label{table:m-svt-range}
    \centering
    \begin{tabular}{|l|l|l|l|l|l|}
    \hline
        m & vars & \# states & \# transitions & wt calc. time & running time \\ \hline
        1 & 2 & 4 & 5 & 0.001s & 0.227s \\ \hline
        2 & 4 & 7 & 10 & 0.001s & 0.225s \\ \hline
        3 & 6 & 10 & 15 & 0.002s & 0.242s \\ \hline
        4 & 8 & 13 & 20 & 0.003s & 0.234s \\ \hline
        5 & 10 & 16 & 25 & 0.003s & 0.258s \\ \hline
        6 & 12 & 19 & 30 & 0.004s & 0.3s \\ \hline
        7 & 14 & 22 & 35 & 0.005s & 0.354s \\ \hline
        8 & 16 & 25 & 40 & 0.006s & 0.392s \\ \hline
        9 & 18 & 28 & 45 & 0.007s & 0.467s \\ \hline
        10 & 20 & 31 & 50 & 0.008s & 0.611s \\ \hline
        11 & 22 & 34 & 55 & 0.009s & 0.735s \\ \hline
        12 & 24 & 37 & 60 & 0.01s & 0.894s \\ \hline
        13 & 26 & 40 & 65 & 0.012s & 1.061s \\ \hline
        14 & 28 & 43 & 70 & 0.012s & 1.259s \\ \hline
        15 & 30 & 46 & 75 & 0.014s & 1.467s \\ \hline
        16 & 32 & 49 & 80 & 0.015s & 1.737s \\ \hline
        17 & 34 & 52 & 85 & 0.017s & 2.085s \\ \hline
        18 & 36 & 55 & 90 & 0.018s & 2.47s \\ \hline
        19 & 38 & 58 & 95 & 0.02s & 2.911s \\ \hline
        20 & 40 & 61 & 100 & 0.021s & 3.469s \\ \hline
        25 & 50 & 76 & 125 & 0.029s & 7.332s \\ \hline
        30 & 60 & 91 & 150 & 0.04s & 13.466s \\ \hline
        35 & 70 & 106 & 175 & 0.049s & 22.775s \\ \hline
        40 & 80 & 121 & 200 & 0.062s & 35.894s \\ \hline
        45 & 90 & 136 & 225 & 0.077s & 56.14s \\ \hline
        50 & 100 & 151 & 250 & 0.09s & 83.05s \\ \hline
        55 & 110 & 166 & 275 & 0.12s & 114.16s \\ \hline
        60 & 120 & 181 & 300 & 0.145s & 156.80s \\ \hline
        65 & 130 & 196 & 325 & 0.167s & 207.30s \\ \hline
        70 & 140 & 211 & 350 & 0.224s & 286.87s \\ \hline
        75 & 150 & 226 & 375 & 0.284s & 402.43s \\ \hline
        80 & 160 & 241 & 400 & 0.259s & 506.33s \\ \hline
    \end{tabular}
\end{table}

\newpage

\begin{table}[ht]\setlength\tabcolsep{3.5pt}
       
\caption{Experimental result of $k$-{\minmax} examples. The columns vars, states, and transitions give the number of variables of states and transitions in the example. The column wt calculation time gives the average running time in seconds, and running time gives the average running time in seconds. The running times are averaged over six executions. In each case, {\toolaut}, returns that the automaton is differentially private with weight $1.$ }
 \label{table:kminmax}
    \centering
    \begin{tabular}{|l|l|l|l|l|l|}
    \hline
         k & vars & \# states & \# transitions & wt. calc time & running time \\ \hline
        2 & 2 & 4 & 7 & 0.001s & 0.22s \\ \hline
        3 & 2 & 5 & 10 & 0.001s & 0.22s \\ \hline
        4 & 2 & 6 & 13 & 0.001s & 0.223s \\ \hline
        5 & 2 & 7 & 16 & 0.001s & 0.223s \\ \hline
        6 & 2 & 8 & 19 & 0.001s & 0.226s \\ \hline
        7 & 2 & 9 & 22 & 0.002s & 0.227s \\ \hline
        8 & 2 & 10 & 25 & 0.002s & 0.228s \\ \hline
        9 & 2 & 11 & 28 & 0.002s & 0.23s \\ \hline
        10 & 2 & 12 & 31 & 0.002s & 0.23s \\ \hline
        11 & 2 & 13 & 34 & 0.002s & 0.234s \\ \hline
        12 & 2 & 14 & 37 & 0.003s & 0.234s \\ \hline
        13 & 2 & 15 & 40 & 0.003s & 0.236s \\ \hline
        14 & 2 & 16 & 43 & 0.003s & 0.238s \\ \hline
        15 & 2 & 17 & 46 & 0.003s & 0.24s \\ \hline
        16 & 2 & 18 & 49 & 0.004s & 0.241s \\ \hline
        17 & 2 & 19 & 52 & 0.004s & 0.244s \\ \hline
        18 & 2 & 20 & 55 & 0.004s & 0.246s \\ \hline
        19 & 2 & 21 & 58 & 0.004s & 0.247s \\ \hline
        20 & 2 & 22 & 61 & 0.004s & 0.248s \\ \hline
        30 & 2 & 32 & 91 & 0.007s & 0.264s \\ \hline
        40 & 2 & 42 & 121 & 0.01s & 0.282s \\ \hline
        50 & 2 & 52 & 151 & 0.012s & 0.299s \\ \hline
        60 & 2 & 62 & 181 & 0.015s & 0.317s \\ \hline
        70 & 2 & 72 & 211 & 0.019s & 0.335s \\ \hline
        80 & 2 & 82 & 241 & 0.022s & 0.365s \\ \hline
        90 & 2 & 92 & 271 & 0.025s & 0.386s \\ \hline
        100 & 2 & 102 & 301 & 0.029s & 0.409s \\ \hline
        110 & 2 & 112 & 331 & 0.034s & 0.44s \\ \hline
        120 & 2 & 122 & 361 & 0.038s & 0.448s \\ \hline
        130 & 2 & 132 & 391 & 0.044s & 0.476s \\ \hline
        140 & 2 & 142 & 421 & 0.047s & 0.492s \\ \hline
        150 & 2 & 152 & 451 & 0.052s & 0.515s \\ \hline
        160 & 2 & 162 & 481 & 0.057s & 0.54s \\ \hline
        170 & 2 & 172 & 511 & 0.062s & 0.565s \\ \hline
        180 & 2 & 182 & 541 & 0.068s & 0.583s \\ \hline
        190 & 2 & 192 & 571 & 0.074s & 0.609s \\ \hline
        200 & 2 & 202 & 601 & 0.08s & 0.643s \\ \hline
    \end{tabular}

\end{table}

\fi
\end{document}